\documentclass[preprintnumbers,nofootinbib,noshowpacs,eqsecnum,prd,superscriptaddress]{revtex4}

\usepackage{graphicx,epsfig}
\usepackage{amsmath,amssymb,bbold,amsbsy,psfrag,color,axodraw,dsfont,ulem}

\graphicspath{ {Plots/} }

\makeatletter       

\renewcommand{\thesection}{\arabic{section}}
\renewcommand{\p@subsection}{}

\makeatother

\newcommand{\onehalf}{\ensuremath{\frac{1}{2}}}

\newcommand{\M}{\mathcal{M}}

\newcommand{\qw}{\ensuremath{\tilde{q}}}
\newcommand{\fw}{\ensuremath{\tilde{f}}}

\newcommand{\cw}{\ensuremath{\tilde{\chi}}}
\newcommand{\ww}{\ensuremath{\tilde{W}}}
\newcommand{\bw}{\ensuremath{\tilde{B}}}

\def\dis{\displaystyle}
\def\beq{\begin{equation}}
\def\eeq{\end{equation}}
\def\barr{\begin{array}}
\def\earr{\end{array}}

\begin{document}
\def\lsim{\:\raisebox{-0.5ex}{$\stackrel{\textstyle<}{\sim}$}\:}
\def\gsim{\:\raisebox{-0.5ex}{$\stackrel{\textstyle>}{\sim}$}\:}

\title{ \hfill{\footnotesize{CERN-PH-TH/2010-041}}      \\[9mm]
        DIRAC NEUTRALINOS AND ELECTROWEAK SCALAR BOSONS \\
        OF \boldmath{$N{=}1/N{=}2$} HYBRID SUPERSYMMETRY AT COLLIDERS \\[3mm]
}

\author{S.~Y.~Choi}
\address{Department of Physics and RIPC, Chonbuk National University,
               Jeonju 561-756, Korea}
\author{D.~Choudhury}
\address{Department of Physics and Astrophysics, University of Delhi,
               Delhi 110007, India}
\author{A.~Freitas}
\address{Department of Physics \& Astronomy, University of Pittsburgh,
               3941 O'Hara St, PA 15260, USA}
\author{J.~Kalinowski}
\address{Faculty of Physics, University of Warsaw, 00681 Warsaw,
               Poland, and \\
               Theory Division, CERN, CH-1211 Geneva 23, Switzerland}
\author{J.~M.~Kim}
\address{Physikalisches Institut and Bethe Center for Theoretical
               Physics, Universit\"at Bonn, Nussallee 12, D-53115
               Bonn, Germany}
\author{P.~M.~Zerwas}
\address{Institut f\"ur Theoretische Teilchenphysik und Kosmologie,
               RWTH Aachen University, D-52074 Aachen, Germany, and \\
               Deutsches Elektronen-Synchrotron DESY, D-22603 Hamburg, Germany}

\date{\today}

\begin{abstract}
  {\it In the $N{=}1$ supersymmetric extension of the Standard Model,
       neutralinos associated in supermultiplets with the neutral
       electroweak gauge and Higgs bosons are, as well as gluinos,
       Majorana fermions. They can be paired with the Majorana fermions
       of novel gaugino/scalar supermultiplets, as suggested by extended $N{=}2$
       supersymmetry, to Dirac particles.
       Matter fields are not extended beyond the standard $N{=}1$ supermultiplets
       in $N{=}1$/$N{=}2$ hybrid supersymmetry to preserve the chiral character
       of the theory. Complementing earlier analyses in  the color sector,
       central elements of such an electroweak scenario are analyzed in the
       present study. The decay properties of the Dirac fermions
       ${\tilde{\chi}}_D$ and of the scalar bosons $\sigma$ are worked out,
       and the single and pair production-channels of the new particles are
       described for proton collisions at the LHC, and electron/positron
       and $\gamma\gamma$ collisions at linear colliders.
       Special attention is paid to modifications of the Higgs sector,
       identified with an $N{=}2$ hypermultiplet,
       by the mixing with the novel electroweak scalar sector.
  }
\end{abstract}

\maketitle

\mbox{ } \\[-1.5cm]

\section{Introduction}
\label{sec:introduction}

\noindent
Neutral electroweak gauge bosons are described by self-conjugate fields
with two degrees of freedom before symmetry breaking. In $N{=}1$
supersymmetry \cite{Wess,Nilles,Drees} the gaugino partners ${\tilde{G}}$
of the gauge bosons $G_\mu$ in the supermultiplets
$\hat{G}=\{G_\mu,\tilde{G}\}$ are correspondingly self-conjugate
Majorana fields with two independent components for
the two helicities. They mix with neutral higgsinos to form the
neutralino fields $\tilde{\chi}^0$. However, in $N{=}2$ extended
supersymmetric scenarios, {\it cf.} Ref.~\cite{Fayet,Gaum,no,us1,us2,Plehn,
Benakli,Goodsell}, gauginos ${\tilde{G}'}$ with scalar partners
$\sigma$ are introduced in novel $N{=}1$ chiral supermultiplets
${\hat{\Sigma}}=\{{\tilde{G}}', \sigma\}$, which together with the original
$N{=}1$ gauge supermultiplets constitute the $N{=}2$ gauge hypermultiplets
${\mathcal{G}} = \{{\hat{G}},{\hat{\Sigma}} \}$.  For suitable mass
matrices, the new gauginos can be combined with the original ones to
form Dirac fields ${\tilde{G}}_D = \tilde{G} \oplus {\tilde{G}}'$.
Including the higgsinos, the neutralino fields can thus be identified
with Dirac fields ${\tilde{\chi}}^0_D$.

The transition from Majorana to Dirac fields renders the theory
[partially] $R$-symmetric \cite{Rsym}. $R$-symmetry, a continuous
extension of the $R$-parity concept, is associated with global
transformations of the fermionic coordinates, $\theta\to e^{i\alpha}
\theta$ and $\bar{\theta}\to e^{-i\alpha}\bar{\theta}$. All Standard
Model (SM) fields carry vanishing $R$-charge. Assigning the
$R$-charge $+1$ to $\theta$,  the gauge superfields and the matter
chiral superfields carry $R$-charges $0$ and $+1$, respectively. As
a result, the kinetic part of the action is $R$-symmetric. In gauge
superfields the $R$-charges of the gaugino components $\tilde{G}$
are $+1$, and equally for the scalar components of the matter lepton
and quark superfields. Higgs superfields are assigned $R$-charges 0,
giving rise to $\{ 0,-1 \}$ for the $R$-charges of the Higgs fields
themselves and the higgsino fields. Thus the tri-linear Yukawa terms
in the superpotential carry $R$-charge $+2$ and the corresponding
action is $R$-invariant, unlike the $\mu$-term for which the
associated action, with $R = -2$, is not $R$-invariant. Soft Majorana
mass terms of gauginos and the tri-linear scalar coupling terms,
which break supersymmetry, carry $R = +2$ so that the corresponding
Lagrangians are not $R$-invariant. However, assigning $R = 0$ to the
chiral superfields $\hat{\Sigma}$, the new gaugino $\hat{G}'$ fields
carry $R$-charge $-1$. Thus, Dirac mass terms, combining the old and
the new gaugino fields, are $R$-invariant.

The conservation of $R$-charges, initially motivated by the
transition from Majorana to Dirac gauginos, has important physical
implications. The theory naturally suppresses
the baryon and lepton number violating operators and the $\mu$ term
in the superpotential. Since it also forbids soft SUSY breaking gaugino
Majorana masses in the Lagrangian and Higgs couplings to sfermion
pairs, SUSY flavor-changing and CP-violating contributions, for instance,
are reduced significantly, widening the potential parameter space
for supersymmetric theories \cite{Rflav}. The more restrictive Dirac
gaugino masses, on the other hand, are allowed. Moreover, since the
scalar components $\sigma$ of the chiral superfields $\hat{\Sigma}$
have $R$-charge 0, they can couple to SM particles so that ${\sigma}$
particles can be produced singly in standard particle collisions. In
addition, they can decay to pairs of SM particles [and, similarly,
to pairs of supersymmetric particles].

The $N{=}1$ chiral supermultiplets within the $N{=}2$ gauge hypermultiplets
contain scalar sigma fields $\sigma$ in the adjoint representations
of the gauge groups SU(1)$_C$, SU(2)$_I$ and U(1)$_Y$.  In the
electroweak SU(2)$_I$ and U(1)$_Y$ sectors the scalar fields can
acquire non-zero vacuum expectation values and they can mix with the
original Higgs fields. As a result, the properties of the Higgs
particles are modified in this scenario.

In $N{=}2$ supersymmetric theories the standard $N{=}1$ $L/R$ matter
supermultiplets are complemented with new $L/R$ matter
multiplets \cite{Polonsky:2000rs}.
To keep the theory chiral, in agreement with experimental observations,
the masses of the new multiplets must be chosen very large so
that $N{=}2$ supersymmetry is effectively reduced to $N{=}1$ supersymmetry
in this sector. Exceptions are the two Higgs doublets which can be
associated with the two supermultiplets within a Higgs hypermultiplet.
Since the gauge and Higgs sectors are framed in the $N{=}2$ formalism,
but the matter sector in $N{=}1$ is not, the theory is conventionally
termed $N{=}1$/$N{=}2$ hybrid theory.

The transition from the Majorana-type Minimal Supersymmetric Standard Model
(MSSM) to a Dirac theory by expanding the gauge sector can be formulated
in a smooth way by suitable transitions of the parameters in the
$\{\tilde{G},{\tilde{G}}'\}$ mass matrix.
We start with an infinitely large ${\tilde{G}}'$ Majorana mass at
the beginning of the path, which is congruent with the original
MSSM. Lowering the Majorana masses to zero and generating
non-diagonal entries in the mass matrix at the end of the path, the
two Majorana fields can be combined to a Dirac field if the two mass
eigenvalues have equal moduli but opposite signs. In this way, the
characteristics of the Majorana theory can systematically be tagged
in the evolution to the Dirac theory, and implications of the Dirac
theory can be connected with experimental analyses.

The Dirac theory, including the scalar sigma fields, has been
analyzed in two earlier studies \cite{us1,us2} primarily in the
colored sector, and experimental consequences have been discussed
for the proton collider LHC. Basic elements of the electroweak
sector, including the interaction of the Higgs field with the novel
scalar fields, have been presented in Ref.~\cite{Belanger:2009wf},
and implications for the relic density in the Universe have been
discussed for such a Dirac theory [see also \cite{Park,hsieh,Chun}].
In the present study we will
focus on collider signatures of the electroweak chargino/neutralino
and the novel sigma sectors at LHC and $e^+e^-$ colliders. In
addition, modifications of the properties of the Higgs particles by
interactions with the novel scalars will be discussed. While the
theoretical basis of the $N{=}1$/$N{=}2$ hybrid theory is summarized in the
next section, phenomenological consequences are worked out for the
chargino/neutralino and scalar/Higgs sectors thereafter.

\section{THEORETICAL BASIS: $N{=}1$/$N{=}2$ HYBRID THEORY}
\label{sec:theoretical_base}

\subsection{Hyper/Superfields and Interactions}
\label{sec:lagrangian}

\noindent The $N{=}1$/$N{=}2$ hybrid model, which can be evolved from the
MSSM continuously to a  Dirac gaugino theory, includes a large
spectrum of fields. The $N{=}2$ gauge hypermultiplets $\mathcal{G} =
\{\hat{G},\hat{\Sigma}\}$ can be decomposed into the usual $N{=}1$
vector supermultiplets of gauge and gaugino fields $\hat{G} =
\{G_\mu,\tilde{G}\}$, complemented by chiral supermultiplets of
novel gaugino and scalar fields ${\hat{\Sigma}} =
\{{\tilde{G}}',\sigma\}$. The new gauge/gaugino/scalar fields, together
with the MSSM fields, are shown explicitly for the color SU(3)$_C$
and the electroweak isospin SU(2)$_I$ and hypercharge U(1)$_Y$ gauge
groups in Tab.$\,$\ref{tab:hyper}.

\begin{table}[t]
\begin{tabular}{|l|c||c|c|c|}
\hline $\;$superfields & SU(3)$_C$,\,SU(2)$_I$,\,U(1)$_Y$   &
Spin 1  & Spin 1/2 & Spin 0         \\[1ex]
\hline\hline $\; \hat{G}_C$ / color    & ${\mathbf{ 8,1,0}}$  & $g^a$
& $\tilde{g}^a $
                          &      \\[1ex]
$\; \hat{G}_I$ / isospin  & ${\mathbf{ 1,3,0}}$  & $W^i$
                          & ${\tilde{W}}^i$ &  \\[1ex]
$\; \hat{G}_Y$ / hypercharge$\,$ & ${\mathbf{ 1,1,0}}$   & $B$
                          & $\tilde{B}$  &    \\[1ex]
\hline $\; \hat{\Sigma}_C$ / color   & ${\mathbf{ 8,1,0}}$  &
                          &  ${\tilde{g}}'^a$  & $\sigma_C^a$     \\[1ex]
$\; \hat{\Sigma}_I$ / isospin & ${\mathbf{ 1,3,0}}$  & &
            ${\tilde{W}}^{\prime i}$ &
                                    ${\sigma}_I^i$ \\[1ex]
$\; \hat{\Sigma}_Y$ / hypercharge$\,$ & ${\mathbf{ 1,1,0}}$   &
                         & ${\tilde{B}'}$   & $\sigma^0_Y$     \\[1ex]
\hline
\end{tabular}
\caption{\it The $N{=}2$ gauge hypermultiplets for the color SU(3)$_C$,
             isospin SU(2)$_I$ and hypercharge U(1)$_Y$ groups.
             The superscripts $a=1$-$8$ and $i=1$-$3$ denote the
             SU(3)$_C$ color and SU(2)$_I$ isospin indices,
             respectively. }
\label{tab:hyper}
\end{table}

In parallel to the gauge fields, the neutral gaugino fields
$\tilde{G}$ are self-conjugate Majorana fields with two helicity
components, analogously the novel gaugino fields ${\tilde{G}'}$.
[The notation $G_\mu,\tilde{G},\tilde{G}',\sigma$ is used
generically for gauge, gaugino and $\sigma$ fields; when specific gauge
groups are referred to, the notation follows Table~\ref{tab:hyper}.]
To match the two gaugino degrees of freedom in the new chiral
supermultiplet, the components of the scalar fields $\sigma$ are
complex. Suitable mass matrices provided, the two gaugino Majorana
fields $\tilde{G}$ and ${\tilde{G}}'$ can be combined to a Dirac
field ${\tilde{G}}_D$.

In a similar way, the two Higgs-doublet superfields ${\hat{H}}_d$
and ${\hat{H}}_u^\dagger$ of the MSSM can be united to an $N{=}2$
hyperfield $\mathcal{H} = \{ {\hat{H}}_d,{\hat{H}}_u^\dagger \}$
\cite{fnw,ant}. It may be noted that, after diagonalizing the off-diagonal
2$\times$2 mass matrix, the two neutral higgsinos can be interpreted
as a Dirac field.

In contrast, the observed chiral character of the Standard Model
precludes the extension of the usual (s)lepton and (s)quark
supermultiplets $\hat{Q}$ to hypermultiplets of $L/R$ symmetric
particles and mirror-particles. Moreover, including such a large
number of new matter fields would make the entire theory
asymptotically non-free. Equivalent to introducing very heavy
masses, the mirror fields can just be eliminated from the system of
matter fields {\it ad hoc}. This supposition generates the $N{=}1$/$N{=}2$
hybrid character of the theory.

Corresponding to the complex spectrum of fields, the sum of
a set of actions with different bases and characteristics describes
the $N{=}1$/$N{=}2$ hybrid theory. The $N{=}2$ action of the gauge
hypermultiplet $\mathcal{G} = \{ \hat{G},{\hat{\Sigma}} \}$ consists
of the usual $N{=}1$ action of the gauge supermultiplet $\hat{G}$
plus the action of the chiral supermultiplet ${\hat{\Sigma}}$ which
couples the new gaugino and scalar fields to the gauge superfield:
\begin{eqnarray}
{\mathcal{A}}_{G}
   &=& \sum \frac{1}{16g^2k}\, \int d^4x \, d^2 \theta \  \;
       {\rm tr} \, {\hat{G}}^\alpha {\hat{G}}_{\alpha}\,,  \\
{\mathcal{A}}_{\Sigma}
   &=& \, \sum \int d^4x \, d^2 \theta d^2\,\bar{\theta}\
       \; {\hat{\Sigma}}^\dagger \exp[\hat{G}] \, {\hat{\Sigma}}\,,
\end{eqnarray}
with the sums running over the gauge groups SU(3)$_C$, SU(2)$_I$ and
U(1)$_Y$. $g$ are the gauge couplings (denoted by $g_s,\,g$ and $g'$
for color, isospin and hypercharge)  and $k$ are the corresponding
quadratic Casimir invariants $C_2(G)$. $\hat{G}_{\alpha} =
2g\hat{G}^a_{\alpha} T^a$ are the gauge superfield-strengths, $T^a$
the generators in the adjoint representation; the traces run over
the gauge-algebra indices. To this class
of actions belongs also the standard (s)lepton/(s)quark gauge action
\begin{eqnarray}
{\mathcal{A}}_{Q}   &=& \,\sum \int d^4x \, d^2 \theta d^2
\bar{\theta} \
 \; {\hat{Q}}^\dagger \exp[ \hat{G}] \, {\hat{Q}}    \,,
\end{eqnarray}
summed over the standard matter chiral superfields, denoted
generically as ${\hat{Q}}$.

These actions are complemented by gauge-invariant $N{=}1$ supersymmetric
Majorana mass terms $M$ for the new gauge superfields and Dirac mass
terms $M^D$ coupling the original and new gauge superfields:
\begin{eqnarray}
\label{eq:newMajoMass}
{\mathcal{A}}_{M}
     &=& \int d^4x \, d^2\theta \, M \,
         {\rm tr} \, {\hat{\Sigma}} \, {\hat{\Sigma}}   \,,
         \label{eq:AM}\\
\label{eq:DiracMass}
{\mathcal{A}}_{D}
     &=& \int d^4x \, d^2 \theta\, M^D \, \theta^\alpha\, {\rm tr} \,
         {\hat{G}}_\alpha {\hat{\Sigma}} \,.
         \label{eq:AD}
\end{eqnarray}
${\mathcal{A}}_{M}$, which is bi-linear in the $\Sigma$ fields, is part of
the superpotential of the theory and contributes to the masses
of the chiral supermultiplets. The Dirac mass term can be generated,
$e.\,g.$, by the interaction $\sqrt{2}\hat{X}^\alpha
\hat{G}_\alpha \hat{\Sigma}/M_X$ when a hidden-sector $U(1)'$
spurion superfield acquires a $D$-component vacuum expectation value
$\hat{X}^\alpha=\theta^\alpha D_X$, giving rise to the Dirac mass
$M^D=D_X/M_X$ \cite{Benakli:2009mk}.

According to the general rules, this set of actions generates
$D$-terms bi-linear in the usual slepton and squark fields and linear
in the new scalar sigma field with a coefficient given by the Dirac
mass $M^D$. When the auxiliary fields $D$ are eliminated through
their equations of motion, the sigma fields get coupled to bi-linears
of the slepton and squark fields with strength $M^D$.

The Higgs sector is rendered more complicated by the interactions with the
non-colored scalar sigma fields. The Higgs supermultiplets
${\hat{H}}_d$ and ${\hat{H}}^\dagger_u$ are coupled
to the SU(2)$_I\times$U(1)$_Y$ supergauge fields in the usual way,
\begin{eqnarray}
{\mathcal{A}}_{H}
   &=& \, \sum_{i=u,d}\int d^4x \, d^2 \theta d^2\, \bar{\theta}
       \, {\hat{H}_i}^\dagger \exp[ \hat{G}_I+\hat{G}_Y] \, {\hat{H}_i}
       \,.
\end{eqnarray}
The part of the superpotential which includes Higgs fields, consists
of the standard $N{=}1$ bi-linear $\mu$-term,
\begin{equation}
{\mathcal{A}}_{\mu}
      = \, \int d^4x\, d^2\theta\;
        \mu {\hat{H}}_u \cdot {\hat{H}}_d  \,,
\label{eq:mu}
\end{equation}
and the tri-linear Higgs Yukawa terms involving the matter fields,
which can be adopted from the $N{=}1$ theory:
\begin{equation}
{\mathcal{A}}'_{Q} = \, \int d^4x\, d^2\theta\
                      \, \sum g_Q \, {\hat{q}}^c \hat{Q} \cdot {\hat{H}}_q
                      \,,
\end{equation}
the dots denoting the asymmetric contraction of the SU(2)$_I$ doublet
components. New tri-linear interactions are predicted in $N{=}2$
supersymmetry~\cite{Gaum} which couple the two supercomponents
of the Higgs hypermultiplet with the new chiral superfields in the
superpotential:
\begin{eqnarray}\label{eq:higgsSup}
{\mathcal{A}}'_{H}
   &=& \int d^4x\, d^2\theta \,\frac{1}{\sqrt{2}}
       \hat{H}_u \cdot (\lambda_I\hat{\Sigma}_I+\lambda_Y\hat{\Sigma}_Y)
       \hat{H}_d\,.
\end{eqnarray}
In $N{=}2$ supersymmetry the couplings $\lambda_I,\lambda_Y$ are identified
with the SU(2)$_I$ and U(1)$_Y$ gauge couplings,
\begin{eqnarray}
\lambda_I = g/\sqrt{2}\quad\mbox{and}\quad\lambda_Y = - g'/\sqrt{2} \,.
\label{eq:N=2_couplings}
\end{eqnarray}
In our phenomenological analyses we will treat them generally as independent
couplings.

It may be noticed that the Majorana action $\mathcal{A}_M$, the $\mu$-term
${\mathcal{A}}_\mu$ and the tri-linear Higgs-sigma term ${\mathcal{A}}'_H$
are manifestly not $R$-invariant.

Finally, the bi-linear and tri-linear soft supersymmetry breaking terms
must be added to the gauge, Higgs and matter Lagrangians:
\begin{eqnarray}
{\mathcal{L}}_{\rm gauge,soft} &=&
   -  \tfrac{1}{2}M_{\tilde{B}} \, \tilde B \, \tilde B
   -  \tfrac{1}{2} M_{\tilde{W}} \, \left(\tilde W^+ \, \tilde W^-
                             + \tilde W^- \, \tilde W^+
                             + \tilde W^0 \, \tilde W^0\right)
   - \tfrac{1}{2}M_{\tilde{g}} \, \tilde g^a \, \tilde g^a
   + {\rm h.c.} \nonumber\\[1ex]
  &&-  \tfrac{1}{2} M'_{\tilde{B}} \, \tilde B' \, \tilde B'
   -  \tfrac{1}{2} M'_{\tilde{W}} \, \left(\tilde W'^+ \, \tilde W'^-
                             + \tilde W'^- \, \tilde W'^+
                             + \tilde W'^0 \, \tilde W'^0\right)
   - \tfrac{1}{2} M'_{\tilde{g}} \, \tilde g'^a \, \tilde g'^a
   + {\rm h.c.} \nonumber \\[1ex]
 && -  m_Y^2 \, |\sigma^0_Y|^2
        - \tfrac{1}{2} \left(m_Y^{\prime 2} (\sigma_Y^0)^2 + {\rm h.c.} \right)
  -  m_I^2 \, \left|\sigma_I^i\right|^2
        - \tfrac{1}{2} \left(m_I^{\prime 2} (\sigma_I^i)^2 + {\rm h.c.} \right)
   \nonumber\\[1ex]
  && -  m_C^2 \, \left|\sigma_C^a\right|^2
        - \tfrac{1}{2} \left( m_C^{\prime 2} (\sigma_C^a)^2 + {\rm h.c.} \right)
        \,, \label{eq:gauge_soft}
        \\[1ex]
{\mathcal{L}}_{\rm Higgs,soft}   &=&
    - m^2_{H_u} \, \left(\left|H_u^+\right|^2 + \left|H_u^0\right|^2 \right)
    - m^2_{H_d} \, \left(\left|H_d^-\right|^2 + \left|H_d^0\right|^2 \right)
    \nonumber\\[1ex]
   && - \left[ B_\mu \,  \left(H_u^+ \, H_d^- - H_u^0 \, H_d^0 \right)
    + {\rm h.c.} \right] \nonumber
   \\[1ex]
   && - \left[ A_Y \, \lambda_Y \, \sigma^0_Y \,
                    \left(H_u^+ \, H_d^- - H_u^0 \, H_d^0 \right)
           + A_I \, \lambda_I \, \sigma_I^i \,
                \left(H_u \cdot \, \tau^i \, H_d\right)
       + {\rm h.c.} \right]\,,
       \label{eq:higgs_soft}
\end{eqnarray}
with $i$ and $a$ being the SU(2)$_I$ and SU(3)$_C$ indices, $\tau^i$ the
Pauli matrices, and moreover,
\begin{eqnarray}
{\mathcal{L}}_{\rm matter,soft}    &=&
    - \left(m^2_{\tilde Q} \right)_{ij} \,
                  \left(\tilde u^*_{iL} \tilde u_{jL} +
                \tilde d^*_{iL} \tilde d_{jL} \right)
    - \left(m^2_{\tilde u} \right)_{ij} \,\tilde u^*_{iR} \tilde u_{jR}
    - \left(m^2_{\tilde d} \right)_{ij} \,\tilde d^*_{iR} \tilde d_{jR}
\nonumber\\[1ex]
   & & -
   \left(m^2_{\tilde L} \right)_{ij} \,
                  \left(\tilde \nu^*_{iL} \tilde \nu_{jL} +
                \tilde e^*_{iL} \tilde e_{jL} \right)
    - \left(m^2_{\tilde e} \right)_{ij} \,\tilde e^*_{iR} \tilde e_{jR}
\nonumber\\[1ex]
   & & -\, (A_u f_u)_{ij} \tilde{u}^*_{iR}
          (\tilde{d}_{jL} H^+_u - \tilde{u}_{jL} H^0_u)
       -\, (A_d f_d)_{ij} \tilde{d}^*_{iR}
         (\tilde{u}_{jL} H^-_d - \tilde{d}_{jL} H^0_d) +{\rm h.c.}
\nonumber\\[1ex]
   & & - \, (A_e f_e)_{ij} \tilde{e}^*_{iR}
         (\tilde{\nu}_{jL} H^-_d - \tilde{e}_{jL} H^0_d) + {\rm h.c.}\,,
\end{eqnarray}
with $i, j$ now denoting the matter generations.
Here, the convention is adopted to use subscripts $C,I,Y$ for
parameters corresponding to color, isospin and hypercharge gauge
groups, respectively. Capitalized mass parameters $M$ are the
Majorana gaugino masses [$M^D$ for Dirac], while lower-case $m$
denotes soft scalar masses. The Majorana mass terms,
$M'_{\tilde{B}}$, $M'_{\tilde{W}}$ and $M'_{\tilde{g}}$, for the
new gauge adjoint fermions are soft $N{=}1$ SUSY breaking parameters
and add to the Majorana mass parameters, $M_Y, M_I$ and $M_C$,
introduced in Eq.$\,$(\ref{eq:AM}) as part of the $N{=}1$
supersymmetric superpotential.

From this set of actions and Lagrangians, and after eliminating the
auxiliary $D^a$ fields through their equations of motion, the masses
and mixings of the Higgs and gauge-adjoint scalar particles and
their interactions can be read off, and correspondingly those of
their superpartners as will be detailed below.
The final form of the Lagrangians are collected in the
following list which, in general, includes only interactions of the
new fields:\footnote{Many
of the mass parameters and couplings defining the $N{=}1$/$N{=}2$
hybrid model can be complex in general. Nevertheless, for the sake
of simplicity all the parameters are assumed to be real throughout
this paper.}

\noindent
{\it (i) \uline{\it{SU(3)$_C\times$SU(2)$_I\times$U(1)$_Y$ gauge
                  boson/sigma sector:}}}

\noindent
The gauge interactions of the adjoint sigma fields are determined from the
scalar kinetic term $(D_\mu \sigma)^\dagger (D^\mu \sigma)$ with the
covariant derivative $ D_\mu = \partial_\mu + i g_s T^a g^a_\mu
+ i g T^i W^i_\mu$. In addition to their kinetic terms, the term
generates the Lagrangian for the derivative three-point and seagull
four-point interaction terms:
\begin{eqnarray}
{\cal L}_{\sigma_C, gauge}
  &=&
   - g_s f^{abc} g^a_\mu\, (\sigma^{b*}_C
    \overleftrightarrow{\partial^\mu} \sigma^c_C)
   + g^2_s f^{ace}f^{bde} g^a_\mu g^{\mu b} \sigma^{*c}_C \sigma^d_C\,,
\\[1ex]
{\cal L}_{\sigma_I, gauge}
  &=& -g \epsilon_{ijk} W^i_\mu\, (\sigma^{j*}_I
    \overleftrightarrow{\partial^\mu}\sigma^k_I)
   + g^2 \epsilon_{ikm}\epsilon_{jlm} W^i_\mu W^{\mu j}
         \sigma^{*k}_I\sigma^l_I\,,
\end{eqnarray}
where
$f^{abc}$ and $\epsilon_{ijk}$ are the SU(3)$_C$ and SU(2)$_I$ structure
constants, respectively, and $A\overleftrightarrow{\partial^\mu} B
\equiv A \partial^\mu B - (\partial^\mu A) B$.

\vskip 0.3cm
\noindent
{\it (ii) \uline{\it{SU(3)$_C$ sfermion/gaugino/sigma sector:}}}

\noindent
The interaction Lagrangian of the sigma field $\sigma_C$ with the squarks
is given by
\begin{eqnarray}
{\cal L}_{\sigma_C (\sigma_C) \tilde q\tilde q} =  - \sqrt{2} \, g_s \, M^D_{C}
\, (\sigma_C^a + \sigma_C^{a*}) \,
    \left(  \tilde q_L^* \, \frac{\lambda^a}{2} \, \tilde q_L
          - \tilde q_R^* \, \frac{\lambda^a}{2} \, \tilde q_R \right)
 + i \, g^2_s \, f^{a b c} \, \sigma^{a*}_C \sigma^b_C\,\,
            \tilde{q}^\dagger \frac{\lambda^c}{2}\,\tilde{q}\,,
 \label{sigma_sigma_tildeq_tildeq'}
\end{eqnarray}
where $\lambda^a$ ($a=1$--$8$) are the Gell-Mann matrices. Therefore,
the $L$-- and $R$--chiral squarks contribute with opposite signs as
demanded by the general form of the super-QCD $D$-terms.
On the other hand, the interactions of the two gluino fields, $\tilde{g}$ and
$\tilde{g}'$, with the SU(3)$_C$ sigma field $\sigma_C^{}$
and with the squark and quark fields are described by the Lagrangians:
\begin{eqnarray}
\label{sigma_tildeg_tildeg'}
{\cal L}_{\tilde g  \tilde g' \sigma_C^{}}
    & = & \dis - \sqrt{2} \, i \, g_s \, f^{a b c} \,
                \overline{\tilde{g}^{\prime a}_L} \,\tilde{g}_R^b \,
        \sigma_C^c + {\rm h.c.}\,, \\[1ex]
\label{tildeg_q_tildeq}
{\cal L}_{\tilde g  q \tilde q}
   &=& -\sqrt{2} \, g_s \,
        \left( \overline{q_L} \, \frac{\lambda^a}{2} \,
               \tilde g^a_R \; \tilde q_L \,
              -\overline{q_R} \, \frac{\lambda^a}{2} \, \tilde g^a_L \;
               \tilde q_R \right)   + {\rm h.c.}\,,
\end{eqnarray}
Only the standard gluino couples to squark fields since, as required
by $N{=}2$ supersymmetry, the new gluino $\tilde{g}'$ couples only to
mirror matter fields, which in the hybrid model are assumed to be
absent.

\noindent
{\it (iii) \uline{\it{SU(2)$_I \times$U(1)$_Y$ sfermion/gaugino/sigma sector:}}}

\noindent
In the weak basis, the $R$-chiral sfermions $\tilde{f}_R$
are SU(2)$_I$ singlets so that only the $L$-chiral sfermions
$\tilde{f}_L$ interact with the SU(2)$_I$ sigma field $\sigma_I$
through the interaction Lagrangians:
\begin{eqnarray}
{\cal L}_{\sigma_I (\sigma_I) \tilde f\tilde f}
    &=&  - \sqrt{2}\, g\,\,
    M^D_I \, (\sigma_I^i + \sigma^{i*}_I)
    \,\, \tilde{f}^\dagger_L \, \frac{\tau^i}{2} \, \tilde f_L
+ i g^2\, \epsilon_{ijk}\, \sigma^{j*}_I \sigma^k_I\,
    \, \tilde{f}^\dagger_L \, \frac{\tau^i}{2} \, \tilde f_L\,,
\end{eqnarray}
where $\tilde f_L$ is any matter SU(2)$_I$-doublet field. On the
other hand, the Lagrangians governing the interactions of the winos,
$\tilde{W}$ and $\tilde{W}'$, with the SU(2)$_I$ sigma field
$\sigma_I$ and the (s)fermion fields are given by
\begin{eqnarray}
\label{sigmai_tildew_tildew'}
{\cal L}_{\sigma_I \tilde{W} \tilde {W'}}
    & = &  - \sqrt{2} \, i \, g \, \epsilon^{i j k} \,
          \overline{\tilde W^{\prime i}_L} \,\tilde{W}_R^j \, \sigma_I^k
          + {\rm h.c.}
\,,\\[1ex]
\label{f_tildef_tildew}
{\cal L}_{\tilde{W}  f \tilde{f}}
    & = &
- \sqrt{2} \, g \, \overline{f_L} \, \frac{\tau^i}{2} \,
             \tilde W^a_R\;\tilde f_L  + {\rm h.c.}\,.
\end{eqnarray}
Only the $L$-chiral sfermions $\tilde{f}_L$ couple to the
standard wino $\tilde{W}$.

The U(1)$_Y$ sigma field $\sigma_Y$ is essentially a SM
singlet state with no tree--level gauge interaction to any of the
gauge bosons, gauginos and higgsinos. The singlet scalar field couples
only to the Higgs bosons and the (s)fermion fields, with the latter
being given by the Lagrangian:
\begin{eqnarray}
{\cal L}_{\sigma_Y \tilde f\tilde f}
  =  - \sqrt{2} \, g' \, M^D_Y (\sigma^0_Y + \sigma^{0*}_Y) \,\,
      (Y_{f_L} |\tilde{f}_L|^2 - Y_{f_R} |\tilde{f}_R|^2 )\,,
\end{eqnarray}
and the standard bino $\tilde{B}$ (but not the new bino $\tilde{B}'$)
couples to the (s)fermion fields through the interaction Lagrangian:
\begin{eqnarray}
\label{f_tildef_tildeb}
{\cal L}_{\tilde{B}  f \tilde f}
     = - \sqrt{2}  g'\,
       ( Y_{f_L} \overline{f_L} \tilde{B}_R \tilde f_L
        -Y_{f_R} \overline{f_R} \tilde{B}_L \tilde f_R) + {\rm h.c.}\,,
\end{eqnarray}
where $Y_{f_L}$ and $Y_{f_R}$ are the hypercharges of the $L$-chiral and
$R$-chiral fermions, $f_L$ and $f_R$, respectively.

\noindent
{\it (iv) \uline{\it{SU(2)$_I \times$U(1)$_Y$ higgsino/sigma sector:}}}

\noindent
The superpotential (\ref{eq:higgsSup}) coupling the new
SU(2)$_I\times$U(1)$_Y$ chiral superfields with the Higgs hypermultiplets
leads to Yukawa-type interactions of the electroweak sigma fields
with the higgsino fields. In the weak basis, the interactions are
described by the Lagrangian
\begin{eqnarray}
{\cal L}_{\sigma\tilde{H}\tilde{H}}
   &=&  -\lambda_Y \sigma^0_Y   (\overline{\tilde{H}^-_{uR}}\tilde{H}^-_{dL} -
         \overline{\tilde{H}^0_{uR}}\tilde{H}^0_{dL})
      +\lambda_I \sigma^0_I (\overline{\tilde{H}^-_{uR}}\tilde{H}^-_{dL}
      + \overline{\tilde{H}^0_{uR}}\tilde{H}^0_{dL})
     +{\rm h.c.}\nonumber\\[1ex]
    &&  -\sqrt{2}\lambda_I\,
       ( \sigma^-_1 \overline{\tilde{H}^-_{uR}} \tilde{H}^0_{dL}
        -\sigma^+_2 \overline{\tilde{H}^0_{uR}} \tilde{H}^-_{dL} )
        + {\rm h.c.}\,,
\label{sigma_tildeh_tildeh}
\end{eqnarray}
where we have introduced two charged scalars and one neutral
scalar defined as
\begin{eqnarray}
\sigma^-_1 = \frac{1}{\sqrt{2}} \left(\sigma^1_I + i \sigma^2_I\right)\,,
             \quad
\sigma^+_2 = \frac{1}{\sqrt{2}} \left(\sigma^1_I - i \sigma^2_I\right)\,,
             \quad
\sigma^0_I = \sigma^3_I\,,
\end{eqnarray}
with each of $\sigma_I^i$ being complex.

Combining the above Lagrangian (\ref{sigma_tildeh_tildeh}) with the Lagrangian
(\ref{sigmai_tildew_tildew'}) for the interactions of the sigma fields with
gauginos will enable us to derive the vertices for the interactions of the sigma
fields with charginos and neutralinos in the mass eigenstate basis.

\noindent
{\it (v) \uline{\it{SU(2)$_I\times$U(1)$_Y$ Higgs/sigma sector :}}}

\noindent
The potential for the neutral and charged electroweak Higgs and scalar fields
receives contributions from three different sources:
the gauge kinetic terms, the superpotential, and the
soft-breaking terms. Complementing the neutral field
interactions noted in Ref.~\cite{Belanger:2009wf}
by the charged fields, the potential $V_{\sigma H}$
for the charged and neutral electroweak Higgs and adjoint scalars
can be written as a sum over four characteristic contributions:
\begin{eqnarray}
V_{\sigma H|1} &=& m^2_{H_u} (|H^+_u|^2 + |H^0_u|^2)
               + m^2_{H_d} (|H^0_d|^2 + |H^-_d|^2)
               + [ B_\mu (H^+_u H^-_d - H^0_u H^0_d) +{\rm h.c.} ]\,,
               \\
V_{\sigma H|2} &=& \frac{1}{2} [\sqrt{2} M^D_{Y}
                (\sigma^0_Y + \sigma^{0*}_Y)
              + \frac{1}{2} g' (|H^+_u|^2 - |H^-_d|^2
                              + |H^0_u|^2 - |H^0_d|^2)]^2
              \nonumber\\
   && + \frac{1}{2} | 2 M^D_{I} (\sigma^+_1 + \sigma^+_2)
             +\sqrt{2}g(\sigma^+_1 \sigma^0_I - \sigma^+_2 \sigma^{0*}_I)
             + g (H^+_u H^{0*}_u + H^0_d H^+_d) |^2
             \nonumber\\
   && + \frac{1}{2} [\sqrt{2} M^D_{I} (\sigma^0_I + \sigma^{0*}_I)
             + g (|\sigma^+_2|^2 - |\sigma^-_1|^2)
             + \frac{1}{2} g (|H^+_u|^2 - |H^-_d|^2 - |H^0_u|^2 + |H^0_d|^2)
                    ]^2\,, \\
V_{\sigma H|3} &=&  |(\mu+\lambda_Y \sigma^0_Y - \lambda_I \sigma^0_I) H^-_d
              + \sqrt{2} \lambda_I \sigma^-_1 H^0_d |^2
              + | (\mu+\lambda_Y \sigma^0_Y - \lambda_I \sigma^0_I) H^+_u
              - \sqrt{2} \lambda_I \sigma^+_2 H^0_u |^2
             \nonumber\\
   && + |(\mu+\lambda_Y \sigma^0_Y + \lambda_I \sigma^0_I) H^0_d
              + \sqrt{2} \lambda_I \sigma^+_2 H^-_d |^2
              + | (\mu+\lambda_Y \sigma^0_Y + \lambda_I \sigma^0_I) H^0_u
              - \sqrt{2} \lambda_I \sigma^-_1 H^+_u |^2
             \nonumber\\
   && + |M_Y \sigma^0_Y + \lambda_Y (H^+_u H^-_d - H^0_u H^0_d)|^2
      + |M_I \sigma^0_I - \lambda_I (H^+_u H^-_d + H^0_u H^0_d)|^2
             \nonumber\\
   && + |M_I \sigma^-_1 - \sqrt{2} \lambda_I H^0_u H^-_d|^2
      + |M_I \sigma^+_2 + \sqrt{2} \lambda_I H^+_u H^0_d|^2\,,\\
V_{\sigma H|4} &=&  m^2_Y |\sigma^0_Y|^2
      + m^2_I (|\sigma^0_I|^2+ |\sigma^-_1|^2 + |\sigma^+_2|^2)
      + \frac{1}{2} ( m^{\prime 2}_Y (\sigma^0_Y)^2 + {\rm h.c.})
      + \frac{1}{2} [ m^{\prime 2}_I ((\sigma^0_I)^2 + 2 \sigma^+_2\sigma^-_1)
                      + {\rm h.c.} ]
             \nonumber\\
   && + A_Y \lambda_Y \sigma^0_Y (H^+_u H^-_d - H^0_u H^0_d)
      - A_I\lambda_I\, \sigma^0_I (H^+_u H^-_d + H^0_u H^0_d)
      + {\rm h.c.} \nonumber\\
   && + \sqrt{2} A_I \lambda_I  (\sigma^-_1 H^+_u H^0_d - \sigma^+_2 H^-_d H^0_u)
        + {\rm h.c.}\,.
\end{eqnarray}
After shifting the neutral fields by their vacuum expectation values, the
physical scalar masses and the tri- and quadri-linear interaction
vertices can be read off.

\subsection{Masses, Mixings and Dirac Fields}

\noindent
Introducing the vacuum expectation values of the scalar/Higgs fields in the
Lagrangians of the previous subsection,
their values are determined by the absence of terms linear in the
fields, while from the terms bi-linear in the fields the mass matrices
for the scalars/Higgs, the charginos and neutralinos can be read
off. The vacuum expectation values ({\it vevs}) of the neutral Higgs
and the neutral sigma fields\footnote{Throughout the paper, we restrict ourselves
to the case of a CP preserving and neutral vacuum with real
vacuum expectation values.} are defined as
\begin{eqnarray}
\langle H^0_{u/d} \rangle  &=& \frac{1}{\sqrt{2}}\, v_{u/d}\,, \\
\langle {\sigma}^0_{Y/I} \rangle  &=& \frac{1}{\sqrt{2}} v_{Y/I} \,.
\end{eqnarray}
As usual, the $vevs$ of the Higgs sector can be rewritten as
\begin{eqnarray}
v         = \sqrt{v_u^2+v_d^2} \quad \mbox{and}\quad
\tan\beta = \frac{v_u}{v_d}                      \,.
\label{eq:vevs_tanb}
\end{eqnarray}
The masses of the electroweak vector bosons $W,Z$ are generated by the
interactions of the fields with the ground states of the neutral Higgs
$H^0_u,H^0_d$ and the neutral scalar iso-triplet field $\sigma^0_I$
(while the hyper-singlet field $\sigma^0_Y$ does not couple)
\begin{equation}
m^2_Z = \frac{1}{4}(g'^2+g^2) v^2 \,,  \qquad
m^2_W = \frac{1}{4} g^2 v^2 + g^2 v^2_I  \,.
\label{eq:vevs_mzw}
\end{equation}
The iso-triplet {\it vev}  shifts the tree-level
$\rho$-parameter away from unity by the amount
\begin{equation}
\Delta\rho =\rho -1 =  4 v_I^2 / v^2\,.
\end{equation}
Allowing a maximum value $\Delta\rho \leq 10^{-3}$ for the
shift \cite{Amsler:2008zzb}, it turns out that the vacuum expectation value
of the iso-triplet field must be very small, $v_I \leq 3\,$GeV,
{\it cf.} \cite{Belanger:2009wf}. We will assume that the soft
supersymmetry breaking scalar $\sigma_I$ mass parameter $m_I$ of
order TeV drives $v_I$ to the small value. As a result, the Higgs
{\it vev} $v$ is close to the standard value $v = 246\,$GeV, and
$\tan\beta$ may be identified approximately with the corresponding
MSSM parameter. And while almost any value for $v_Y$ is phenomenologically
quite consistent, a large $m_Y $ would drive $v_Y$ to small values.

\subsubsection{Charginos}

\noindent
Defining the current bases,
$\{\tilde{W}'^+_R,\tilde{W}^+_R,{\tilde{H}}^+_{uR}\}$ and
$\{\tilde{W}'^-_L,\tilde{W}^-_L,{\tilde{H}}^-_{dL}\}$
for the two charged winos and the charged higgsino, the chargino mass
matrix can be written as
\begin{equation}
{\mathcal{M}}_C =
\begin{pmatrix}
      M'_2           & M^D_{2}- g v_I  & -\lambda_I v_u \\
      M^D_{2}+g v_I       & M_2            & \frac{1}{\sqrt{2}} g v_d \\
      \lambda_I v_d  & \frac{1}{\sqrt{2}} g v_u & \mu_c
      \end{pmatrix}\,,
\label{eq:chargino_mass_matrix}
\end{equation}
where
\begin{eqnarray}
M_2  = M_{\tilde{W}}, \quad M'_2 = M'_{\tilde{W}} + M_I, \quad
M^D_{2} = M^D_{I} \quad \mbox{and}\quad \mu_c = \mu+(\lambda_Y v_Y
-\lambda_I v_I)/\sqrt{2}\,.
\label{eq:chargino_parameter_defined}
\end{eqnarray}
Three charginos, $i.\,e.$ one degree of freedom more than in MSSM
and related iso-singlet extensions like NMSSM or USSM, are predicted
in the general $N{=}1$/$N{=}2$ hybrid model, labeled
${\tilde{\chi}}_1^\pm,{\tilde{\chi}}_2^\pm,{\tilde{\chi}}_3^\pm$
(ultimately for ascending mass values). The MSSM case is reached in
the limit $M'_2\to -\infty$ which corresponds to infinitely heavy
$\tilde{W}'$. By raising the magnitude of the $\tilde{W}'$ gaugino
mass parameter $M'_2$ from $-\infty$ to $0$ and lowering at the same
time  $M_2$ to $0$ the Dirac limit is obtained. Though the $3 \times
3$ mass matrix can be diagonalized analytically for arbitrary
parameters, we study instead the evolution of the eigenvalues
analytically in the limit of small couplings, and numerically
by varying $-\infty \leq M'_2 \leq 0$ from the MSSM to the Dirac limit.

For small gaugino/higgsino mixings in the area where the
supersymmetry mass parameters $M'_2,M_2, M^D_{I}, \mu$ [and the size
of their mutual differences] are much larger than the electroweak
parameter $v$, the eigenvalues and mixing parameters can be calculated
easily. This approximation leaves us with one higgsino mass
eigenvalue
\begin{eqnarray}
\overline{m}^\pm_3\, =\, \mu_c\,,
\end{eqnarray}
and a $2\times 2$ gaugino mass submatrix with two eigenvalues
\begin{eqnarray}
\overline{m}^\pm_{1,2}\,=\,
  \frac{1}{2} \left|\,\gamma_2 \mp \delta_2 \right|\quad
  \mbox{where}\quad
  \gamma_2 = \sqrt{(M'_2+M_2)^2+4 g^2 v^2_I}\quad
  \mbox{and}\quad
  \delta_2 = \sqrt{(M'_2-M_2)^2+ 4 (M^D_2)^2}\,,
\end{eqnarray}
and the two mixing angles for the positive and negative states
\begin{eqnarray}
\cos\theta_\pm &\equiv& c_\pm
      = \frac{1}{\sqrt{2}}
        \sqrt{1-(M'^2_2-M^2_2\mp 4 g v_I M^D_2)/\gamma_2 \delta_2}\,,\\
\sin\theta_\pm &\equiv& s_\pm
      = \frac{1}{\sqrt{2}}
        \sqrt{1+(M'^2_2-M^2_2\mp 4 g v_I M^D_2)/\gamma_2 \delta_2}\,,
\label{eq:noewmix}
\end{eqnarray}
With $M'_2 = -\infty$ in the MSSM limit we get  $c_\pm = 0$
and $s_\pm = 1$, while $c_+=s_- = 1$ and $c_-=s_+ = 0$ in the
Dirac limit with $M'_2=M_2=0$ and $M^D_2, v_I > 0$.

Switching on the weak couplings among the gaugino and higgsino
sectors, the chargino mass eigenvalues and the mixing
parameters derived from
\begin{equation}
{\mathcal{M}}^{\rm diag}_C\, =\, U^T_+\, \mathcal{M}_C\, U_-\,,
\label{eq:chtoapp}
\end{equation}
can be calculated analytically in simple form.\footnote{In Appendix A we
provide an analytic prescription for the singular value decomposition of a
general $2\times 2$ matrix and in Appendix B the small-mixing approximation.}
The results are presented in Appendix C.

\begin{figure}[t]
\epsfig{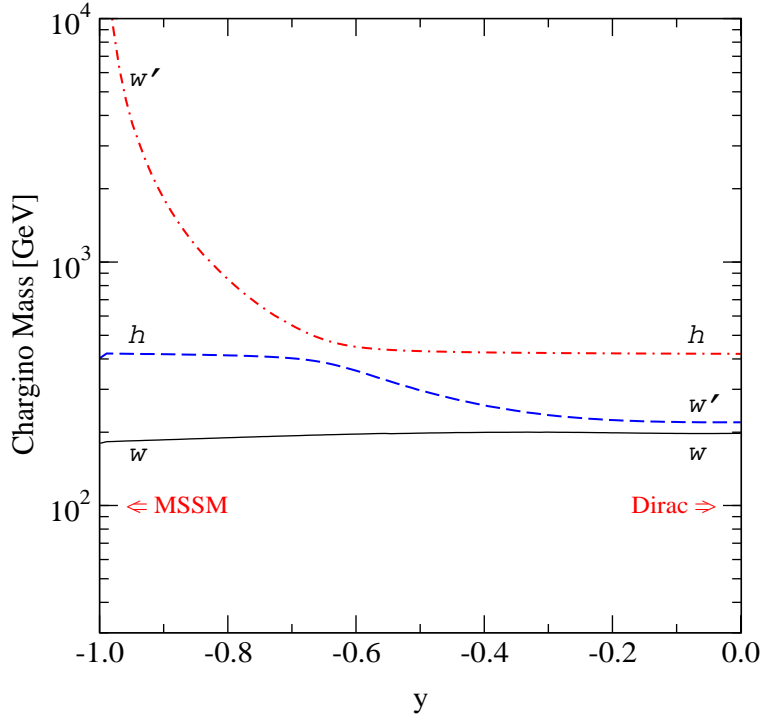}
\caption{\it Evolution of the chargino masses as a function of the control
             parameter $y$ from the MSSM doublet ($y=-1$) to the Dirac
             ($y=0$) triplet along the path ${\cal P}_C$ in
             Eq.$\,$(\ref{eq:chargino_path}) for $m=200$~GeV,
             $\tan\beta=5$, $v_I=v_Y=3$ GeV and the $N{=}2$ values for the
             couplings $\lambda_{Y,I}$ in Eq.$\,$(\ref{eq:N=2_couplings}).}
\label{fig:chargino}
\end{figure}

In analogy to the color sector in Ref.$\,$\cite{us1}
we study the evolution of the eigenvalues in Fig.$\,$\ref{fig:chargino}
numerically by varying the mass parameters along the path
\begin{eqnarray}
{\mathcal{P}}_C \; :&& \; M'_2 = my/(1+y)\,,  \nonumber \\
                    && \; M_2  = -my\,,       \nonumber \\
                    && \; M^D_{2} = m\,,      \nonumber \\
                    && \; \mu    = 2 m\,,
\label{eq:chargino_path}
\end{eqnarray}
for a fixed value of $m=200$ GeV with the control parameter $-1 \leq
y \leq 0$ running from the MSSM [$y=-1$] to the Dirac limit [$y=0$].
This set corresponds to mass parameters giving rise to
$m_{\tilde\chi^\pm_1} \approx m$ (fixed),
$m_{\tilde\chi^\pm_2} \approx m [y+1/(1+y)]$, moving from $\infty$ to $m$,
and $m_{\tilde\chi^\pm_3} \approx \mu$ (fixed) in the decoupled
wino and higgsino sectors and for very small $v_I$. The other parameters
in the chargino mass matrix (\ref{eq:chargino_mass_matrix}) are chosen
as $\tan\beta=5$, $v_I=v_Y=3$ GeV and the $N{=}2$ values for the couplings
$\lambda_{Y,I}$ are adopted.

For the parameters chosen, the descending order of the physical masses
in the figure reflects, in obvious notation, the pattern $w' \gg h > w$
in the MSSM limit. At some medium $y$, the states $w'$ and $h$ cross over
to $h > w'$, keeping the ordering $h > w' >w$ until the Dirac limit is
reached. The physical masses in the cross-over zone of the states $w'$
and $h$ cannot be described by the standard analytical expansion applied
above. They must either be obtained numerically or by analytical expansions
tailored specifically for cross-over phenomena, see Ref.$\,$\cite{Haber}.

\subsubsection{Neutralinos}

\noindent
Six neutral electroweak Majorana fields are incorporated in the $N{=}1$/$N{=}2$
hybrid model. The mass matrix can be extracted from the bi-linear terms of the
gaugino, gaugino$'$ and higgsino fields in the Lagrangian of the preceding
subsection, written in the current basis
$\{\tilde{B'},\tilde{B},{\tilde{W'}}^0,\tilde{W}^0,{\tilde{H}}^0_u,
{\tilde{H}}^0_d\}$ as
\begin{equation}
{\mathcal{M}}_N =
 \begin{pmatrix}
     M'_1  & M^D_{1} & 0  & 0  & -\frac{1}{\sqrt{2}} \lambda_Y v_d &
            -\frac{1}{\sqrt{2}} \lambda_Y v_u \\
     M^D_{1} & M_1 &  0  &  0 & \frac{1}{2} g' v_u &
              -\frac{1}{2} g' v_d \\
     0  &  0  & M'_2 & M^D_{2} & -\frac{1}{\sqrt{2}} \lambda_I v_d &
            -\frac{1}{\sqrt{2}} \lambda_I v_u \\
     0  &  0  & M^D_{2} & M_2  & -\frac{1}{2} g v_u &
            \frac{1}{2} g v_d \\
    -\frac{1}{\sqrt{2}} \lambda_Y v_d &  \frac{1}{2} g' v_u &
    -\frac{1}{\sqrt{2}} \lambda_I v_d & -\frac{1}{2} g v_u &
          0 & -\mu_n \\
    -\frac{1}{\sqrt{2}} \lambda_Y v_u & -\frac{1}{2} g' v_d &
    -\frac{1}{\sqrt{2}} \lambda_I v_u &  \frac{1}{2} g v_d &
         -\mu_n & 0
     \end{pmatrix}\,,
 \label{eq:neutralino_mass_matrix}
\end{equation}
where
\begin{eqnarray}
M_1= M_{\tilde{B}},\quad M'_1 = M'_{\tilde{B}} + M_Y,\quad M^D_{1} = M^D_{Y},
\quad \mu_n = \mu+(\lambda_Y v_Y + \lambda_I v_I)/\sqrt{2}\,.
\label{eq:neutralino_mass_parameter}
\end{eqnarray}
and $M_2, M'_2$ are defined in Eq.$\,$(\ref{eq:chargino_parameter_defined}).
This $6 \times 6$ mass matrix is diagonalized by the unitary transformation
\begin{equation}
{\mathcal{M}}^{\rm diag}_N = U_N^T\, \mathcal{M}_N\, U_N\,.
\label{eq:neutralino_mixing_matrix}
\end{equation}
Six neutralinos, $i.\,e.$ two degrees of freedom more than in MSSM, are
predicted in the general $N{=}1$/$N{=}2$ hybrid model, labeled
${\tilde{\chi}}_{1\cdots 6}^0$ (ordered according to
ascending mass values). They evolve from the MSSM by raising the
magnitude of the gaugino mass parameters $M'_{1,2}$ from $-\infty$
to finally $0$ in the Dirac limit.

In general, the diagonalization of the $6\times 6$ neutralino mass matrix
cannot be carried out in analytic form.
However, as before, in the limit in which the supersymmetry masses are
much larger than the electroweak scale, approximate solutions can be found
analytically. First switching off the electroweak mixings among the
bino, wino and higgsino sectors leaves us with two bino mass eigenvalues,
two wino mass eigenvalues and two higgsino mass eigenvalues:
\begin{eqnarray}
\overline{m}^0_{1,2}
   &=& \frac{1}{2} \left|  |M_1+M'_1| \mp \delta_1 \right|\,,
             \\
\overline{m}^0_{3,4}
   &=& \frac{1}{2} \left|  |M_2+M'_2| \mp \delta_2 \right|\,,
             \\[1mm]
\overline{m}^0_{5,6}
   &=& \mu\,,
\end{eqnarray}
with $\delta_{1,2} = \sqrt{(M'_{1,2}-M_{1,2})^2+ 4 (M^D_{1,2})^2}
\,$, and the block-diagonal mixing matrix
\begin{eqnarray}
\overline{U}_N
  = {\rm diag}\left( \overline{U}_1, \overline{U}_2, \overline{U}_h\right)\quad
  \mbox{with}\quad
\overline{U}_{1,2} = \left(\begin{array}{cc}
            c_{1,2}  & - i s_{1,2} \\
            s_{1,2}  & \phantom{-}i c_{1,2}
                \end{array}\right)\ \ \mbox{and} \ \
\overline{U}_h = \left(\begin{array}{cc}
           i/\sqrt{2} & -1/\sqrt{2} \\
           i/\sqrt{2} & \phantom{-}1/\sqrt{2}
            \end{array}\right)\,,
\label{eq:UNnovev}
\end{eqnarray}
with the mixing angles $c_{1,2}/s_{1,2}
= \sqrt{[1\pm (M'_{1,2}-M_{1,2})/\delta_{1,2}]/2}$.

Switching on the weak couplings among the bino, wino gaugino
sectors and the higgsino sector, the mass eigenvalues  and mixing
parameters are calculated using the block-diagonalization method
described in Appendix B.
The results of this procedure are relegated to Appendix C.

\begin{figure}[t]
\epsfig{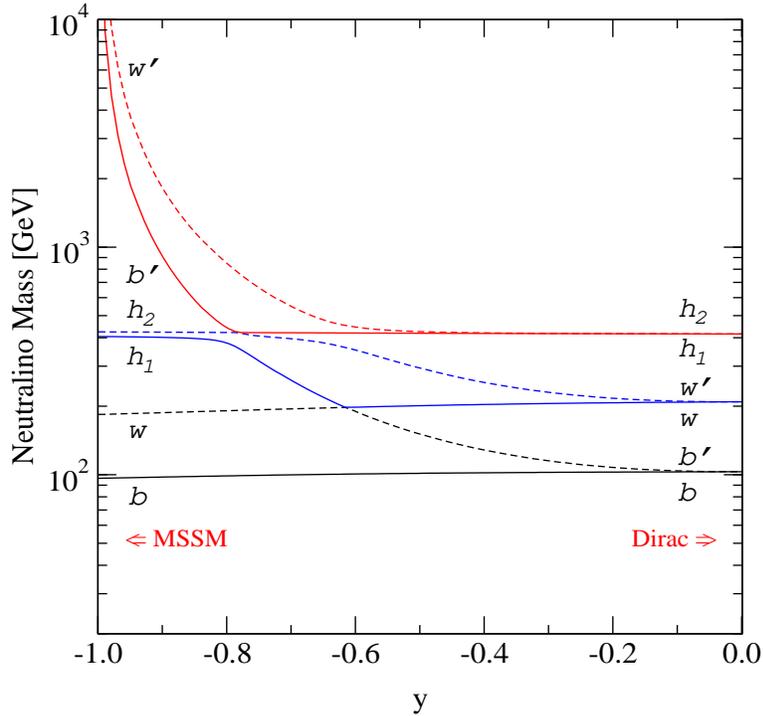}
\caption{\it Evolution of the neutralino masses as a function of the control
             parameter $y$ from the MSSM ($y=-1$) quartet to the Majorana
             sextet, merging to the Dirac triplet in the ($y=0$)
             limit, for the same path (\ref{eq:chargino_path}) as in the
             chargino sector with bino/wino relations chosen as
             $M^{(D)}_1 / M^{(D)}_2 = 1/2$.}
\label{fig:neutralino}
\end{figure}

The numerical evolution of the neutralinos in the hybrid model is
displayed in Fig.~\ref{fig:neutralino} as a function of the control
parameter $y$  for the same path and parameter set as in the
chargino sector, Eq.$\,$(\ref{eq:chargino_path}), and supplemented
by the bino/wino mass relations $M^{(D)}_1 \approx M^{(D)}_2 /2$, and
setting $\tan\beta=5$ and $v_I=v_Y=3$ GeV.

The evolution of the neutralino masses follows the same pattern as
the charginos, though being more complex due to the increased number
of states. Starting from the mass pattern $w' > b' \gg h_1 \sim h_2 >
w >b$ of the neutral states in the MSSM limit for the parameters
chosen above, the first cross-over is observed for $b' \leftrightarrow
h_1$, followed by $w' \leftrightarrow h_2$ and $b' \leftrightarrow w$
at roughly the same position. The mass system moves to the final pattern
$h_1 = h_2 > w' = w > b' = b$ in the Dirac limit.

The transition from the Majorana to the Dirac theory in the limits
$M'_2, M_2$ and $M'_1, M_1 \to 0$ can easily be studied by analyzing
the mass matrix ${\mathcal{M}}_N$ for vanishing gaugino/higgsino
mixing. The eigenvalues of the matrix come in pairs of opposite
signs: $\pm m_j$ for $j = 1,2,3$. The Majorana fields in each pair,
denoted by ${\tilde{\chi}}_\pm$ according to the sign of the
eigenvalue, can be combined to one Dirac field,
\begin{equation}
{\tilde{\chi}}_D
 =  \left(\tilde{\chi}_+ + i \tilde{\chi}_- \right)/\sqrt{2}\,,
\end{equation}
the superposition giving rise to vanishing contractions
$\langle {\tilde{\chi}}_D {\tilde{\chi}}_D \rangle =0$, as required
for Dirac fields.

The $\pm$ pairing of the eigenstates is not restricted to the hybrid neutralino
mass matrix with vanishing gaugino/higgsino mixing but it is also realized
if the gaugino/higgsino mixing is switched on and the couplings
$\lambda_{Y,I}$ are given by the $N{=}2$ relations, $\lambda_Y =- g'/\sqrt{2}$
and $\lambda_I = g/\sqrt{2}$ \cite{Belanger:2009wf}. The key is the
vanishing of the coefficients of odd powers of the eigenvalues in the
characteristic eigenvalue equation:
\begin{equation}
\det ({\mathcal{M}}_N - m) =
 r_0 + r_2 m^2 + r_4 m^4 + r_6 m^6
 = 0\,,
\end{equation}
with the coefficients given by
\begin{eqnarray}
r_6 &=& 1 \,,                                       \nonumber \\
r_4 &=& - \frac{1}{2} {\rm tr} ({\mathcal{M}}_N^2)
     = - [(M^D_{Y})^2 + (M^D_{I})^2+\mu^2_n + m^2_Z] \,, \nonumber\\
r_2 &=&   \frac{1}{8} [ {\rm tr} ({\mathcal{M}}_N^2) ]^2
        - \frac{1}{4} {\rm tr} ({\mathcal{M}}_N^4)  \nonumber \\
    &=& (M^D_{Y} M^D_{I})^2 + [(M^{D}_{Y})^2 + (M^{D}_{I})^2] \mu^2_n
       +2 m^2_Z [(M^{D}_{Y})^2 c^2_W + 2 (M^{D}_{I})^2 s^2_W]
       -2 m^2_Z (M^D_{Y} s^2_W + M^D_{I} c^2_W) \mu_n c_{2\beta}
       + m^4_Z\,,\nonumber\\[2mm]
r_0 &=& - \frac{1}{48} [ {\rm tr} ({\mathcal{M}}_N^2) ]^3
        + \frac{1}{12} [ {\rm tr} ({\mathcal{M}}_N^3) ]^2
        + \frac{1}{8} {\rm tr} ({\mathcal{M}}_N^4)\,
                      {\rm tr} ({\mathcal{M}}_N^2)
        - \frac{1}{6} {\rm tr} ({\mathcal{M}}_N^6)\nonumber\\
     &=& -(M^{D}_{Y} M^{D}_{I} \mu_n)^2
         + 2 m^2_Z M^D_{Y} M^D_{I} \mu_n (M^D_{Y} c^2_W
         + M^D_{I} s^2_W) c_{2\beta}
         - m^4_Z (M^D_{Y} c^2_W + M^D_{I} s^2_W)^2 \,.
\end{eqnarray}
The odd coefficients $r_{2j+1}$ are linear in traces of odd powers
of ${\mathcal{M}}_N$ which vanish. While this is obvious for ${\rm
tr} ({\mathcal{M}}_N)$ in the Dirac limit $M_{1,2},M'_{1,2} \to 0$,
it can easily be proven also for odd powers of
the mass matrix if the submatrix that mixes the mass submatrix of
the gauginos with the mass submatrix of the higgsinos is orthogonal.
This is satisfied in Eq.$\,$(\ref{eq:neutralino_mass_matrix}), a
sufficient [but not necessary] condition for orthogonality being the
$N{=}2$ symmetry of the basic Lagrangian.

If the lightest neutralino is the lightest supersymmetric particle
(LSP) and stable, its Dirac or Majorana  nature  has important
consequences for cold dark matter phenomenology. This is most
clearly seen by inspecting, for example, the neutralino annihilation cross
section into an electron-positron pair in the MSSM and Dirac
limits. Assuming for simplicity a
pure bino-type MSSM neutralino state $\tilde{\chi}^0_1=\tilde{B}^0$,
we obtain
\begin{eqnarray}
\frac{d\sigma}{d\cos\theta}[\tilde{\chi}^0_1\tilde{\chi}^0_1\to e^-e^+]
   &=&\frac{\pi^2\alpha}{16 c_W^2 s}\, \beta^3\,
      \left[ \frac{\eta^2_{1L} + (\eta^2_{1L}-4\eta_{1L}+2-\beta^2)\cos^2\theta
                   + \beta^2\cos^4\theta}{
                   (\eta^2_{1L} - \beta^2\cos^2\theta)^2} \right.\nonumber\\
  &&\left.\qquad\quad
      +16\, \frac{\eta^2_{1R} + (\eta^2_{1R}-4\eta_{1R}+2-\beta^2)\cos^2\theta
                   + \beta^2\cos^4\theta}{
                   (\eta^2_{1R} - \beta^2\cos^2\theta)^2}\right]\,,
                   \label{eq:DMmaj}
\end{eqnarray}
due to the $t$- and $u$-channel $L$- and $R$-chiral selectron exchange
where $c_W=\cos\theta_W$, $\theta$ is the c.m. scattering angle,
$\beta=(1-4m_{\tilde{\chi}^0_1}^2/s)^{1/2}$ and $\eta_{1L,R}=1+
2(m^2_{\tilde{e}_{L,R}}-m^2_{\tilde{\chi}^0_1})/s$.
On the other hand, in the Dirac theory with the pure bino-type
Dirac neutralino state $\tilde{\chi}^0_{D1}= \tilde{B}'^0_L + \tilde{B}^0_R$
we obtain for the annihilation cross section
\begin{eqnarray}
 \frac{d\sigma}{d\cos\theta}
 [\tilde{\chi}^0_{D1}\tilde{\chi}^{0c}_{D1}\to e^-e^+]
&=&\frac{\pi\alpha^2}{32 c_W^2 s}\,\beta\,
   \left[\frac{(1-\beta\cos\theta)^2}{(\eta_{1L}-\beta\cos\theta)^2}
      +16\frac{(1+\beta\cos\theta)^2}{(\eta_{1R}+\beta\cos\theta)^2}\right]\,.
\label{eq:DMdir}
\end{eqnarray}
In other words, in the limit of $\beta\to 0$, the annihilation cross section
(\ref{eq:DMmaj}) in the MSSM shows a $P$-wave suppression behavior
$\sim \beta^3$, while in the Dirac case, the cross section (\ref{eq:DMdir})
shows only a $S$-wave suppression behavior $\sim \beta$. As a result, the
$P$--wave suppression of the MSSM LSP annihilation cross sections requires
a significant fine-tuning of the spectra to be consistent with the WMAP
observations \cite{Dunkley:2008ie}. In contrast, the annihilation of Dirac
gauginos into a fermion and anti-fermion pair has a non-vanishing $S$-wave
contribution even in the limit of vanishing fermion masses.  Thus, the
annihilation to fermions does not require the chirality flip in the final
state, giving rise to enhanced decay branching fractions to leptons.
This opens the parameter space that fits the WMAP measurements
\cite{hsieh,Belanger:2009wf}. Moreover, in contrast to the Majorana
case, Dirac gauginos with non-vanishing higgsino fraction can lead
to spin-independent scattering cross sections off nuclei via the
$Z$-boson exchange \cite{Chun}, thus significantly altering the prospects
for dark matter detection experiments.

\subsubsection{Scalar/Higgs Particles}

\noindent
The scalar/Higgs sector involves various components in the basic Lagrangian:
terms derived from the $N{=}2$ Higgs-Higgs-scalar interactions, the superpotential,
the $D$-terms and the soft breaking terms. Expanding the scalar/Higgs potential
about the vacuum expectation values of the neutral fields, $v_{u/d},v_{Y/I}$,
linear and bi-linear terms of the physical fields associated with the masses
are generated, while tri- and quadri-linear terms describe the
self-interactions of the physical scalar/Higgs fields.

To stabilize the system, the coefficients of the linear terms must vanish;
this condition connects the vacuum expectation values with the basic parameters
of the Lagrangian:
\begin{eqnarray}
v_Y &=& \frac{v^2}{ 4\tilde{m}^2_Y \tilde{m}^2_I - \lambda^2_Y\lambda^2_I v^4}
    \left\{ 2 \tilde{m}^2_I \left[g' M^D_{Y} c_{2\beta} -\sqrt{2}\lambda_Y \mu
                          +(M_Y + A_Y)\lambda_Y s_{2\beta}/\sqrt{2}\right]
                           \right.\nonumber\\
      && \left. \hskip 2.9cm  +\lambda_Y \lambda_I v^2 \left[ g M^D_{I}
                               c_{2\beta}
                          + \sqrt{2}\lambda_I \mu -(M_I + A_I)\lambda_I
                            s_{2\beta}/\sqrt{2}\right]\right\}\nonumber\\
      &\sim& \frac{v^2}{2\tilde{m}^2_Y}
       \left[g' M^D_{Y} c_{2\beta} -\sqrt{2}\lambda_Y \mu
                          +(M_Y + A_Y)\lambda_Y s_{2\beta}/\sqrt{2}\right]
             \quad\mbox{for}\quad \tilde{m}_{Y,I}\gg \lambda_{Y,I} v
                          \,,
\end{eqnarray}
\begin{eqnarray}
v_I &=& \frac{v^2}{4\tilde{m}^2_Y \tilde{m}^2_I - \lambda^2_Y\lambda^2_I v^4}
    \left\{ 2 \tilde{m}^2_Y \left[-g M^D_{I} c_{2\beta} -\sqrt{2}\lambda_I \mu
                          +(M_I + A_I)\lambda_I s_{2\beta}/\sqrt{2}\right]
                           \right.\nonumber\\
      && \left. \hskip 2.9cm -\lambda_Y \lambda_I v^2 \left[g' M^D_{Y}
                             c_{2\beta} -\sqrt{2}\lambda_Y \mu
                          +(M_Y + A_Y)\lambda_Y s_{2\beta}/\sqrt{2}\right]
                          \right\}\nonumber\\
      &\sim& -\frac{v^2}{2\tilde{m}^2_I}
       \left[g M^D_{I} c_{2\beta} +\sqrt{2}\lambda_I \mu
                          -(M_I + A_I)\lambda_I s_{2\beta}/\sqrt{2}\right]
            \quad\mbox{for}\quad \tilde{m}_{Y,I}\gg \lambda_{Y,I} v\,,
\label{eq:v_Y and v_I}
\end{eqnarray}
with the abbreviations $c_{2\beta}=\cos 2\beta$ and $s_{2\beta}=\sin 2\beta$, and
\begin{eqnarray}
\tilde{m}^2_Y &=& m^2_Y + m'^2_Y + M^2_Y + 4 (M^{D}_{Y})^2
                 + \tfrac{1}{2}\lambda^2_Y v^2\,,\\
\tilde{m}^2_I &=& m^2_I + m'^2_I + M^2_I + 4 (M^{D}_{I})^2
                 + \tfrac{1}{2}\lambda^2_I v^2\,.
\end{eqnarray}
The Higgs {\it vevs} $v_{u,d}$ are determined by
\begin{eqnarray}
0 &=& (m^2_{H_u}+\mu^2) v_u - B_\mu v_d +\frac{1}{8}(g'^2+g^2)(v^2_u-v^2_d) v_u
      +\frac{1}{2} (\lambda^2_Y +\lambda^2_I) v_u v^2_d\nonumber\\
  && +(\sqrt{2}\lambda_Y \mu + g' M^D_{Y}) v_Y v_u
     +(\sqrt{2}\lambda_I \mu - g M^D_{I}) v_I v_u \nonumber\\
  && -\frac{1}{\sqrt{2}} (M_Y+A_Y) \lambda_Y v_Y v_d
     -\frac{1}{\sqrt{2}} (M_I+A_I) \lambda_I v_I v_d
     +\frac{1}{2}(\lambda_Y v_Y + \lambda_I v_I)^2 v_u\,,\\
0 &=& (m^2_{H_d}+\mu^2) v_d - B_\mu v_u -\frac{1}{8}(g'^2+g^2)(v^2_u-v^2_d)v_d
      +\frac{1}{2} (\lambda^2_Y +\lambda^2_I) v^2_u v_d\nonumber\\
  && +(\sqrt{2}\lambda_Y \mu - g' M^D_{Y}) v_Y v_d
     +(\sqrt{2}\lambda_I \mu + g M^D_{I}) v_I v_d \nonumber\\
  && -\frac{1}{\sqrt{2}} (M_Y+A_Y) \lambda_Y v_Y v_u
     -\frac{1}{\sqrt{2}} (M_I+A_I) \lambda_I v_I v_u
     +\frac{1}{2}(\lambda_Y v_Y + \lambda_I v_I)^2 v_d\,,
\end{eqnarray}
after inserting the {\it vevs} $v_{Y,I}$ from Eq.$\,$(\ref{eq:v_Y and v_I}).
The values of $v_{u,d}$ and $v_I$ can be determined
phenomenologically in terms of the observables
$\tan\beta$ and $m^2_W, m^2_Z$, {\it vide}  Eqs.$\,$(\ref{eq:vevs_tanb})
and (\ref{eq:vevs_mzw}).

The terms in the Lagrangian which are bi-linear in the fields build up the
scalar/Higgs mass matrices. Decomposing the neutral fields
into ground-state values, real and imaginary parts,
\begin{eqnarray}
H^0_u &=& \frac{1}{\sqrt{2}} \left[s_\beta (v+  h) + c_\beta H
                            + i (c_\beta A -s_\beta a)\right] \,,
        \qquad
H^+_u = c_\beta\, H^+ - s_\beta\, a^+    \,,  \\
H^0_d &=& \frac{1}{\sqrt{2}} \left[c_\beta (v+  h) - s_\beta H
                            + i (s_\beta A + c_\beta a)\right]\,,
        \qquad
H^-_d = s_\beta\, H^- + c_\beta\, a^-\,,
\end{eqnarray}
with the abbreviations $c_\beta=\cos\beta$ and $s_\beta=\sin\beta$,
and
\begin{eqnarray}
\sigma^0_Y   &=& \frac{1}{\sqrt{2}} (v_Y + s_Y + i a_Y)      \,, \\
\sigma^3_I &=& \frac{1}{\sqrt{2}} (v_I + s_I + i a_I)      \,, \;\;\;
\sigma^1_I  = \frac{1}{\sqrt{2}} (\sigma^+_2 + \sigma^-_1) \,, \;\;\;
\sigma^2_I  = \frac{i}{\sqrt{2}} (\sigma^+_2 - \sigma^-_1) \,,
\end{eqnarray}
it can be ascertained that the matrix of the imaginary fields involves a
massless Goldstone field $a$, and likewise the charged fields involve
$a^\pm_G = [v a^\pm + \sqrt{2} v_I
  (\sigma^\pm_1+\sigma^\pm_2)]/\sqrt{v^2 + 4 v^2_I}$.  These
are absorbed to provide masses to the neutral and charged gauge
bosons. The neutral fields $h,H,s_Y,s_I$ are parity-even scalars
while $A,a_Y,a_I$ are parity-odd pseudoscalars; the charged fields
$H^\pm$ mix with the associated charged scalar fields $s^\pm_{1,2}$,
defined later in Eq.$\,$(\ref{eq:charged_sigma_scalar}).
These elements build up the neutral pseudoscalar $3 \times 3$ mass matrix,
the neutral scalar $4 \times 4$ mass matrix and the charged scalar
$3\times 3$ mass matrix:

\noindent
{\it (i) \uline{neutral pseudoscalars:}}

\noindent
In the $\{A, a_Y, a_I\}$ basis, the $3\times 3$ real and symmetric pseudoscalar
mass matrix squared is given by
\begin{equation}
{\mathcal{M}}^2_P =
\begin{pmatrix}
     M^2_A   & -\frac{1}{\sqrt{2}} (M_Y-A_Y) \lambda_Y v
             & -\frac{1}{\sqrt{2}} (M_I-A_I) \lambda_I v \\
    -\frac{1}{\sqrt{2}} (M_Y-A_Y) \lambda_Y v
             & \tilde{m}'^2_Y
             & \frac{1}{2} \lambda_Y \lambda_I v^2 \\
    -\frac{1}{\sqrt{2}} (M_I-A_I) \lambda_I v
             &  \frac{1}{2} \lambda_Y \lambda_I v^2
             & \tilde{m}'^2_I
    \end{pmatrix}\,,
\end{equation}
where
\begin{eqnarray}
M^2_A &=& 2 \left[B_\mu + \lambda_Y v_Y (M_Y+A_Y)/\sqrt{2} + \lambda_I v_I (M_I+A_I)/\sqrt{2}
            \right]/s_{2\beta}\,,\\  
\tilde{m}'^2_Y &=&  m^2_Y - m'^2_Y + M^2_Y + \tfrac{1}{2}\lambda^2_Y v^2\,,
\\
\tilde{m}'^2_I &=& m^2_I-m'^2_I + M^2_I + \tfrac{1}{2}\lambda^2_I v^2\,.
\end{eqnarray}

\noindent
This matrix can easily be diagonalized in approximate form
in the limit of the genuine supersymmetry parameters, $m_{Y,I}$ being much
larger than the electroweak scale $v$, $i.\,e.$ $v/m_{Y,I} \ll 1$. This
leaves us with three approximately unmixed states with their masses
\begin{eqnarray}
\overline{M}^2_{A_1}\, = \, M^2_A\,,\qquad
\overline{M}^2_{A_2}\, = \, \tilde{m}^{\prime 2}_Y\,,\qquad
\overline{M}^2_{A_2}\, = \, \tilde{m}'^2_I\,.
\end{eqnarray}
The expressions for the mass eigenvalues and mixing elements, when
the weak coupling among the $\{A, a_Y,a_I\}$ states is retained,
are given in Appendix C.

\noindent
{\it (ii) \uline{neutral scalars:}}

\noindent
In the $\{h, H, s_Y, s_I\}$ basis, the real and symmetric $4\times 4$ scalar
mass matrix squared ${\mathcal{M}}^2_S$ is given by
\begin{eqnarray}
{\mathcal{M}}^2_S =
\begin{pmatrix}
 m^2_Z + \delta_H s_{2\beta} & \delta_H c_{2\beta}
   & -  \frac{v_Y}{v} (2\tilde{m}^2_Y - \lambda^2_Y v^2)
   & -  \frac{v_I}{v} (2\tilde{m}^2_I - \lambda^2_I v^2)  \\
 \delta_H c_{2\beta} & M^2_A - \delta_H s_{2\beta}
   & \Delta_Y & \Delta_I \\
 -  \frac{v_Y}{v} (2\tilde{m}^2_Y-\lambda^2_Y v^2)  & \Delta_Y
   & \tilde{m}^2_Y
   & \frac{1}{2} \lambda_Y \lambda_I v^2 \\
 -  \frac{v_I}{v} (2\tilde{m}^2_I-\lambda^2_I v^2)
   & \Delta_I & \frac{1}{2} \lambda_Y \lambda_I v^2
   & \tilde{m}^2_I
      \end{pmatrix}\,,
\label{eq:scalar_mass_matrix}
\end{eqnarray}
where
\begin{eqnarray}
\delta_H &=& \left[(\lambda^2_Y + \lambda^2_I) v^2 - 2 m^2_Z\right] s_{2\beta}/2\,,\\
\Delta_Y &=& g' M^D_{Y} v\, s_{2\beta}
            - \tfrac{1}{\sqrt{2}} \lambda_Y (M_Y + A_Y) v\, c_{2\beta}\,,\\
\Delta_I &=&- g M^D_{I} v\,  s_{2\beta}
            - \tfrac{1}{\sqrt{2}} \lambda_I (M_I + A_I) v\, c_{2\beta}\,.
\label{eq:Delta_I}
\end{eqnarray}
Note that $\delta_H$ vanishes in the $N{=}2$ SUSY limit. Thus, in this limit,
the
eigenvalues of the Higgs submatrix $\{h,H\}$ are just $m^2_Z$ and
$M^2_A$~\cite{ant}, with no dependence on $\tan\beta$,
a feature markedly different from the
MSSM. This submatrix receives several radiative
corrections, the most important one accruing from stop/top loops,
due to their large Yukawa couplings. As a result, the Higgs
submatrix is modified to
\begin{equation}
\begin{pmatrix}
m^2_Z + \delta_H s_{2\beta} & \delta_H c_{2\beta}  \\
\delta_H c_{2\beta} & M^2_A - \delta_H s_{2\beta}
      \end{pmatrix}\ \ \to\ \
\begin{pmatrix}
  m^2_Z + \delta_H s_{2\beta}+\epsilon_H
    & \delta_H c_{2\beta} + \epsilon_H/t_\beta  \\
\delta_H c_{2\beta} + \epsilon_H /t^2_\beta
    & M^2_A - \delta_H s_{2\beta} + \epsilon_H /t_\beta
      \end{pmatrix}\,,
\end{equation}
where
\begin{eqnarray}
\epsilon_H \simeq \frac{3G_F m^4_t}{\sqrt{2}\pi^2}
                  \ln \frac{m_{\tilde{t}_1} m_{\tilde{t}_2}}{m^2_t}\,.
\end{eqnarray}
The transition from the current basis to the diagonal $2 \times 2$ Higgs matrix
with eigenvalues
\begin{eqnarray}
\overline{M}^2_{S_1}
   &\approx& m^2_Z + \delta_H s_{2\beta} +\epsilon_H
            -\frac{(\delta_H c_{2\beta} + \epsilon_H/t_\beta)^2}{
                    M^2_A- m^2_Z}\,, \\
\overline{M}^2_{S_2}
   &\approx& M^2_A - \delta_H s_{2\beta}+\epsilon_H/t^2_\beta
            +\frac{(\delta_H c_{2\beta} + \epsilon_H/t_\beta)^2}{
                    M^2_A- m^2_Z}\,,
\end{eqnarray}
is carried out by an orthogonal transformation  with
the mixing element given by
\begin{equation}
\tan\theta_h
 = \frac{m^2_Z + \delta_H s_{2\beta} +\epsilon_H-\overline{M}^2_{S_1}}
        {|\delta_H c_{2\beta} + \epsilon_H/t_\beta|} \,,
\end{equation}
with $0\leq \theta_h \leq \pi/2$.

In the limit of $m_{Y,I}$ being much larger than the electroweak scale $v$,
the $\{s_Y, s_I\}$ submatrix leads to the two  approximate mass eigenvalues
\begin{eqnarray}
\overline{M}^2_{S_3}\, = \, \tilde{m}^2_Y\,,\qquad
\overline{M}^2_{S_4}\, = \, \tilde{m}^2_I\,.
\end{eqnarray}

The $\{h,H\}$ and $\{s_Y, s_I\}$ systems are weakly coupled at the
order $v/M_A,v/m_{Y,I}$, and the block diagonalization allows to derive
the results given in Appendix C.

\noindent
{\it (iii) \uline{charged scalars:}}

\noindent
After the charged Goldstone bosons $a^\pm_G$ are absorbed into the charged
gauge bosons, there remain three physical charged scalar states $\{H^\pm,
s^\pm_1, s^\pm_2\}$ with the second and third states defined by
\begin{eqnarray}
s^\pm_1 = (\sigma^\pm_1-\sigma^\pm_2)/\sqrt{2}
                   \quad  \mbox{and}\quad
s^\pm_2 = \frac{v (\sigma^\pm_1+\sigma^\pm_2)/\sqrt{2}
                   -2 v_I a^\pm}{\sqrt{v^2+ 4 v^2_I}}\,.
\label{eq:charged_sigma_scalar}
\end{eqnarray}
The real symmetric $3\times 3$ charged scalar mass matrix squared
${\mathcal{M}}^2_{H^\pm}$ is then given in the $\{H^\pm, s^\pm_1, s^\pm_2\}$
basis by
\begin{equation}
{\mathcal{M}}^2_{H^\pm} =
\begin{pmatrix}
 \tilde{M}^2_{H^\pm} & \Delta_\pm
                     & -\sqrt{\rho}\, \Delta_I \\
 \Delta_\pm & \tilde{m}'^2_I + g^2 v^2_I
     & \frac{1}{2}\sqrt{\rho}
        \left(\lambda^2_I-\frac{1}{2} g^2\right) v^2 c_{2\beta}\\
-\sqrt{\rho}\, \Delta_I & \frac{1}{2}\sqrt{\rho}
                \left(\lambda^2_I-\frac{1}{2} g^2\right) v^2 c_{2\beta}
              & \rho\, \tilde{m}^2_I
      \end{pmatrix}\,,
\end{equation}
where
\begin{eqnarray}
\tilde{M}^2_{H^\pm} &=& M^2_A + m^2_W+\frac{1}{2}(\lambda^2_I-\lambda^2_Y) v^2
          - 4\frac{v^2_I}{v^2} \tilde{m}^2_I + 4 \lambda^2_I v^2_I
          -4\sqrt{2} \mu_n \lambda_I v_I\,, \nonumber \\
\Delta_\pm &=& \left(g^2/2-\lambda^2_I\right) v_I v s_{2\beta}
                       -(M_I-A_I) \lambda_I v/\sqrt{2} \,,
\end{eqnarray}
and $\Delta_I$ is introduced in Eq.$\,$(\ref{eq:Delta_I}). We note in passing that
in the $N=$ SUSY scenario with $\lambda_I = g/\sqrt{2}$ the charged
states $s^\pm_{1,2}$ do not mix.

Assuming that $\tilde{m}^2_I > \tilde{m}'^2_I > \tilde{M}^2_{H^\pm}$
and $M_I, A_I\sim M_A$, and observing that, again, the charged Higgs/scalar
states are weakly coupled at the order of $v/\tilde{m}_I$
or $v/\tilde{m}'_I$, the block-diagonalization procedure provides
approximate solutions as given in Appendix C.

\begin{figure}[t]
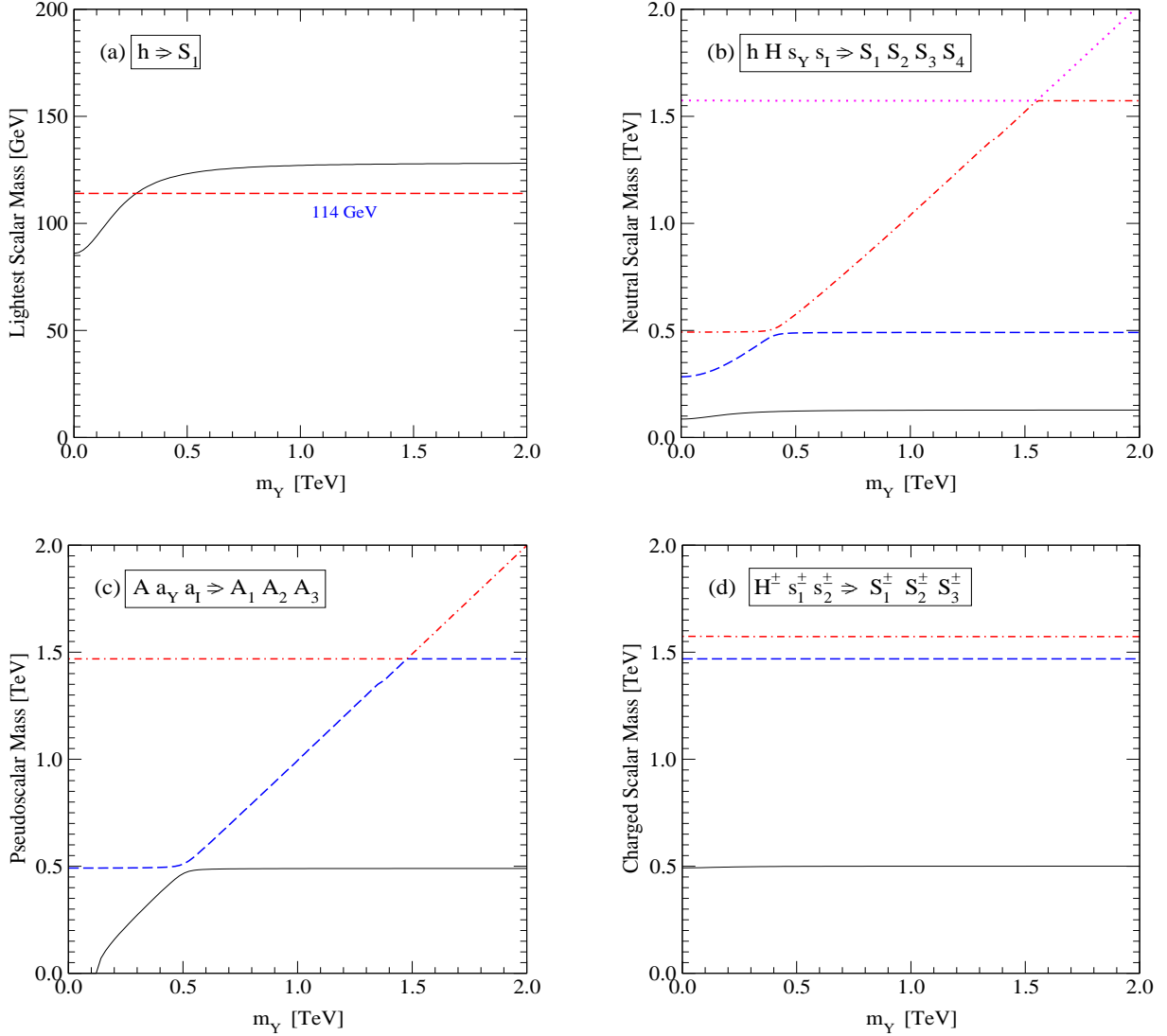

\begin{center}
\includegraphics[width=7.5cm,height=7.cm]{ssmass11.eps}
\hskip 1.0cm
\includegraphics[width=7.5cm,height=7.cm]{ssmass21.eps}\\[5mm]
\includegraphics[width=7.5cm,height=7.cm]{spmass1.eps}
\hskip 1.0cm
\includegraphics[width=7.5cm,height=7.cm]{scmass1.eps}
\caption{\it (a) The lightest neutral scalar boson including one-loop
           top/stop radiative corrections. The red dashed line indicates the
           present experimental lower bound on the mass $M_{S_1}\gtrsim 114$
           GeV; (b) the neutral scalar masses; (c) the neutral pseudoscalar
           masses; (d) the charged scalar masses, as a function of the
           hypercharge soft scalar mass $m_Y$; the isospin soft scalar mass
           is set to $m_I=6v$ to accommodate the small $\rho$-parameter and
           the $N{=}2$ values for the couplings $\lambda_{Y,I}$ in
           Eq.$\,$(\ref{eq:N=2_couplings}) are adopted. The other
           parameters are fixed to $\tan\beta=5$,
           $m_{\tilde{t}_1}=m_{\tilde{t}_2}= 1$~TeV, $M_A=2v$,
           $m'_Y=M^D_Y=v/2$, $m'_I=M^D_I=\mu=v$, $A_Y=A_I=2v$ and
           $M_Y=M_I=0$ for Dirac gauginos.
}
\label{fig:sigma}
\end{center}
\end{figure}

The extension of the Higgs sector by the novel SU(2)$_I\times$U(1)$_Y$
adjoint sigma fields has two important consequences:
\begin{itemize}
\item Each of the neutral pseudoscalar/scalar and charged sectors are
  extended by two new states with masses of the order of the
  characteristic scalar parameters $m_Y$ and $m_I$. As a result, one
  of the new pseudoscalar/scalar states may acquire mass between a few
  hundred GeV up to several TeV, while the other will be heavy, {\it i.e.},
  ${\cal O}$(TeV);
  both the new charged states will be heavy likewise.
\item The mass matrix of the Higgs system is modified compared to the
  MSSM. As pointed out before, the tree-level Higgs masses are
  independent of the mixing parameter $\tan\beta$. In addition, the lower
  bound on the [lightest] charged Higgs mass is not guaranteed
  to exceed the $W$ mass any more [experimentally of course, any charged
  Higgs boson with mass below $\sim 100$ GeV is excluded by direct
  searches \cite{Amsler:2008zzb}.]
\end{itemize}

The tableau in Fig.$\,$\ref{fig:sigma} illustrates the evolution of the
three neutral pseudoscalar, four neutral scalar, and three charged masses with the
hyper-singlet mass parameter $m_Y$ introduced in the soft SUSY
breaking Lagrangian while all other parameters are kept fixed.
These parameters have been chosen as indicated in the figure caption.

\section{CHARACTERISTIC PHYSICAL PROCESSES}

\noindent New colored particles like gluinos and squarks are
expected to be generated, and detected, at the LHC
for masses up to 2--3 TeV, and the strikingly different phenomenology of
novel Dirac gluinos and colored adjoint scalars have been discussed
in Refs.$\,$\cite{us1,us2}. In contrast, the mass window for generating
non-colored states like charginos/neutralinos directly in
quark-antiquark collisions is much smaller as a result of the small
electroweak production cross sections. Cascade decays of colored
states, however, provide a copious source of non-colored particles
with large masses {\it i.e.} through the decay $\tilde{q} \to q +
\tilde{\chi}$. Pair production of
non-colored states at TeV $e^+e^-$ and $e^-e^-$ lepton colliders
ILC/CLIC, on the other hand, gives access to the non-colored sector
up to masses close to half the c.m. energy, $i.\,e.$ about 0.5 TeV
and 1.5 TeV at the ILC and CLIC, respectively, while non-colored
adjoint scalars can be produced with high masses in $\gamma\gamma$
collisions. Without specifying the
relative size of the masses of the new particles, a myriad of
possible cascade decays would be predicted, which can, nevertheless,
be analyzed phenomenologically by applying quite similar techniques.
To present a transparent overview we, therefore, focus on
representative chains in which sigma masses generally exceed the
chargino/neutralino masses, as motivated already earlier.

\subsection{Charginos and Neutralinos}

\noindent
Below, explicit formulae will be given for the $N{=}1$ MSSM and the
$N{=}2$ Dirac limit, while scenarios interpolating between the MSSM and
the Dirac limit could be obtained by summing up the two individual
chargino/Majorana neutralino contributions after the proper
diagonalization of the hyper-system.

In the hybrid theory, only the original $N{=}1$ chargino and
neutralino fields couple to the matter fields.  The analysis is
simplified considerably by restricting ourselves to interactions with
first and second generation (s)fermions. In this sector, which is most
relevant experimentally, only the gauge components of charginos and
neutralinos couple to the matter fields.

In the limit of large supersymmetry scales (in relation to
the electroweak scale), the Dirac chargino fields and their
charge conjugates are given by
\begin{eqnarray}
&& \cw^-_{D1} = \ww'^-_L + \ww^-_R\,, \quad\qquad
   \cw^+_{D1} =-\ww^+_L - \ww'^+_R\,,  \label{chaD1} \\
&& \cw^-_{D2} = \ww^-_L + \ww'^-_R\,, \quad\qquad
   \cw^+_{D2} =-\ww'^+_L - \ww^+_R\,, \label{chaD2} \\
&& \cw^-_{D3} =\tilde{H}^-_{dL} + \tilde{H}^-_{uR}\,,\quad\qquad
   \cw^+_{D3} = \phantom{+}\tilde{H}^+_{uL} + \tilde{H}^+_{dR}\,,
   \label{chaD3}
\end{eqnarray}
whereas  the Dirac neutralino fields and their charge conjugates are
\begin{eqnarray}
&& \cw^0_{D1} = \bw'_L + \bw_R\,, \qquad\qquad\ \
   \cw^{0c}_{D1} = -\bw_L - \bw'_R\,,\label{neuD1}\\
&& \cw^0_{D2} = \ww'^0_L + \ww^0_R\,, \qquad\qquad\,
   \cw^{0c}_{D2} = -\ww^0_L - \ww'^0_R\,,\label{neuD2}\\
&& \cw^0_{D3} = i(\tilde{H}^0_{dL}-\tilde{H}^0_{uR})\,,\qquad\ \
   \cw^{0c}_{D3} = i(\tilde{H}^0_{uL}-\tilde{H}^0_{dR})\,, \label{neuD3}
\end{eqnarray}
up to terms of order $v/M_{\rm SUSY}$.
Expressed in terms of these fields, the
Lagrangians for matter-chargino/neutralino interactions in the MSSM Majorana
limit and in the Dirac theory can be written as
\begin{align}
{\mathcal{L}}^{\rm C}_{\rm Majo}  &=
 g \, \overline{\rule{0mm}{1.5ex}u_L} \, \cw^+_1 \tilde{d}_L
+ g \, \overline{\cw^+_1} \, u_L\, \tilde{d}_L^*
- g \, \overline{d_L} \, \cw^-_1 \tilde{u}_L
- g \, \overline{\cw^-_1} \, d_L\, \tilde{u}_L^* \,,
\\
{\mathcal{L}}^{\rm C}_{\rm Dirac} &=
 g \, \overline{\rule{0mm}{1.5ex}u_L} \, \cw^+_{D2} \tilde{d}_L
+ g \, \overline{\cw^+_{D2}} \, u_L\, \tilde{d}_L^*
- g \, \overline{d_L} \, \cw^-_{D1} \tilde{u}_L
- g \, \overline{\cw^-_{D1}} \, d_L\, \tilde{u}_L^* \,,
\end{align}
and
\begin{align}
{\mathcal{L}}^{\rm N}_{\rm Majo}  &=
- g_{Li}\, \overline{f_L} \, \cw^0_i \, \fw_L
- g_{Li}^*\, \overline{\cw^0_i} \, f_L\, \fw_L^*
+ g_{Ri}\, \overline{f_R}\, \cw^0_i \, \fw_R
+ g_{Ri}^*\, \overline{\cw^0_i} \, f_R\, \fw_R^* \,,
\\
{\mathcal{L}}^{\rm N}_{\rm Dirac} &=
- g_{Li}\, \overline{f_L} \, \cw^0_{Di} \, \fw_L
- g_{Li}^*\, \overline{\cw^0_{Di}} \, f_L\, \fw_L^*
+ g_{Ri}\, \overline{f_R}\, \cw_{Di}^{0c} \, \fw_R
+ g_{Ri}^*\, \overline{\cw_{Di}^{0c}} \, f_R\, \fw_R^* \,,
\end{align}
where
\begin{equation}
g_{Li} = \sqrt{2}\left[ g'Y_{f_L} \delta_{i1}
  + gI^3_f \delta_{i2}
  \right]
\quad {\rm and} \quad
g_{Ri} = \sqrt{2}\, g'Y_{f_R} \delta_{i1}
  \,.
\end{equation}
Here $u/\tilde{u}$ correspond to up-type (s)quarks or (s)neutrinos, whereas
$d/\tilde{d}$ denote down-type (s)quarks or charged (s)leptons.
As mentioned above, mixings from electroweak symmetry breaking as well as from
the CKM matrix have been neglected.

In the approximation described by the Dirac Lagrangians a Dirac charge
$D$ \cite{us1} can be defined which is conserved in all processes:
\begin{align}
D[\qw^{1,2}_L] &= D[\tilde{\ell}^{1,2}_L] = D[\tilde{\nu}^{1,2}]
    = D[\cw^{0c}_D] = D[\cw^+_{D1}] = D[\cw^-_{D2}]=-1 \,, \\
D[\qw^{1,2}_R] &= D[\tilde{\ell}^{1,2}_R] \phantom{\,\,= D[\tilde{\nu}^{1,2}] }
    = D[\cw^{0}_D] = D[\cw^-_{D1}] = D[\cw^+_{D2}] = +1  \,.
\end{align}
Antiparticles carry the corresponding opposite Dirac charges $-D$.
The Dirac charges of all SM particles vanish. The squarks
${\tilde{q}}^{1,2}$, sleptons ${\tilde{\ell}}^{1,2}$, and sneutrinos
${\tilde{\nu}}^{1,2}$ belong to the first and second generation.
$L,R$ mixing and large couplings to higgsinos preclude the extension
of this approximate scheme to the third generation. Nevertheless,
the scheme proves useful for a quick overview of allowed and
forbidden processes in the first two generations. For example, in
the Dirac limit, the production processes $e^-_L e^-_L \to
{\tilde{e}}^-_L {\tilde{e}}^-_L$  and $e^-_R e^-_R \to
{\tilde{e}}^-_R {\tilde{e}}^-_R$ with equal helicities are forbidden
while the opposite-helicity process $e^-_L e^-_R \to {\tilde{e}}^-_L
{\tilde{e}}^-_R$ is allowed.

\subsubsection{Squark Cascade Decays at LHC}

\noindent
Cascade decays, see {\it e.g.} Ref.~\cite{barr}, are crucial for the analysis
of the non-colored supersymmetry
sector at LHC. Following the rules discussed earlier, we will study invariant
masses of quark-jets with charged leptons in squark cascade decays:
\begin{align}
\hspace{5em}
&\text{\uline{Charginos:}}
 &&\text{MSSM:}
   && {\tilde{u}}_L \to d \, \cw^+_1
    \to d \, \nu_l \, \tilde{l}^+_L ,\;
        d \, l^+ \, \tilde{\nu}_l
    \to d \, l^+ \, \nu_l \, \cw^0_1 \,, \hspace{10em} \nonumber \\
&&&&& {\tilde{d}}_L \to u \, \cw^-_1
    \to u \, \bar{\nu}_l \, \tilde{l}^-_L ,\;
        u \, l^- \, \tilde{\nu}_l^*
    \to u \, l^- \, \bar{\nu}_l \, \cw^0_1 \,, \\
&&&\text{Dirac:}
   && {\tilde{u}}_L \to d \, \cw^+_{D1}
    \to d \, l^+ \, \tilde{\nu}_l
    \to d \, l^+ \, \nu_l \, \cw^{0c}_{D1} \,, \nonumber \\
&&&&& {\tilde{d}}_L \to u \, \cw^-_{D2}
    \to u \, \bar{\nu}_l \, \tilde{l}^-_L
    \to u \, l^- \, \bar{\nu}_l \, \cw^{0c}_{D1} \,,
\\[1em]
&\text{\uline{Neutralinos:}}
 &&\text{MSSM:}
   && \qw_L \to q \, \cw^0_2
    \to q \, l^\pm \, \tilde{l}^\mp_L
    \to q \, l^\pm \, l^\mp \, \cw^0_1 \,, \\
&&&\text{Dirac:}
   && \qw_L \to q \, \cw^{c0}_{D2}
    \to q \, l^+ \, \tilde{l}^-_L
    \to q \, l^+ \, l^- \, \cw^0_1 \,.
\end{align}
Due to CP invariance, the charge conjugated versions of these
processes are obtained simply by flipping the gauge/Dirac charges
and chiralities at each step.

\begin{figure}[t]
\begin{tabular}{c@{\hspace{5em}}c}
\epsfig{figure=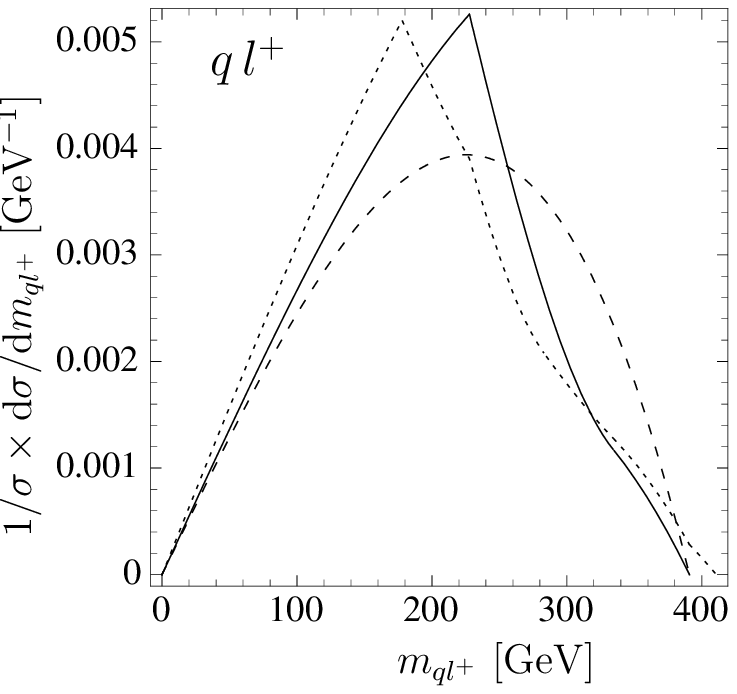, width=7.35cm} &
\epsfig{figure=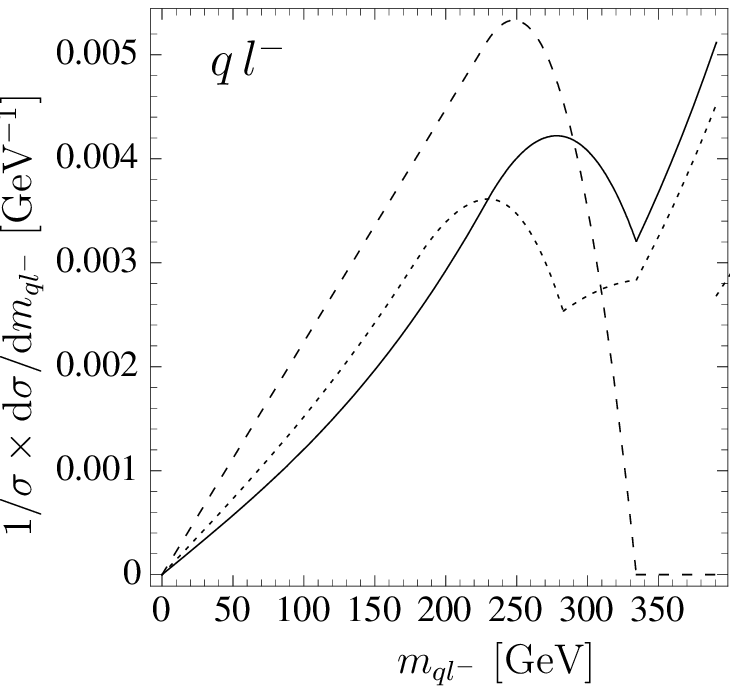, width=7.35cm} \\
\begin{tabular}{l@{\hspace{1em}}l@{\hspace{1em}}l}
 -------- &$\tilde{u} \to d \tilde{\chi}_1^+ \to ...$ &[MSSM] \\[.5ex]
 -- -- -- &$\tilde{u} \to d \tilde{\chi}_{D1}^+ \to ...$
  & [Dirac model] \\[.5ex]
 $\cdot\cdot\cdot\cdot\cdot$
  &$\tilde{u} \to d \tilde{\chi}_{1,2}^+ \to ...$ &[hybrid model]
\end{tabular} &
\begin{tabular}{l@{\hspace{1em}}l@{\hspace{1em}}l}
 -------- &$\tilde{d} \to u \tilde{\chi}_1^- \to ...$ &[MSSM] \\[.5ex]
 -- -- -- &$\tilde{d} \to u \tilde{\chi}_{D2}^- \to ...$
  & [Dirac model] \\[.5ex]
 $\cdot\cdot\cdot\cdot\cdot$
  &$\tilde{d} \to u \tilde{\chi}_{1,2}^- \to ...$ &[hybrid model]
\end{tabular}
\end{tabular}
\caption{\it Quark-lepton invariant mass distributions for squark decay chains
             with intermediate charginos, comparing the $N{=}1$ MSSM (solid lines)
             with the $N{=}2$ Dirac gaugino theory (dashed lines) and the
             intermediate hybrid theory (dotted lines). Numerical inputs for
             the plots are $m_{\qw} = 565$~GeV, $m_{\cw^\pm_1} = m_{\cw^\pm_{D1}}
             = m_{\cw^\pm_{D2}} = 184$~GeV, $m_{\tilde{l}}= m_{\tilde{\nu}}
             =125.3$~GeV, and $m_{\cw^0_1} = m_{\cw^0_{D1}} = 97.7$~GeV. For the
             case of the hybrid model, the second chargino mass is $m_{\cw^\pm_2}
             = 199$~GeV, corresponding to a mixing angle $\cos\theta_2=0.6$.
             Electroweak symmetry breaking effects on the chargino and neutralino
             mixing matrices have been neglected.}
\label{fig:ql}
\end{figure}

As evident from the list above, the decay chains differ in their
chirality structure between the MSSM and the Dirac theory, which will leave a
characteristic imprint on the angular distributions of visible decay jets and
leptons. For the squark-chargino cascades  this is illustrated by the
quark-lepton invariant mass distributions shown in Fig.$\,$\ref{fig:ql}.

Also shown in the figure is an example of the general 2-Majorana hyper-system
away from the Dirac limit. In this case one obtains two wino-like charginos
$\chi^\pm_{1,2}$ with distinct masses. The dotted lines in the plots corresponds to
a scenario with relatively small departure from the Dirac limit, so that the
two chargino masses are of the same order and the $\ww/\ww'$ mixing angle is close to
maximal mixing.

Nevertheless, the distributions of the 2-Majorana hyper-system are closer
to the MSSM in the plots, while
the Dirac limit leads to drastically different distributions. This can be
understood from the fact that the two independent charginos $\cw^\pm_1$ and
$\cw^\pm_2$ in the hybrid model become degenerate in the exact Dirac limit.
Interference effects lead to large mixing between the two states in this limit.
However, a slight deviation from the Dirac limit is already sufficient to effectively
turn off these interference contributions, since the width of both charginos is
relatively small.

\begin{figure}[t]
\centering
\epsfig{figure=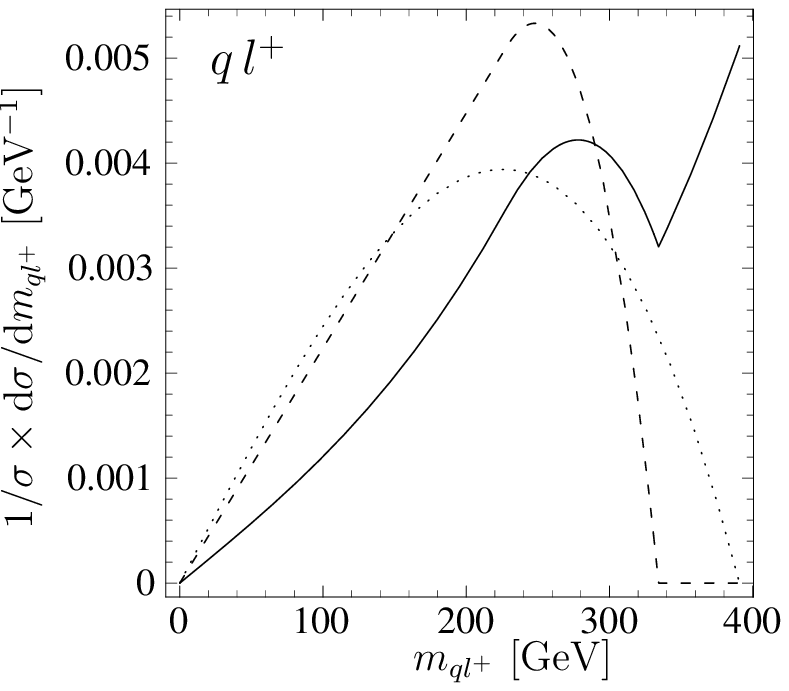, width=7.35cm}
\hspace{4em}
\epsfig{figure=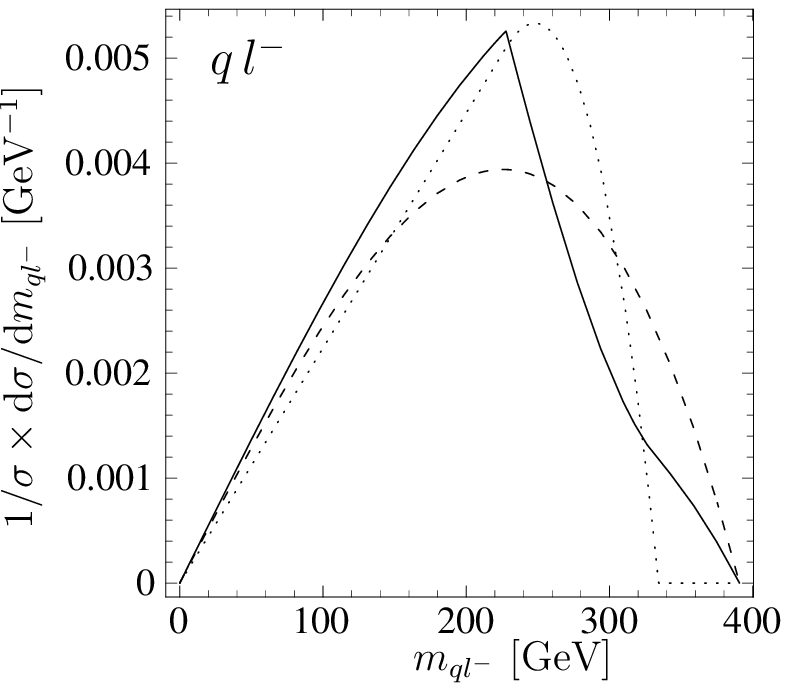, width=7.35cm}
\\
\makebox[17.5cm]{\scriptsize --------
$\;\; \tilde{q}_L \to q \tilde\chi^0_2 \to q  l^\mp_n  \tilde l_R^\pm
               \to q l^\mp_n l^\pm_f  \tilde\chi^0_1$
\hfill
-- -- --
$\;\;\; \tilde{q}_L \to q \tilde\chi^{c0}_{D2}
               \to q l^-_n \tilde l^+_R
               \to q l^-_n l^+_f \tilde\chi^{c0}_{D1}$
\hfill
$\cdot\cdot\cdot\cdot\cdot$
$\;\; \tilde{q}_L^* \to \bar{q} \tilde\chi^0_{D2}
               \to \bar{q} l^+_n \tilde l^-_R
               \to \bar{q} l^+_n l^-_f \tilde\chi^0_{D1}$}
\caption{\it $ql$ invariant mass distributions for squark decay chains involving
             Majorana or Dirac neutralinos. In the $N{=}1$ MSSM (solid lines) the
             squark and anti-squark decay chains lead to identical distributions,
             in contrast to the $N{=}2$ Dirac gaugino theory (dashed and dotted
             lines).  Numerical inputs for the plots are $m_{\qw} = 565$~GeV,
             $m_{\cw^0_2} = m_{\cw^0_{D2}} = 184$~GeV, $m_{\tilde{l}}= 125.3$~GeV,
             and $m_{\cw^0_1} = m_{\cw^0_{D1}} = 97.7$~GeV. Electroweak symmetry
             breaking effects on the chargino and neutralino mixing matrices
             have been neglected.}
\label{fig:qll}
\end{figure}

The squark-neutralino cascades have been worked out in Ref.$\,$\cite{us1},
and are reproduced in Fig.$\,$\ref{fig:qll}. Again, the plots show distinct
differences between the MSSM and Dirac limits, which can be exploited to
experimentally distinguish the two cases at the LHC.

\subsubsection{Selectron Pair-Production in $e^- e^-$ and $e^+ e^-$ Collisions}

\noindent
Conservation of the Dirac charge $D$ in the first generation forbids the
production of selectrons in equal-helicity $e^- e^-$ collisions but allows
the production in opposite-helicity collisions in the Dirac theory, while all
three helicity combinations are non-trivially realized in Majorana theories:
\begin{eqnarray}
e^-_L e^-_L &\to& {\tilde{e}}^-_L {\tilde{e}}^-_L \,, \;
e^-_R e^-_R \to {\tilde{e}}^-_R {\tilde{e}}^-_R\,,        \\
e^-_L e^-_R &\to& {\tilde{e}}^-_L {\tilde{e}}^-_R\,.
\end{eqnarray}
Three other independent processes are possible in $e^- e^+$ collisions:
\begin{eqnarray}
e^-_L e^+_L &\to& {\tilde{e}}^-_L {\tilde{e}}^+_R \,, \\
e^-_L e^+_R &\to& {\tilde{e}}^-_L {\tilde{e}}^+_L \,, \;
e^-_R e^+_L \to {\tilde{e}}^-_R {\tilde{e}}^+_R   \,.
\end{eqnarray}
Noting that $(\psi_{L/R})^c = {(\psi}^c)_{R/L}$, the additional process
$e^-_R e^+_R \to {\tilde{e}}^-_R {\tilde{e}}^+_L$ in the second group is
the CP-conjugate of the first process and needs not be analyzed separately.
Since non-zero $s$-channel $\gamma, Z$ exchange requires opposite lepton
helicities, the first electron/positron process is driven only by neutralino
exchanges while the other two processes are mediated by both $t$-channel
neutralino and $s$-channel vector-boson exchanges. Moreover, the first
process violates conservation of the $D$ charge and thus is forbidden in
the Dirac theory. Simulations of some processes have been presented in
Refs.~\cite{sese,sese1}.

\noindent
{\it (i) \uline{\it{$e^-e^-$ collisions :}}}

\noindent
Recalling the definitions introduced in Ref.~\cite{us1},
the $e^- e^-$ scattering amplitudes for selectron pair production
in the general hybrid hyper-system on which we have based the detailed
analyses, can be written as
\begin{align}
{\cal A}[e^-_L e^-_L \to {\tilde{e}}^-_L {\tilde{e}}^-_L]
&= -2 e^2 \left[{\cal M}_{LL}(s,t) + {\cal M}_{LL}(s,u)\right]\,,
\\[1ex]
{\cal A}[e^-_R e^-_R \to {\tilde{e}}^-_R {\tilde{e}}^-_R]
&= \phantom{+} 2 e^2 \left[{\cal M}^*_{RR}(s,t) + {\cal M}^*_{RR}(s,u)\right]\,,
\end{align}
for same helicity-pairs and
\begin{align}
{\cal A}[e^-_L e^-_R \to {\tilde{e}}^-_L {\tilde{e}}^-_R]
&= \phantom{+} e^2 \lambda^{1/2}_{LR}\, \sin\theta\, {\cal D}_{LR}(s,t)\,,
\\[1ex]
{\cal A}[e^-_R e^-_L \to {\tilde{e}}^-_L {\tilde{e}}^-_R]
&=-e^2 \lambda^{1/2}_{LR}\, \sin\theta\, {\cal D}_{RL}(s,u) \,,
\end{align}
for opposite helicity-pairs, with the two-body final state kinematic
factor $\lambda_{ab} =\lambda(1, m^2_{\tilde{e}_a}/s,
m^2_{\tilde{e}_b}/s)$ $[a,b=L,R]$ and
\begin{equation}
\lambda(1,x,y)=1+x^2+y^2-2(x+y+xy)\,. \label{eq:lambda}
\end{equation}
Here $\theta$ is the scattering angle, and the dimensionless neutralino
functions ${\cal M}_{ab}$ and ${\cal D}_{ab}$ ($a,b=L,R$) \cite{Peskin:1998jy}
are defined by
\begin{eqnarray}
&& {\cal M}_{ab} (s,t/u)=\sum_{k=1}^6 \frac{m_{\tilde{\chi}^0_k}}{\sqrt{s}}
              {\cal V}_{ak} {\cal V}_{bk}\, D_{kt/u}\,,\\
&& {\cal D}_{ab}(s,t/u) =\sum_{k=1}^6
              {\cal V}_{ak} {\cal V}^*_{bk}\, D_{kt/u}\,,
\end{eqnarray}
They are determined by the normalized neutralino propagators
$D_{kt} = s/(t-m^2_{\tilde{\chi}^0_k})$, and similarly for $D_{ku}$, and
the effective mixing coefficients
\begin{eqnarray}
{\cal V}_{Lk} = U_{N2k}/(2c_W) + U_{N4k}/(2s_W)\,,\qquad
{\cal V}_{Rk} = U_{N2k}/c_W\,.
\end{eqnarray}
The neutralino mixing matrix elements $U_{N\alpha k}$, introduced in
\eqref{eq:neutralino_mixing_matrix},
have a very simple structure if effects from electroweak symmetry breaking
are neglected, see Eq.~\eqref{eq:UNnovev}.

After calculating the polarization averaged squared matrix elements and
including the phase space factor the differential cross sections are
\begin{align}
\frac{d\sigma_{LL}}{d\cos\theta}
&= \frac{\pi\alpha^2}{4 s} \lambda^{1/2}_{LL}
   \left|\, {\cal M}_{LL}(s,t) + {\cal M}_{LL}(s,u)\,\right|^2\,,
\\[1ex]
\frac{d\sigma_{RR}}{d\cos\theta}
&=  \frac{\pi\alpha^2}{4 s} \lambda^{1/2}_{RR}
   \left|\, {\cal M}_{RR}(s,t) + {\cal M}_{RR}(s,u)\,\right|^2\,,
\\[1ex]
\frac{d\sigma_{LR}}{d\cos\theta}
&=  \frac{\pi\alpha^2}{4 s} \lambda^{3/2}_{LR} \sin^2\theta
  \left[ |{\cal D}_{LR}(s,t)|^2 + |{\cal D}_{RL}(s,u)|^2\right]\,.
\end{align}
Finally, the unpolarized total cross sections can be obtained by
performing the remaining integration over the scattering angle
$\theta$. Note that $\sigma_{LR}$ and $\sigma_{RL}$ are not
physically distinguishable in the $e^-e^-$ case, unlike for $e^+e^-$
annihilation. The cross sections reduce, on the one side, to the
familiar MSSM form, see Ref.$\,$\cite{sese}, while in the Dirac theory,
on the other side, they simplify considerably to
\begin{align}
\sigma[e^- e^- \to {\tilde{e}}^-_L {\tilde{e}}^-_L]
&= \sigma[e^- e^- \to {\tilde{e}}^-_R {\tilde{e}}^-_R] = 0\,, \\
\sigma[e^- e^- \to {\tilde{e}}^-_L {\tilde{e}}^-_R]
&= \frac{\pi\alpha^2}{2c_W^4s} \left[
   (1+2m_{\cw^0_{D1}}^2/s-m_{\tilde{e}_L}^2/s-m_{\tilde{e}_R}^2/s)\, L'_{D1}
 -2\beta'\right]\,,
\end{align}
with $\beta' = \lambda^{1/2}_{LR}$ and the logarithmic function defined by
\begin{eqnarray}
L'_i = \log
\frac{ 1+\beta'+(2m_{\cw^0_i}^2 - m_{\tilde{e}_L}^2-m_{\tilde{e}_R}^2)/s}{
       1-\beta'+(2m_{\cw^0_i}^2 - m_{\tilde{e}_L}^2-m_{\tilde{e}_R}^2)/s} \,.
\end{eqnarray}
The vanishing of the $LL$ and $RR$ cross sections is obvious from $D$-charge
conservation. In the absence of higgsino exchanges only the bino-exchange can
drive the $LR$ process.

\begin{figure}[t]
\begin{center}
\epsfig{figure=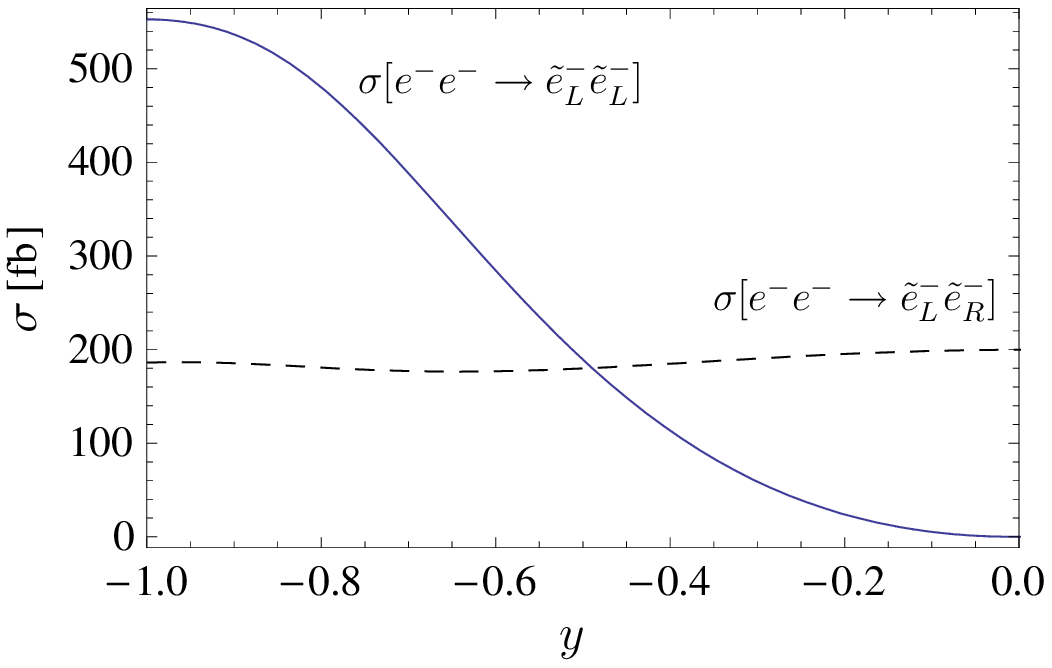, width=8.7cm, height=7.2cm}
\hfill
\epsfig{figure=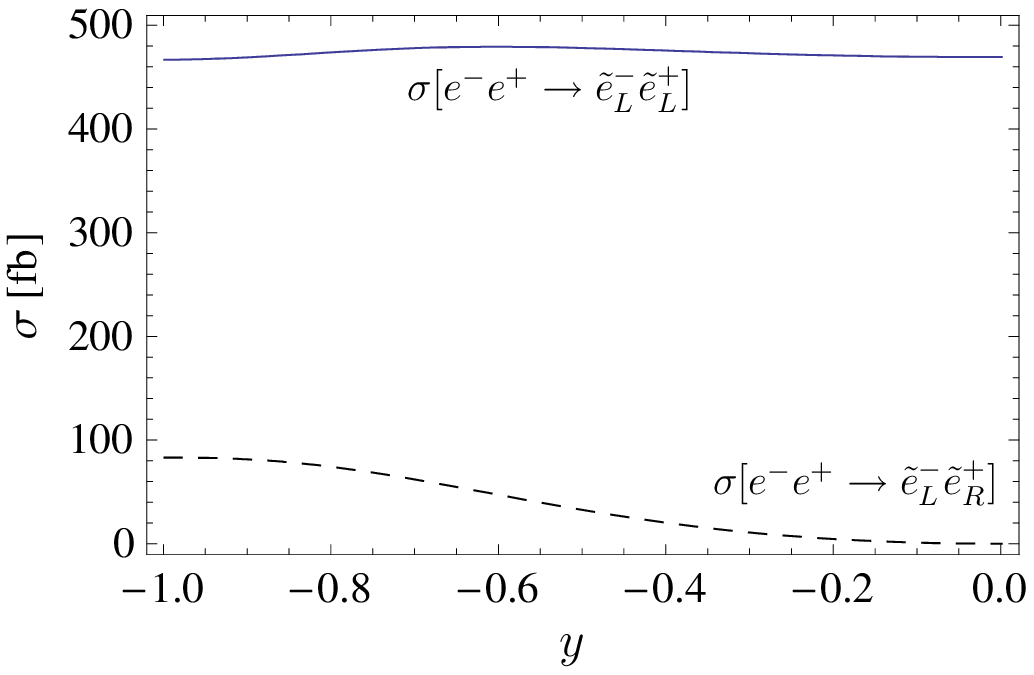, width=8.7cm, height=7.2cm}
\end{center}
\vspace{-3ex}
\caption{\it Dependence of the cross sections for same-sign (left) and
             opposite-sign (right) selectron production on the Dirac/Majorana
             control parameter~$y$, for $\sqrt{s} = 500$ GeV and SPS1a$'$
             parameters \cite{sps}. Not shown are the cross sections for
             $e^-e^\pm \to \tilde{e}^-_R \tilde{e}^\pm_R$, which, apart from
             the different normalization, shows a similar behavior as the cross
             section for $e^-e^- \to \tilde{e}^-_L \tilde{e}^\pm_L$.}
\label{fig:sese}
\end{figure}

The evolution of the total cross section from the MSSM to the Dirac limit is
illustrated  for the two characteristic processes $e^- e^- \to
{\tilde{e}}^-_L {\tilde{e}}^-_L$ and $e^- e^- \to {\tilde{e}}^-_L
{\tilde{e}}^-_R$ in the left panel of Fig.$\,$\ref{fig:sese},
which demonstrates how the first process is switched
off when the Dirac limit is approached.

\noindent
{\it (ii) \uline{\it{$e^+e^-$ collisions :}}}

\noindent
The analysis of the $e^- e^+$ processes follows the same path.
By introducing a normalized $s$-channel $Z$ boson propagator
$D_Z=s/(s-m^2_Z+i m_Z\Gamma_Z)$ and four bi-linear charges
\begin{eqnarray}
&& Z^+_{LL} = 1 + \frac{s^2_W-1/2}{c^2_W} D_Z\,, \quad
   Z^-_{LL} = 1 + \frac{(s^2_W-1/2)^2}{c^2_W s^2_W} D_Z\,, \\
&& Z^+_{RR} = 1 + \frac{s^2_W}{c^2_W} D_Z\,,\ \ \quad \quad\quad
   Z^-_{RR} = 1 + \frac{s^2_W-1/2}{c^2_W} D_Z\,,
\end{eqnarray}
we obtain six non-vanishing helicity amplitudes
\begin{eqnarray}
&& {\cal A}[e^-_L e^+_R \to {\tilde{e}}^-_L {\tilde{e}}^+_L]
  = -e^2 \lambda^{1/2}_{LL} \sin\theta
    \left[{\cal D}_{LL}(s,t) + Z^-_{LL}\right]\,,
\\[1ex]
&& {\cal A}[e^-_R e^+_L \to {\tilde{e}}^-_L {\tilde{e}}^+_L]
  = -e^2 \lambda^{1/2}_{LL} \sin\theta Z^+_{LL}\,,
\\[1ex]
&& {\cal A}[e^-_L e^+_R \to {\tilde{e}}^-_R {\tilde{e}}^+_R]
  = -e^2 \lambda^{1/2}_{RR} \sin\theta Z^-_{RR}\,,
\\[1ex]
&& {\cal A}[e^-_R e^+_L \to {\tilde{e}}^-_R {\tilde{e}}^+_R]
  = -e^2 \lambda^{1/2}_{RR} \sin\theta
    \left[ {\cal D}_{RR}(s,t)+Z^+_{RR}\right]\,,
\\[1ex]
&& {\cal A}[e^-_L e^+_L \to {\tilde{e}}^-_L {\tilde{e}}^+_R]
  = \phantom{+} 2 e^2 {\cal M}_{LR}(s,t)\,,
\\[1ex]
&& {\cal A}[e^-_R e^+_R \to {\tilde{e}}^-_R {\tilde{e}}^+_L]
  =-2 e^2 {\cal M}^*_{RL}(s,t)\,.
\end{eqnarray}
By squaring the helicity amplitudes, the differential cross sections can easily
be derived. The squares are summed incoherently if the initial lepton
helicities are not specified experimentally.

As before, the cross sections reduce to the familiar MSSM limit on one side,
while in the Dirac limit, on the other side, the processes with $LR/RL$ initial
state helicities remain allowed, but the $LL$ and $RR$ processes are excluded
by $D$-charge conservation.  The $D$-charge of the pair ${\tilde{e}}^-_L
{\tilde{e}}^-_R$ vanishes, thus allowing production in $e^- e^-$ collisions,
but the pair ${\tilde{e}}^-_L {\tilde{e}}^+_R$ carries the charge $D = 2$ so
that production of this pair in $e^- e^+$ collisions is forbidden.

The continuous transition from the MSSM to the Dirac limit is illustrated in
the right panel of Fig.$\,$\ref{fig:sese}, for the two representative total cross
sections of $e^- e^+ \to {\tilde{e}}^-_L {\tilde{e}}^+_R$ and $e^- e^+
\to {\tilde{e}}^-_L {\tilde{e}}^+_L$.

\subsubsection{Chargino and Neutralino Production in $e^+e^-$ Collisions}

\noindent

Direct production of
chargino and neutralino pairs in $e^+e^-$ annihilation are ideal laboratories
to study the properties of these particles,
see {\it e.g.} Ref.~\cite{choi}.
As will be shown here, the characteristic differences between the Dirac theory and the
MSSM also become evident in these processes.

The chargino reactions proceed in general through
$s$-channel $\gamma, Z$ and $t$-channel ${\tilde{\nu}}_e$ exchanges.
Focusing on the gaugino sector, doubled in the general hybrid theory
compared to the MSSM, the production cross sections for diagonal and non-diagonal
charged gaugino pairs are given by
\begin{align}
\frac{d\sigma}{d\cos\theta}[e^+ e^- \to \cw^+_1 \cw^-_1]
&=  \begin{aligned}[t]
\frac{\pi\alpha^2\lambda_{11}^{1/2}}{16s_W^4s} \Biggl [
 &\frac{[s^2-4s_W^2m_Z^2s+8s_W^4m_Z^4][2
         -\lambda_{11}\,\sin^2\theta]}{(s-m_Z^2)^2}
\label{c1c1}
 \\
 +2 \,& s_2^2 \, \frac{[s-2s_W^2m_Z^2][1-\lambda_{11}
                   +(1-\lambda_{11}^{1/2}\cos\theta)^2]} {
                   (s-m_Z^2)(\eta_{11}-\lambda^{1/2}_{11} \cos\theta)}
 + 2 s_2^4 \, \frac{(1-\lambda_{11}^{1/2}\cos\theta)^2}{
                   (\eta_{11}-\lambda^{1/2}_{11}\cos\theta)^2}
 \Biggr ]\,,
\end{aligned} \displaybreak[0] \\[1ex]
\frac{d\sigma}{d\cos\theta}[e^+ e^- \to \cw^+_2 \cw^-_2]
&=  \begin{aligned}[t]
\frac{\pi\alpha^2\lambda_{22}^{1/2}}{16s_W^4s} \Biggl [
 &\frac{[s^2-4s_W^2m_Z^2s+8s_W^4m_Z^4][2-
       \lambda_{22} \sin^2\theta]}{(s-m_Z^2)^2}
 \\
 +2 \,& c_2^2 \, \frac{[s-2s_W^2m_Z^2][1-\lambda_{22}
                   +(1-\lambda_{22}^{1/2}\cos\theta)^2]} {
                   (s-m_Z^2)(\eta_{22}-\lambda^{1/2}_{22}\cos\theta)}
 + 2 c_2^4 \, \frac{(1-\lambda_{22}^{1/2}\cos\theta)^2}{
                   (\eta_{22}-\lambda^{1/2}_{22}\cos\theta)^2}
 \Biggr ]\,,
\end{aligned} \displaybreak[0] \\[1ex]
\frac{d\sigma}{d\cos\theta}[e^+ e^- \to \cw^\pm_1 \cw^\mp_2]
&=  \begin{aligned}[t]
\frac{\pi\alpha^2\lambda_{12}^{1/2}}{4s_W^4s}
 \, c_2^2s_2^2 \, \frac{(1-\lambda_{12}\cos\theta)^2 -
   (m_{\cw_1^\pm}^2-m_{\cw_2^\pm}^2)^2/s^2}{
   (\eta_{12}-\lambda^{1/2}_{12}\cos\theta)^2}\,,
\end{aligned}
\end{align}
and the production cross section of a charged higgsino pair by
\begin{eqnarray}
\frac{d\sigma}{d\cos\theta}[ e^+ e^- \to \cw^+_3 \cw^-_3 ]
= \frac{\pi\alpha^2\lambda_{33}^{1/2}}{16s}\,
  \frac{(8 s^4_W-4 s^2_W+1)(s^2_W-1/2)^2}{c^4_W s^4_W}\,
  \frac{s^2(2-\lambda_{33}\sin^2\theta)}{
        (s-m^2_Z)^2}\,,
\end{eqnarray}
with $\eta_{ij} = 1+ (2 m^2_{\tilde{\nu}}-m^2_{\tilde{\chi}^\pm_i}
-m^2_{\tilde{\chi}^\pm_j})/s$,
where we ignore the $Z$ boson width and introduce the usual
K\'allen functions $\lambda_{ij} = \lambda^{1/2}(1,
m^2_{\cw_i^\pm}/s, m^2_{\cw_j^\pm}/s)$. As before electroweak
symmetry breaking effects in the chargino mixing matrix have been
neglected. The mixing angles $c_2$ and $s_2$, derived from
Eq.$\,$\eqref{eq:noewmix} by neglecting $v_I$ and explicitly given
by
\begin{eqnarray}
c_2/s_2 = \sqrt{\left[1\pm(M'_2-M_2)/\delta_2\right]/2}\quad
\mbox{with}\quad \delta_2=\sqrt{(M'_2-M_2)^2 + 4 (M^D_2)^2}\,,
\end{eqnarray}
under the assumption $M'_2+M_2\leq 0$ and $M^{(D)}_2\geq 0$,
only modify the $t$-channel sneutrino amplitude, so that they can
be determined from the angular
distribution of $\cw^+_1 \cw^-_1$ production in a straightforward
manner. The MSSM limit corresponds to \eqref{c1c1} with $c_2=0$ and
$s_2=1$. In the Dirac limit, using the basis
\eqref{chaD1},\eqref{chaD2} for the two degenerate gauginos,
one finds
\begin{eqnarray}
\frac{d\sigma}{d\cos\theta}[e^+ e^- \to \cw^+_{D1} \cw^-_{D1}]
 &=&
 \frac{\pi\alpha^2\lambda_{11}^{1/2}}{16s_W^4s} \Biggl [
 \frac{[s^2-4s_W^2m_Z^2s+8s_W^4m_Z^4][2-\lambda_{11}\sin^2\theta]}{
        (s-m_Z^2)^2}
  \nonumber\\
 & & \hskip 0.7cm
     +2\, \frac{[s-2s_W^2m_Z^2][2-2\lambda_{11}^{1/2}\cos\theta
                                 - \lambda_{11}\sin^2\theta]}
               {(s-m_Z^2)(\eta_{11}-\lambda^{1/2}_{11}\cos\theta)}
        + 2 \frac{(1-\lambda_{11}^{1/2}\cos\theta)^2}{
            (\eta_{11}-\lambda^{1/2}_{11}\cos\theta)^2} \Biggr ]\,,
  \\
\frac{d\sigma}{d\cos\theta}[e^+ e^- \to \cw^+_{D2} \cw^-_{D2}]
 &=&
 \frac{\pi\alpha^2\lambda_{22}^{1/2}}{16s_W^4s} \,
 \frac{[s^2-4s_W^2m_Z^2s+8s_W^4m_Z^4][2-\lambda_{22}\sin^2\theta]}{
      (s-m_Z^2)^2}\,,
  \\
\frac{d\sigma}{d\cos\theta}[e^+ e^- \to \cw^\pm_{D1} \cw^\mp_{D2}]
&=& 0\,,
\end{eqnarray}
while the higgsino production is identical to the MSSM case. It is
noteworthy that unlike the MSSM, three distinct pairs of charginos
can be produced in the Dirac limit, but the cross sections for
$\cw^+_1 \cw^-_1$ ($\cw^+_3 \cw^-_3$) production, in the MSSM limit,
are identical to those for
    $\cw^+_{D1} \cw^-_{D1}$ ($\cw^+_{D3} \cw^-_{D3}$) production in
    the Dirac limit. The latter
characteristic is in obvious contrast to neutralino and gluino
production, which are Majorana particles in one limit and Dirac
particles in the other.

As a characteristic example in the neutralino sector we will
focus on the production of wino pairs, $e^+e^- \to {\tilde{\chi}}_2
{\tilde{\chi}}_2^{(c)}$ in the MSSM and Dirac limits for the comparison of
Majorana and Dirac theories. Neglecting neutralino mixing from electroweak
symmetry breaking, the production mechanisms proceed via exchange of
selectrons in the $t$-channel for Dirac neutralinos, and both the $t,u$-channels
for Majorana neutralinos. The differential cross sections as a function
of the production angle $\theta$ read
\begin{align}
&\text{MSSM:} & \frac{d\sigma}{d\cos\theta}[e^+ e^- \to \cw^0_2 \cw^0_2] &=
     \frac{\pi \alpha^2}{32 s_W^4 s} \, \lambda_{22}^{3/2} \,
     \frac{\eta_{2L}^2+(\eta_{2L}^2-4\eta_{2L}+2-\lambda_{22})\cos^2\theta
           +\lambda_{22} \cos^4\theta}
          {(\eta_{2L}^2-\lambda_{22}\cos^2\theta)^2}\,,  \\[2mm]
&\text{Dirac:} &
  \frac{d\sigma}{d\cos\theta}[e^+ e^- \to \cw^{0c}_{D2} \cw^0_{D2}] &=
     \frac{\pi \alpha^2}{32 s_W^4 s} \, \lambda_{22}^{1/2} \,
     \frac{(1-\lambda_{22}^{1/2}\cos\theta)^2}
          {(\eta_{2L}-\lambda_{22}^{1/2}\cos\theta)^2} \,.
\end{align}
As before, $\lambda_{22}$ denotes the usual 2-body phase space function
and $\eta_{2L} = 1+2(m^2_{{\tilde{e}}_L}-m^2_{{\tilde{\chi}}_2^0})/s$.

\begin{figure}[t]
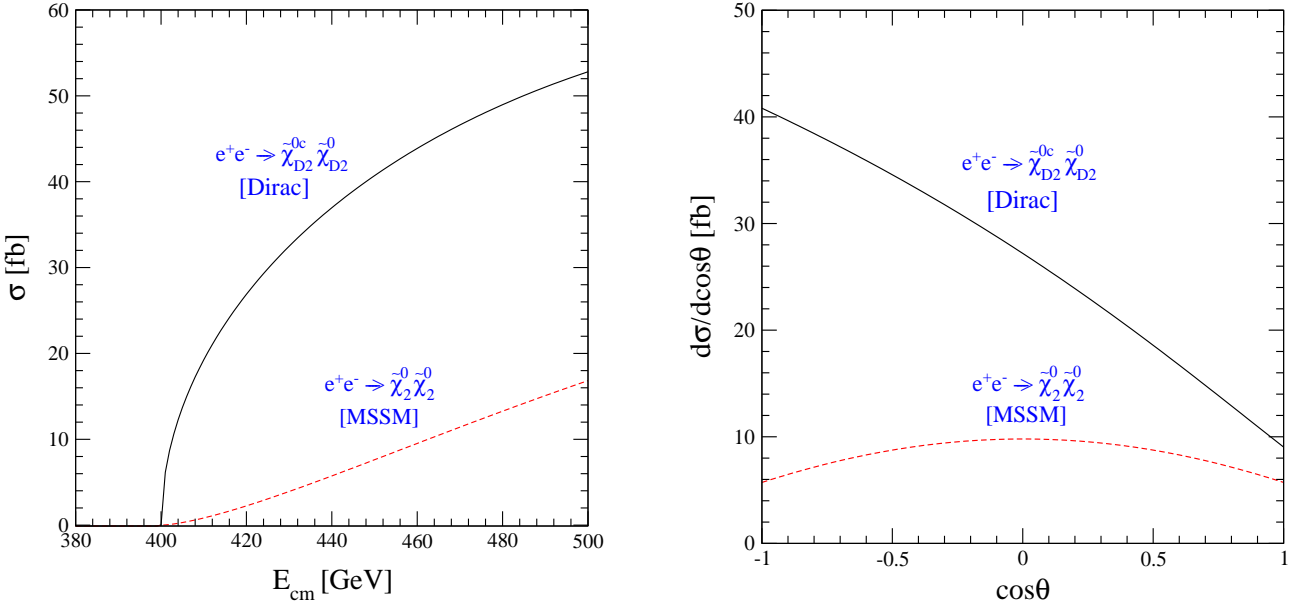

\begin{center}
\epsfig{figure=nn_prod_hybrid.eps, width=8.cm, height=8.cm}
\hskip 1.cm
\epsfig{figure=nn_angle_hybrid.eps, width=8.cm, height=8.cm}
\end{center}
\caption{\it Left: the total cross sections for pair
         production of wino-like neutralinos near threshold in the MSSM and
         the Dirac theory. Right: dependence of the cross sections
         on the production angle $\theta$ for $\sqrt{s}=E_{\rm cm}=500$~GeV.
         The sparticle masses in both plots are $m_{\tilde{\chi}_2^0}
         = m_{\tilde{\chi}_{D2}^0} = 200$~GeV and $m_{\tilde{e}_L}
         = 400$~GeV.}
\label{fig:n2n2}
\end{figure}

Two characteristics distinguish the Dirac from the Majorana
cross section, see Fig.$\,$\ref{fig:n2n2}. Dirac particles are generated
in $S$-waves near threshold, identical Majorana particles in $P$-waves,
giving rise to threshold onsets proportional to the $\tilde{\chi}$ velocity
and its third power, respectively. In contrast to identical Majorana particle
production, Dirac particle production is not forward-backward symmetric
in the production angle $\theta$. The integrated asymmetry is substantial,
for example $\mathcal{A}_{\rm FB} \approx -0.30$ for
$m_{\tilde{\chi}_{D2}^0} = 200$~GeV, $m_{\tilde{e}_L} = 400$~GeV and
$\sqrt{s}=500$~GeV. In practice, the measurable asymmetry is somewhat
reduced by experimental acceptances and cuts, and the fact that the
neutralino cannot be reconstructed fully from its decay products, but
it is nevertheless an important tool to discriminate the Dirac theory
from the MSSM. It should be noted finally that the cross section for
Dirac pair production is equal to the sum of the cross sections for
the corresponding $\{ kl \}$ diagonal and off-diagonal Majorana pairs
as shown explicitly by meticulous accounting of interference effects
for gluino production in Ref.~\cite{us1}.

\subsection{Scalar Particles}

\begin{figure}[b]
\begin{center}
\raisebox{3cm}{\rm (a)}\hskip 0.3cm
\epsfig{figure=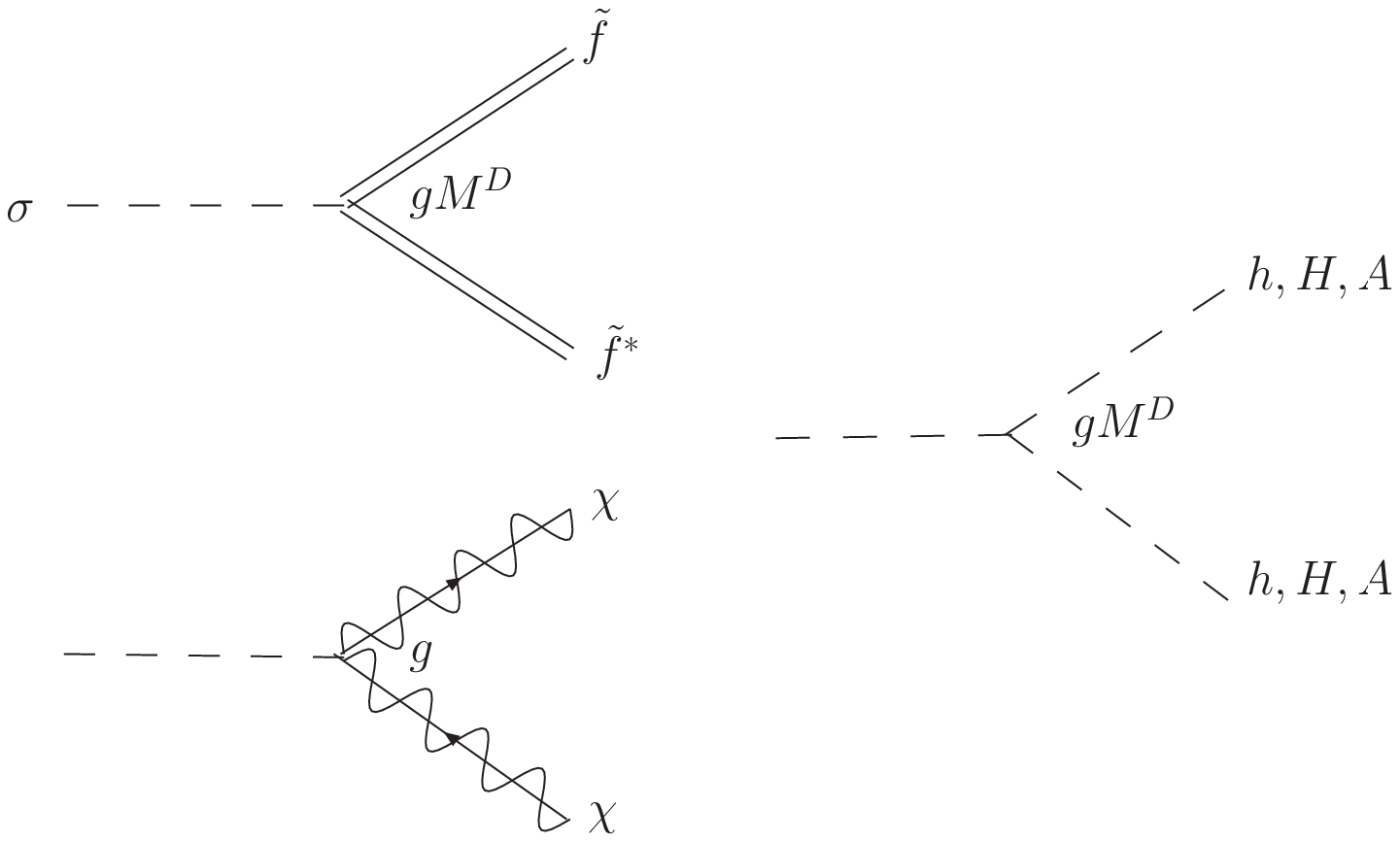, height=6.cm, width=7.5cm, clip=true}
\hskip 1.cm
\raisebox{3cm}{\rm (b)}\hskip 0.3cm
\epsfig{figure=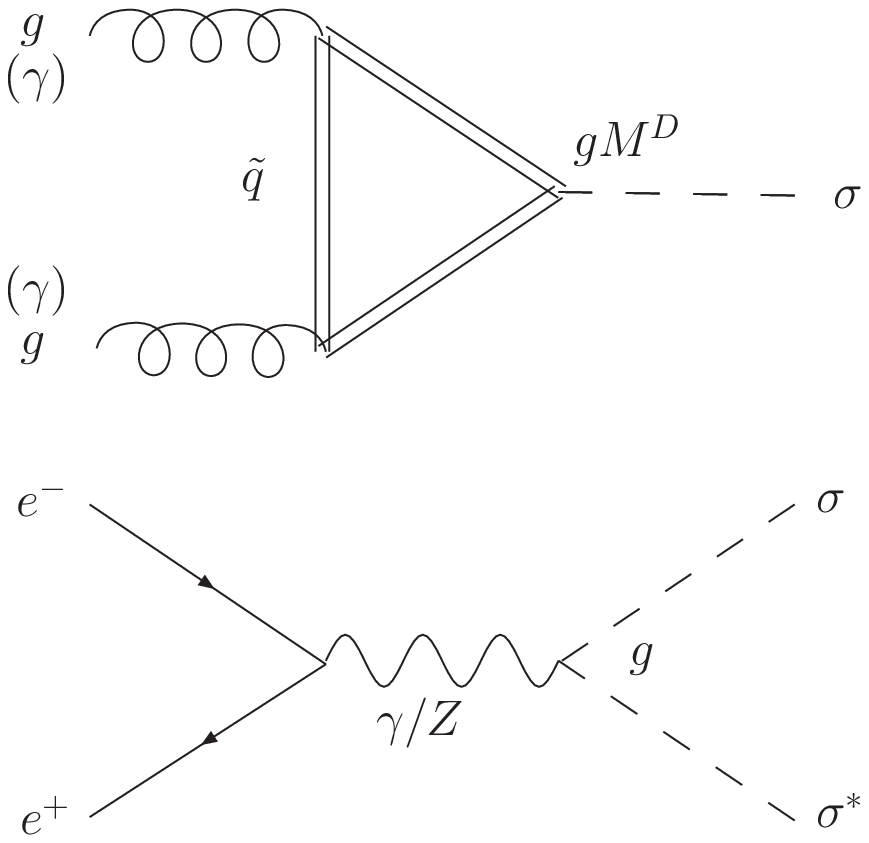, height=6.5cm, width=7.cm, clip=true}
\end{center}
\caption{\it Diagrams relevant for (a) electroweak $\sigma$ decays and (b)
             electroweak $\sigma$ production. The $\gamma\gamma$ couplings to $\sigma$
             include also loops of charginos, W and charged Higgs bosons.
             [Values of the couplings
             denoted at the vertices, are generic.]}
\label{fig:diagram}
\end{figure}

\noindent
At the Born level the iso-triplet and hyper-singlet sigma fields,
$\sigma^i_I$ and $\sigma^0_Y$, couple to sfermions, charginos/neutralinos
and Higgs bosons. The relevant couplings can be derived from the
Lagrangian terms and the scalar potential listed in
subsection \ref{sec:lagrangian}. The set of new Born and
effective loop couplings relevant for the phenomenological analyses of the
dominant $\sigma$ production and decays is displayed in
Fig.$\,$\ref{fig:diagram}. Only the generic form of the couplings
are noted explicitly at the vertices.

\subsubsection{Sigma Decays}

\noindent
Expressed by the effective couplings, $g_B$ and $g_F$, in the Lagrangians
${\cal L} = g_B s \, B^\ast B$ and ${\cal L} = g_F \phi \, \bar{F} [i \gamma_5] F$,
$\phi=s,a$, the partial decay widths can be derived generically for bosons $B$ and
fermions $F$ as
\begin{eqnarray}
\Gamma[s \to B \bar{B}]  &=& \frac{g_B^2}{16 \pi m_s} \beta\,,    \\
\Gamma[s \to F \bar{F}] &=& \frac{g_F^2 m_s}{8 \pi} \beta^3 \,, \\
\Gamma[a \to F \bar{F}] &=& \frac{g_F^2 m_a}{8 \pi} \beta \,,
\end{eqnarray}
with $\beta$ denoting the velocity of the final state particles. The two
standard coefficients $\beta$ and $\beta^3$ correspond to $S$- and
$P$-wave decays.

In the following analyses we will focus on the gross features
of production channels and decay modes of the novel scalar states
so that small block mixing can be neglected. The mass eigenstates
are therefore approximately identified with the unmixed states
$s_{Y,I}$, $a_{Y,I}$ and $s^\pm_{1,2}$, and the MSSM states
$h,H,A,H^\pm$ correspondingly.

For unspecified masses and couplings the following decays are
the leading modes of the particles $s_Y$ and $a_Y$:
\begin{eqnarray}
s_Y &\to& hh,\ \ hH,\ \ HH,\ \ AA,\ \ H^+H^-; \ \
          \tilde{f} {\tilde{f}}^\ast; \ \
          \tilde{\chi}^+ \tilde{\chi}^-, \ \
          \tilde{\chi}^0 \tilde{\chi}^{0(c)}\,,\\
a_Y &\to&  \tilde{\chi}^+ \tilde{\chi}^-, \ \
          \tilde{\chi}^0 \tilde{\chi}^{0(c)}\,.
\end{eqnarray}
The pseudoscalar particle $a_Y$ decays only to (higgsino-type) neutralino
or chargino pairs, with equal probability sufficiently above the threshold region.
If 2-body decays are kinematically forbidden, 3-body decays
to a (higgsino-type) neutralino, (bino-type) neutralino and Higgs
boson, as well as loop-decays to $t\bar t$ pairs and photons are
predicted. It should also be noted that the mass eigenstate proper, $A_2$, may decay
through channels opened by the mixing with the pseudoscalar $A$ Higgs
boson.
As the couplings are of size
$\mathcal{O}(g' M_Y^D)$ and/or $\mathcal{O}(\lambda_Y\mu, \lambda_Y M_Y, \lambda_Y A_Y)$,
the ensuing partial widths are typically of electroweak size above the 2-body
threshold regions.

\begin{figure}[t]
\begin{center}
\epsfig{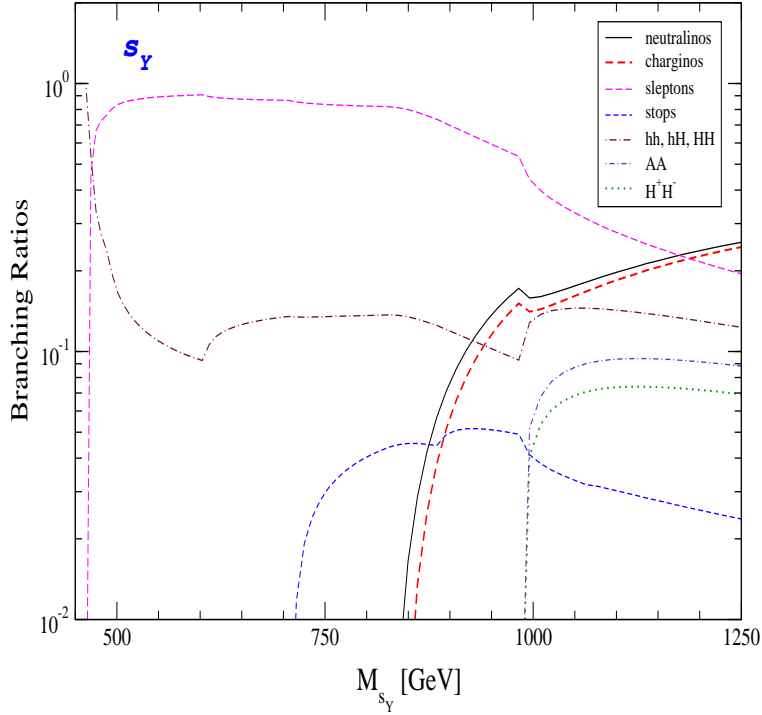}
\end{center}
\caption{\it Dependence of the branching ratios for $s_Y$
             decays on the mass $M_{s_Y}$. The values of the relevant
             SUSY parameters are taken to be tan$\beta$=5, $\mu=400$~GeV,
             $m_Y'=M_Y^D=v/2$, $m_I'=M_I^D=v$, $A_Y=A_I=2v$, together
             with $m_{{\tilde l}_L}=v$, $m_{{\tilde e}_R}
             =0.95 m_{{\tilde l}_L}$, $m_{{\tilde q}_L}=2v$, $m_{{\tilde q}_R}
             =0.95 m_{{\tilde q}_L}$, $m_{{\tilde t}_R}=0.8m_{{\tilde q}_L}$,
             $m_{{\tilde t}_L}=X_t=0.9m_{{\tilde q}_L}$, $m_H=m_A=m_H^\pm=2v$,
             $m_h=114$~GeV. For the charginos and neutralinos the Dirac
             limit with $M^{(\prime)}_{1,2}=0$ is assumed. Only the leading
             2-body decays are shown. }
\label{fig:BR}
\end{figure}

A detailed set of leading decay branching ratios is shown for the
hyper-singlet scalar particle $s_Y$ in Fig.$\,$\ref{fig:BR}.
The relevant couplings $g_B$ and $g_F$ for the scalar $s_Y$ to Higgs
bosons are:
\begin{eqnarray}
g_B[s_Y hh]  &=&-\sqrt{2}\lambda_Y\mu_n +  g'M_Y^D c_{2\beta}
                 +(M_Y+A_Y) \lambda_Y s_{2\beta}/\sqrt{2} \,, \nonumber\\
g_B[s_Y h H] &=& -g' M^D_Y s_{2\beta}
                 + (M_Y +A_Y) \lambda_Y c_{2\beta}/\sqrt{2}
                  \,, \nonumber\\
g_B[s_Y H H] &=& g_B[s_Y AA]
              = -\sqrt{2}\lambda_Y\mu_n -  g'M_Y^D c_{2\beta}
                 -(M_Y+A_Y) \lambda_Y s_{2\beta}/\sqrt{2} \,, \nonumber\\
g_B[s_Y H^+H^-] &=&-\sqrt{2} \lambda_Y \mu_c + g' M^D_Y c_{2\beta}
                   -(M_Y+A_Y)\lambda_Y s_{2\beta}/\sqrt{2} \,,
\end{eqnarray}
those to supersymmetric particles are:
\begin{eqnarray}
g_B[s_Y \tilde{f}_L\tilde{f}^*_L] &=&-2g'M_Y^D Y_{f_L} \,,    \nonumber\\
g_B[s_Y \tilde{f}_R\tilde{f}^*_R]  &=&\phantom{+}2g'M_Y^D Y_{f_R} \,,   \nonumber\\
g_F[s_Y \tilde{H}^+_u \tilde{H}^-_d]
             &=&\phantom{+} g_F[s_Y \tilde{\chi}^+_{D3} \tilde{\chi}^-_{D3}]
             = - \lambda_Y/\sqrt{2} \,, \nonumber\\
g_F[s_Y \tilde{H}^0_u \tilde{H}^0_d]
             &=& -g_F[s_Y \tilde{\chi}^0_{D3} \tilde{\chi}^{0c}_{D3}]
             = \phantom{+}\lambda_Y/\sqrt{2} \,,
\end{eqnarray}
and the relevant couplings for the pseudoscalar $a_Y$ are:
\begin{eqnarray}
g_F[a_Y \tilde{H}^+_u \tilde{H}^-_d]
             &=& \phantom{+} g_F[a_Y \tilde{\chi}^+_{D3} \tilde{\chi}^-_{D3}]
             = \phantom{+} \lambda_Y/\sqrt{2} \,, \nonumber\\
g_F[a_Y\, \tilde{H}^0_u\, \tilde{H}^0_d\,]
             &=& - g_F[a_Y \tilde{\chi}^0_{D3} \tilde{\chi}^{0c}_{D3}]
             = - \lambda_Y/\sqrt{2} \,.
\end{eqnarray}
The Dirac chargino and neutralino, $\tilde{\chi}^\pm_{D3}$ and
$\tilde{\chi}^0_{D3}$, are defined in terms of higgsinos in
Eqs.$\,$(\ref{chaD3}) and (\ref{neuD3}), respectively.
For the specific set of parameters the hyper-singlet scalar $s_Y$ decays
dominantly to Higgs bosons and sleptons.  The decays to gaugino-like
neutralinos are forbidden due to gauge symmetry and the decays to
higgsino-like neutralinos and charginos are kinematically allowed
only when the particle is very heavy.

The iso-triplet scalar states, $s_I, a_I$ and $s^\pm_{1,2}$, have been
assumed very heavy. Several features of the iso-triplet scalar interactions
determine their potential decay modes. In parallel to the hypercharge
states, they do not couple to quarks and leptons, but they couple to
gauginos, higgsinos and scalar pairs, sfermions, as well as Higgs
bosons and/or gauge bosons. Thus, if kinematically allowed, the gauge/Higgs
bosons, sfermions, charginos and neutralinos constitute the dominant decay
channels for the $s_I, a_I$ and $s^\pm_{1,2}$ states:
\begin{eqnarray}
\label{sIdecay}
s_I &\to& h h,\ \ h H,\ \ H H,\ \ AA,\ \  H^+H^-; \ \
          \tilde{f} \tilde{f}^\ast;\ \ \tilde{\chi}^+ \tilde{\chi}^-, \ \
          \tilde{\chi}^0 \tilde{\chi}^{0(c)}\,, \\
a_I &\to& \tilde{\chi}^+ \tilde{\chi}^-,\ \
          \tilde{\chi}^0 \tilde{\chi}^{0(c)}\,, \\
s^\pm_{1,2}
    &\to& H^\pm h, \ \ H^\pm H, \ \ H^\pm A; \ \
          \tilde{f} \tilde{f'}^\ast; \ \
          \tilde{\chi}^+ \tilde{\chi}^0/\tilde{\chi}^- \tilde{\chi}^{0c} \,,
\end{eqnarray}
with partial widths of the electroweak scale. In addition, the small couplings to
electroweak gauge bosons, developed by the small iso-triplet {\it vev} $v_I$,
lead to the two-boson decay
$
s_I \to W^+W^-
$,
albeit at reduced rate. Furthermore, the iso-triplet scalar states may decay
to gluons, photons, electroweak bosons, quarks and leptons through sfermion
and chargino/neutralino loops.

\subsubsection{Stop and Stau Decays to Sigma Particles}

\noindent
Sigma fields carry positive $R$-parity, but they couple preferentially
to gaugino and squark/slepton pairs. Assuming, as before, that they are
heavier than charginos and neutralinos, this leaves us with heavy
sfermion decays as a possible source for neutral sigma particles:
\begin{equation}
\tilde{f}_2\ \ \to\ \ \tilde{f}_1 + s_Y           \,.
\end{equation}
While the $s_I$ channel is likely too heavy to be open, the
pseudoscalars $a_{Y,I}$ do not couple.

Since the mass splitting between the two stops is typically large, let us
examine the process $\tilde{t}_2 \to \tilde{t}_1 + s_Y$ first.
The partial width for this decay mode is given by
\begin{equation}
\Gamma_{\tilde{t}_2}
   = \frac{g^2_{\tilde{t}}}{16 \pi \, m_{\tilde{t}_2}}
      \lambda^{1/2}(1, m^2_{\tilde{t}_1}/m^2_{\tilde{t}_2},
                   M^2_{s_Y}/m^2_{\tilde{t}_2}) \,,
\end{equation}
with $\lambda$ denoting the usual phase space function defined in
\eqref{eq:lambda} and the coupling
\begin{equation}
g_{\tilde{t}} = \frac{5}{6 \sqrt{2}} \, g' M^D_Y\, \sin 2 \theta_{\tilde{t}}\,.
\end{equation}
Even for large values of the stop mixing angle $\theta_{\tilde{t}}$,
$\Gamma_{\tilde{t}}$ is typically less than 1~GeV for sparticle masses of a few
100~GeV, compared to a typical ${\tilde{t}}_2$ total width of a few tens of GeV.
Thus the branching ratio for $\tilde{t}_2 \to \tilde{t}_1 + s_Y$ can amount to a
few per-cent at most, so that experimental discovery of this decay channel will
be very challenging.

In the stau system, the decay mode $\tilde{\tau}_2 \to \tilde{\tau}_1 + s_Y$
only becomes viable for large values of the mass splitting
$m_{\tilde{\tau}_2}-m_{\tilde{\tau}_1}$ and of $\tan\beta$.
In such a scenario, however,
one typically obtains sizable
branching ratios of order 10\%. This can be explained by
the larger hypercharges of the staus compared to the stops, leading to a similar
expression for the partial width as above but with $g_{\tilde{t}}$ replaced by
\begin{equation}
g_{\tilde{\tau}} = \frac{3}{2 \sqrt{2}} \, g' M^D_Y\, \sin 2 \theta_{\tilde{\tau}}\,.
\end{equation}
While heavy staus will be swamped by background at hadron colliders, the process
$\tilde{\tau}_2 \to \tilde{\tau}_1 + s_Y$ could
be a possible discovery mode for the $s_Y$ scalar at high-energy lepton colliders.

\subsubsection{Production of Sigma Particles at LHC}

\noindent The neutral hyper-singlet and iso-triplet scalar particles
$s_Y$ and $s_I$ can, in principle, be generated singly in gluon
fusion processes at the LHC, analogous to Higgs bosons:
\begin{eqnarray}
pp \ \ \to \ \  gg \ \ \to \ \ s_{Y,I} \,.
\end{eqnarray}
Since the pseudoscalar states $a_{Y,I}$ do not couple to gluons
through squark loops, their single formation channel is shut. The
adjoint $s_{Y,I}$ scalar coupling to the gluons are mediated by
squark triangles, the $D$-terms providing the interactions of the
squarks with the sigma fields.

The partonic fusion cross section for $s_Y$ production, with the
Breit-Wigner function in  units of $1/M_{s_Y}^2$ factored off,
\begin{equation}
\hat{\sigma}[gg\to s_Y] = \frac{\pi^2}{8 M_{s_Y}} \Gamma (s_Y \to gg) \,,
\end{equation}
can be expressed in terms of the partial width for $s_Y \to gg$,
\begin{equation}
\Gamma [s_Y \to gg]
    = \frac{\alpha_Y \alpha_s^2}{8 \pi^2} \, \frac{(M^D_a)^2}{M_{s_Y}} \,
      \left|\sum [Y_L \tau_L f(\tau_L) - Y_R \tau_R f(\tau_R)]\right|^2\,.
\end{equation}
with $\alpha_Y = g'^2/4\pi$. The standard triangular function
$f(\tau)$ is defined by
\begin{eqnarray}
f(\tau)
  = \left[ \sin^{-1}(1/\sqrt{\tau}) \right]^2 \quad \mbox{if}\ \ \tau \geq 1,
    \quad \mbox{and}\quad
   -\frac{1}{4} \left[\ln\left(\frac{1+\sqrt{1-\tau}}{1-\sqrt{1-\tau}}\right)
                    -i \pi\right]^2\quad \mbox{if}\ \ 0\leq \tau < 1 \,,
\label{eq:triangular_function}
\end{eqnarray}
with $\tau_{L,R} = 4 M^2_{\tilde{q}_{L,R}} / M^2_{s_Y}$ and $Y_{L,R}$
being the hypercharges of the $L$ and $R$-squarks. It should be noted that
the hypercharges add up to zero for complete generations, but not
individually for up- and down-type states for which the $L/R$ hypercharge
difference amounts to $\mp 1$. While for mass-degenerate complete generations
the sum of the form factors in the partial width vanishes, the cancelation
is lifted for stop states, in particular, with the non-zero difference
enhanced by the different $L/R$ hypercharges.

\begin{figure}[t]
\epsfig{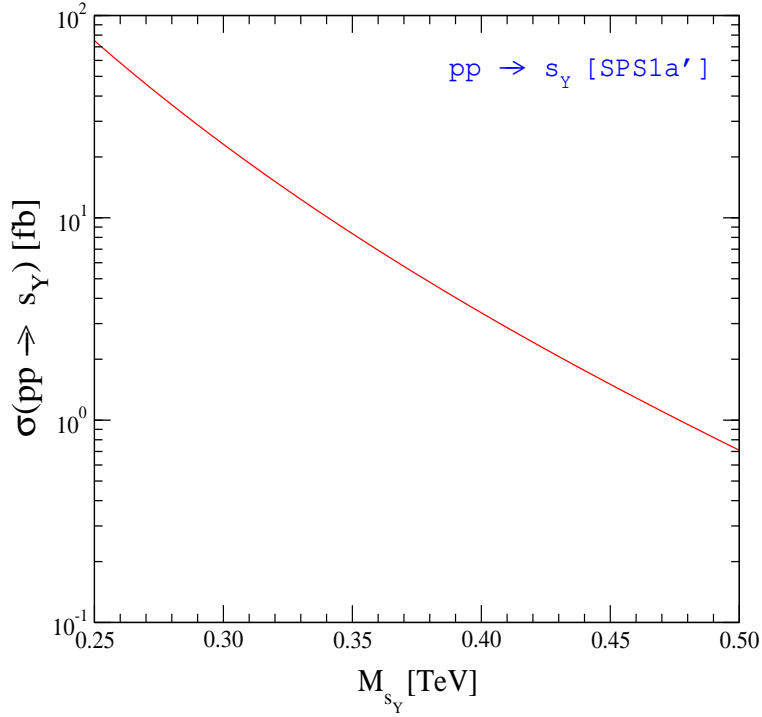} \\
\caption{\it Cross sections for single $s_Y$ production through $gg$ fusion
             in $pp$ collisions at LHC ($\sqrt{s} = 14$~TeV).
             The MSSM benchmark point SPS1a$'$, Ref.~\cite{sps}, is adopted
             for the numerical analysis.
}
\label{fig:singleSY}
\end{figure}

The $pp$ cross section is finally found by convoluting the parton cross section
with the $gg$ luminosity \cite{CTEQ},
\begin{equation}
\sigma [pp \to s_Y] = \frac{\pi^2}{8s} \, \frac{\Gamma(s_Y \to gg)}{M_{s_Y}} \,
                    \int^1_{M^2_{s_Y}/s} \frac{dx}{x} \, g(x;M^2_{s_Y}) \,
                    g(\tau /x;M^2_{s_Y}) \,,
\end{equation}
in the usual notation.
The cross section for $s_Y$ production is shown in Fig.$\,$\ref{fig:singleSY}
as a function of the $s_Y$ mass.

An analogous expression holds for $s_I$ production in $pp$
collisions. However, since the mass of $s_I$ needs to  be very large
due to the $\rho$-parameter constraint, the actual size of the
fusion cross section will be significantly below 1~fb.

Other production channels are offered by Higgs-strahlung and electroweak boson
fusion. Pairs of electroweak gauge bosons couple to the lightest Higgs boson
$h$ and the iso-vector $s_I$. All these couplings involve either the electroweak
$\it vev$ $v$ or the iso-scalar $\it vev$ $v_I$. Rotating the mass eigenstates
$s_i$ to the current eigenstates $h$ {\it etc}, the cross sections for
Higgs-strahlung and vector-boson fusion can easily be expressed by the
corresponding cross section for the production of the SM Higgs boson with
equivalent mass:
\begin{eqnarray}
\sigma [pp \to W \to W s_i]
   &=& \left({\cal O}_{S1i} + 4 \frac{v_I}{v} {\cal O}_{S4i} \right)^2
       \sigma [pp \to W \to W H_{SM}] \,, \\
\sigma [pp\, \to\, Z\, \to\, Z s_i]
   &=& ({\cal O}_{S1i})^2\,
       \sigma [pp \to Z \to Z H_{SM}]\,,
\end{eqnarray}
with ${\cal O}_S$ denoting the $4\times 4$ rotation matrix diagonalizing the
scalar mass matrix squared ${\cal M}^2_S$ in
Eq.$\,$(\ref{eq:scalar_mass_matrix})
as $O^T_S {\cal M}^2_S O_S = {\rm diag}(M^2_{S_1},\ldots, M^2_{S_4})$.
Cross sections for vector boson fusion are related in the same way.

Numerical analyses taking into account the experimental constraint on the
$\rho$ parameter and the mass bound on the lightest neutral
scalar lead to mixing coefficients of $10^{-2}$ and less so that these
channels are presumably of little value in practice.

\subsubsection{Charged Adjoint Scalar Pair-Production in $e^+e^-$
               and $\gamma\gamma$ Collisions}

\noindent
{\it (i) \uline{\it{$e^+e^-$ collisions :}}}

\begin{figure}[t]
\epsfig{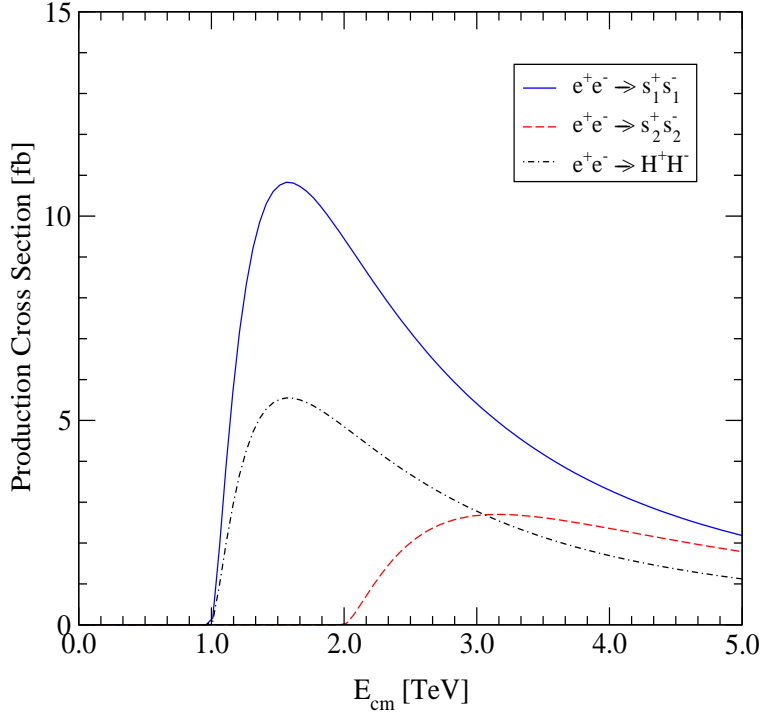} \\
\caption{\it The cross sections for charged $s^\pm_{1,2}$
             pair production in $e^+e^-$ collisions at TeV energies.
             The charged scalar masses are assumed to be $M_{\tilde{s}^\pm_1}=0.5$
             TeV and $M_{\tilde{s}^\pm_2}=1.0$~TeV. For comparison, the cross
             section for charged Higgs pair production is shown with its
             mass $M_{H^\pm}=0.5$~TeV.
}
\label{fig:eesapairs}
\end{figure}

\noindent Resonance production of sigma particles in $e^+ e^-$
collisions is strongly suppressed as the production amplitude scales
with the electron mass. Since the quantum numbers $Q,I_3,Y$ of the
neutral sigma states $\sigma^0_{I,Y}$ all vanish, these particles
cannot be pair-produced in $e^+ e^-$ collisions. However, production
channels open up for diagonal charged scalar pairs $s^\pm_{1,2}$
defined in Eq.$\,$(\ref{eq:charged_sigma_scalar}),
\begin{equation}
e^+ e^-\ \ \to\ \ s^+_n s^-_n\quad [n=1,2]\,,
\end{equation}
through $s$-channel $\gamma, Z$ exchanges.
With effective charges
\begin{equation}
\begin{array}[b]{rll@{$\qquad$}rll}
g_{L\ast}[s^\pm_n]&= 1 - \dfrac{s^2_W - 1/2}{s^2_W} \dfrac{s}{s-m^2_Z} \,
                  &\approx\; 2 \,, &
g_{R\ast}[s^\pm_n] &= 1 - \dfrac{s}{s-m^2_Z}
                 &\approx\; 0 \,, \\[3ex]
g_{L\ast}[H^\pm]&= 1 + \dfrac{(s^2_W - 1/2)^2}{c^2_W s^2_W}
                  \dfrac{s}{s-m^2_Z} \,
                  &\approx\;
                 \dfrac{4}{3} \,, &
g_{R\ast}[H^\pm]&= 1 + \dfrac{s^2_W-1/2}{c^2_W}\dfrac{s}{s-m^2_Z}
                 &\approx\;
                 \dfrac{2}{3} \,,
\end{array}
\end{equation}
with $n{=}1,2$ for $L$ and $R$-chiral electron pairs coupled to the
$s^\pm_{1,2}$ pairs and equivalently to the $H^\pm$ pair, the cross section
reads:
\begin{equation}
\sigma = \frac{\pi \alpha^2}{3 s} \, \frac{g^2_{L\ast}+g^2_{R\ast}}{2} \,
         \beta^3 \,,
\end{equation}
where $s$ is the total c.m. energy squared and $\beta$ the velocity of the
particles $s^\pm_{1,2}$ and $H^\pm$; $s^2_W = \sin^2\theta_W$ denotes the
electroweak mixing parameter. The size of the three production cross sections,
identical in form, is illustrated in Fig.$\,${\ref{fig:eesapairs}}
for two different mass values $M_{s^\pm_1, H^\pm}=0.5$ TeV and
$M_{s^\pm_2}=1.0$ TeV.

\vskip 0.3cm
\noindent
{\it (ii) \uline{\it{$\gamma\gamma$ collisions :}}}

\noindent
Excellent instruments for searching for heavy scalar/pseudoscalar particles
and studying their properties are $\gamma\gamma$
colliders, see Ref.$\,$\cite{MMM}. About 80\% of the incoming electron
energy can be converted to a high-energy photon by Compton back-scattering
of laser light, with the spectrum peaking at the maximal energy by
choosing proper helicities.

Depending on the nature of the neutral scalars/pseudoscalars, their couplings
to the two photons is mediated by charged $W$-bosons, charginos,
and charged scalars and Higgs bosons. As before, the formation cross
sections for the states $\phi = s_Y, s_I$ and $a_Y, a_I$ can be expressed
by the $\gamma\gamma$ widths of the particles
and the $\gamma\gamma$ luminosity:
\begin{eqnarray}
\langle \sigma (\gamma\gamma \to \phi) \rangle
    \, &=& \, 8 \pi^2\, \frac{\Gamma(\phi \to \gamma\gamma)}{M_\phi^3} \,
        \tau_\phi \frac{d{\mathcal{L}}_{\gamma\gamma}}{\tau_\phi}  \nonumber \\
    \, &=& \, \sigma_0 (\gamma\gamma\to \phi)\,
        \tau_\phi \frac{d{\mathcal{L}}_{\gamma\gamma}}{\tau_\phi}\,,
\label{eq:rr_cross_section}
\end{eqnarray}
with $\tau_\phi = M^2_\phi /s$. For qualitative estimates the luminosity
function $\tau_\phi d{\mathcal{L_{\gamma\gamma}}}/d\tau_\phi$
can be approximated by unity after
splitting off the overall $e^+e^-$ luminosity \cite{Asner:2001ia}.

The partial $\gamma\gamma$ widths are parameterized by couplings and loop
functions,
\begin{equation}
\Gamma (s/a \to \gamma\gamma )
   = \frac{\alpha^2}{64 \pi^3} \, M_{s/a} \,
     \left| \sum_i N_{c_i} e_i^2 g^{s/a}_i A^{s/a}_i \right|^2 \,.
\end{equation}
The factor $N_{c_i}$ denotes the color factor of the loop line, while the
couplings $g^{s/a}_i$ are expressed by
\begin{eqnarray}
&& g^{s_Y}_{\tilde{H}^\pm_D} = \lambda_Y/\sqrt{2}\,, \qquad
   g^{s_Y}_{\tilde{f}_{L,R}}  = \pm Y_{f_{L,R}} M^D_Y/M_{s_Y}\,,\nonumber\\
&& g^{s_Y}_{H^\pm}
          = \left(\sqrt{2}\lambda_Y \mu_c-g' M^D_Y c_{2\beta}
                  +\lambda_Y (M_Y+A_Y) s_{2\beta}/\sqrt{2}\right)/M_{s_Y}\,,\\
&& g^{s_I}_{W^\pm} = 2 g^2 v_I/M_{s_I}\,, \qquad
   g^{s_I}_{\tilde{H}^\pm_D} = \lambda_I/\sqrt{2}\,, \qquad
   g^{s_I}_{\tilde{W}^\pm_{1,2}} = \mp g\,, \qquad
   g^{s_I}_{\tilde{f}_{L,R}} = \pm g I^f_3 M^D_I/M_{s_I},\nonumber\\
&& g^{s_I}_{H^\pm}
          = -\left(\sqrt{2} \lambda_I \mu_c - g M^D_I c_{2\beta}
                  +\lambda_I(M_I-A_I) s_{2\beta}/\sqrt{2}\right)/M_{s_I}\,,
\end{eqnarray}
for the hyper-singlet scalar $s_Y$ and the iso-triplet scalar $s_I$, and
\begin{eqnarray}
&& g^{a_Y}_{\tilde{H}^\pm_D} = -\lambda_Y/\sqrt{2}\,, \nonumber\\
&& g^{a_I}_{\tilde{H}^\pm_D} = \lambda_I/\sqrt{2}\,, \qquad
   g^{a_I}_{\tilde{W}^\pm_{1,2}} =\pm g\,,
\end{eqnarray}
for the hyper-singlet pseudoscalar $a_Y$. The loop functions
$A^{s/a}_i$, identical in form for the particles of a given spin,
include the standard triangular function $f(\tau)$
in Eq.$\,$(\ref{eq:triangular_function}) as
\begin{eqnarray}
A^{s}_0 &=&  1-\tau f(\tau)\,, \qquad
A^{s}_{1/2} =  -2\sqrt{\tau} \left[ 1 + (1-\tau) f(\tau)\right]\,, \qquad
A^{s}_1 = 2/\tau + 3 + 3 (2-\tau) f(\tau)\,,
\end{eqnarray}
for the scalar $s$, and
\begin{eqnarray}
A^{a}_{1/2} &=& -2\sqrt{\tau} f(\tau)\,,
\end{eqnarray}
for the pseudoscalar $a$, where the subscripts $0, 1/2$ and $1$ stand
for spin-0, spin-1/2 and spin-1 intermediate particles.

\begin{figure}[t]
\epsfig{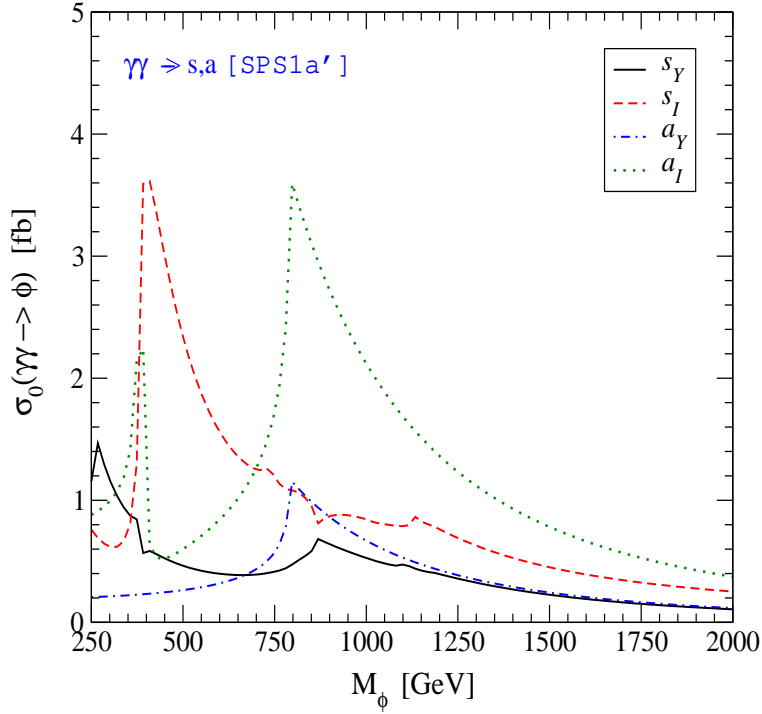} \\
\caption{\it The reduced production cross sections
             $\sigma_0(\gamma\gamma\to\phi)$ for $\phi=s_{Y,I}$ and
             $a_{Y,I}$. The SPS1a$'$ parameter set \cite{sps} for the
             (s)particle masses, couplings and mixing parameters are
             adopted for the numerical analysis.
}
\label{fig:rrphi}
\end{figure}

Adopting the SPS1a$'$ parameters in the SUSY sector \cite{sps}, the reduced
$\gamma\gamma$ cross section $\sigma_0(\gamma\gamma\to\phi)$ for
$\phi=s_{Y,I}$ and $a_{Y,I}$, defined in Eq.$\,$(\ref{eq:rr_cross_section}),
amounts to order 1 fb as shown in Fig.$\,$\ref{fig:rrphi} so that for
an overall luminosity of several hundred fb$^{-1}$ a sizable sample of
neutral scalars $s_{Y,I}$ and pseudoscalars $a_{Y,I}$ can be generated
in $\gamma\gamma$ collisions.

\section{SUMMARY}

\noindent
In minimal supersymmetric extensions of the Standard Model the fermionic
partners of color and electroweak gauge bosons are self-conjugate Majorana
fields. Their properties are characteristically distinct from Dirac fields.
To investigate this point quantitatively, we have adopted an $N{=}1$/$N{=}2$
hybrid supersymmetry model in which the gauge and Higgs sectors are
extended to $N{=}2$ while the matter sector remains restricted to $N{=}1$.
This extension is wide enough to allow the joining of Majorana to Dirac
fields while keeping the matter sector chiral. By properly varying gaugino
mass matrices, the original MSSM Majorana theory can be transformed smoothly
to the Dirac theory.

The transition from $N{=}1$ to $N{=}2$ expands the gauge sector by a matter
supermultiplet composed of a new gaugino and adjoint scalar multiplet.

\noindent
\begin{itemize}
\item[{\bf (i)}] The doubling of the gauginos in $N{=}2$ gives rise to new
     particles along the path from the MSSM to the Dirac theory:
\begin{itemize}
\item 8 Majorana gluinos $\to$ 16 Majorana gluinos $\to$ 8 Dirac gluinos
\item 2 charginos $\to$ 3 charginos
\item 4 Majorana neutralinos $\to$ 6 Majorana neutralinos $\to$
      3 Dirac neutralinos
\end{itemize}
%
\item[{\bf (ii)}] The adjoint scalars expand also the number of states
     originally present in the MSSM scalar sector; the new
     SU(2)$_I\times$U(1)$_Y$ scalars mix with the Higgs fields in the
     electroweak sector:
\begin{itemize}
\item 8 octet complex scalar gluons, generally termed sgluons
\item 1 pseudoscalar state $\to$ 3 pseudoscalar states
\item 2 scalar states $\to$ 4 scalar states
\item 1 charged scalar $[\pm]$ pair $\to$ 3 charged scalar $[\pm]$ pairs
\end{itemize}
\end{itemize}

The scale of the new degrees of freedom is strongly restricted by the
experimentally allowed deviation of the $\rho$ parameter from unity. Since
the new electroweak SU(2)$_I$ and U(1)$_Y$ scalars acquire vacuum
expectation values, the SU(2)$_I$ iso-triplet {\it vev} must be small,
and the iso-triplet scalar mass parameter is driven into the TeV region.
The U(1)$_Y$ hyper-singlet {\it vev}, on the other hand, is not restricted
by the $\rho$ parameter and the hyper-singlet scalar mass may still be
characterized by a fraction of TeV.

The new degrees of freedom are coupled to the original MSSM fields rather
weakly at the order $g v/{\tilde{M}}$. This allows us to solve the
complicated chargino, neutralino and scalar systems analytically in
a systematic expansion, {\it i.e.} mass eigenvalues and mixings. This
leads in a straightforward way to the prediction of production channels
and decay modes.

The Majorana or Dirac character of gluinos and neutralinos can nicely be
discriminated in sfermion-sfermion production. While allowed by Majorana
exchanges, equal $L$- or $R$-sfermion pair production is forbidden in Dirac
theories, {\it i.e.} ${\tilde{q}}_L {\tilde{q}}_L$ and ${\tilde{e}}_L^-
{\tilde{e}}_L^-$ production in $pp$ and $e^-e^-$ collisions. In addition,
cascade decays involving Majorana neutralinos give rise to charge-symmetric
lepton final states, while Dirac neutralinos predict charge sequences as
governed by the conservation of Dirac charges. Total cross sections and
angular distributions in pair production of charginos and neutralinos in
$e^+e^-$ collisions depend on the Majorana or Dirac nature of the underlying
gaugino theory.

A variety of production channels are predicted for the scalar
states,
depending on the nature of the particles:\\[-0.8cm]
\begin{itemize}
\item[{$-$}] Scalar and pseudoscalar sgluons can be produced in pairs
      in $pp$ collisions, and scalars singly via
      gluon-gluon fusion.
\item[{$-$}] The U(1)$_Y$ scalar state can be generated in $pp$ collisions via
      gluon fusion or in stop or stau decays, but not the corresponding
      pseudoscalar partner.
\item[{$-$}] $\gamma\gamma$ collisions offer production channels for
      all neutral scalar and pseudoscalar, iso-scalar and iso-vector
      states.
\item[{$-$}] The charged iso-vector states can be generated pairwise
      in $e^+ e^-$ collisions.
\end{itemize}

Thus in contrast to the color sector it is quite difficult to cover
experimentally the electroweak hyper-singlet and iso-triplet states,
the main reason being the expected heavy masses of the new scalar and
pseudoscalar particles. Nevertheless, given the joint potential of
hadron and lepton colliders, all the new scalar and pseudoscalar
particles introduced by the $N{=}1$/$N{=}2$ supersymmetric hybrid theory can
in principle be accessed. Only the basics of the processes
have been investigated in this report, while detailed analyses of final
states under realistic detector conditions are far beyond the scope
of this study.

On the other hand, the suppression of reaction channels and cascade
decays in gaugino Dirac theories as opposed to Majorana theories for
color and electroweak gauginos should provide unique signatures for
the Majorana or Dirac nature of gluinos and electroweak gauginos.

\acknowledgments{}

\noindent The work was partially supported by the Korea Research
Foundation Grant funded by the Korean Government (MOERHRD, Basic
Research Promotion Fund) (KRF-2008-521-C00069), by Bundesministerium
f\"ur Bildung und Forschung under contract no. 05HT6PDA, by the
Marie Curie Training Research Networks ``UniverseNet'' under
contract no. MRTN-CT-2006-035863, ``ForcesUniverse'' under contract
no. MRTN-CT-2004-005104,  ``Tools and Precision Calculations for
Physics Discoveries at Collider'' under contract no.
MRTN-CT-2006-035505, ``The Quest for Unification'' under contract
no. MRTN-CT-2004-503369,  ``Particle Physics and Cosmology: the
Interface'' under contract no. MTKD-CT-2005-029466, by the Polish
Ministry of Science and Higher Education Grant no~N~N202~230337, the
Department of Science and Technology, India, under project number
SR/S2/RFHEP-05/2006, as well as by the National Science Foundation
under grant no. PHY-0854782. PMZ thanks the Institut f\"ur
Theoretische Teilchenphysik und Kosmologie for the warm hospitality
extended to him at RWTH Aachen University. JMK thanks for the
hospitality of Korea Institute for Advanced Study where this work
was partially done. DC is thankful to the Institute of Theoretical
Physics for warm hospitality extended to him during his visit at
University of Warsaw.\

\vskip 1.cm

\setcounter{equation}{0}
\setcounter{section}{1}
\def\thesection{\Alph{section}}
\renewcommand{\theequation}{{\rm \thesection.\arabic{equation}}}

\section*{Appendix A: Analytic analysis of the singular value
           decomposition of a $2\times 2$ matrix}
\label{sec:appendix_a}

For any $n\times n$ complex matrix, there exist two unitary matrices $U_L$ and
$U_R$ such that
\begin{eqnarray}
\label{app:svd}
U_L^T \M U_R= {\M}_D={\rm diag}(m_1,m_2,\ldots,m_n)\,,
\end{eqnarray}
where the diagonal elements $m_k$ are real and non--negative.
This procedure is the singular value decomposition (SVD) of the matrix
$\M$. If the matrix $\M$ is symmetric, there exists a single unitary
matrix $U$ such that $U^T \M \, U = {\M}_D$ with $U_L=U_R=U$,
called the Takagi diagonalization of the symmetric matrix $\M$.

The singular value decomposition of a $2\times 2$ real matrix
can be performed analytically. The result is more involved than the
standard diagonalization of a $2\times 2$ symmetric matrix by a single
orthogonal matrix. The $2\times 2$ matrix be defined as:
\begin{eqnarray}
\M = \left(\begin{array}{cc}
            a      &    c  \\
      \tilde{c}    &    b
          \end{array}\right)\,,
\end{eqnarray}
where at least $c$ or $\tilde{c}$ be non-zero. Generally we can parameterize
two $2\times 2$ unitary matrices $U_L$ and $U_R$ in Eq.\,(\ref{app:svd}) by
\begin{eqnarray}
&& U_L = O_L P
     = \left(\begin{array}{cc}
         \cos\theta_L   &    \epsilon_{_L} \sin\theta_L  \\
     -\epsilon_{_L} \sin\theta_L & \cos\theta_L
             \end{array}\right) \,
       \left(\begin{array}{cc}
          \alpha  &  0 \\
          0  & \beta
             \end{array}\right)\,,\label{eq:ulpl} \\
&& U_R = O_R P
     = \left(\begin{array}{cc}
         \cos\theta_R   &    \epsilon_{_R} \sin\theta_R  \\
     -\epsilon_{_R} \sin\theta_R & \cos\theta_R
       \end{array}\right)\,
       \left(\begin{array}{cc}
          \alpha  &  0 \\
          0  & \beta
             \end{array}\right)\,,\label{eq:urpr}
\end{eqnarray}
where $0\leq \theta_{L,R} \leq \pi/2$, $\epsilon_{L,R}=\pm 1$
and $\alpha, \, \beta=1,i$. The two phase matrices, which map
the singular values onto non-negative values, can be identified
without loss of generality, as only their product is fixed.

If two singular values $m_{1,2}$ of the matrix $\M$ are non-degenerate,
they can be determined by taking the positive square root of the non-negative
eigenvalues $m^2_{1,2}$ of the orthogonal matrix ${\M}^T \M$:
\begin{eqnarray}
m_{1,2}= \frac{1}{2}\, \left|\, \sigma_+\mp \sigma_- \right|\quad
\mbox{with}\quad
\sigma_{\pm} = \sqrt{(a\pm b)^2+(c\mp \tilde{c})^2}\,.
\end{eqnarray}
Two eigenvalues become identical only if $a=\pm b$ and $c=\mp \tilde{c}$.
The smaller eigenvalue, $m_1$, vanishes if the invariant determinant vanishes,
${\rm det}\M = ab - c\tilde{c}=0$.

Explicitly performing the diagonalization of ${\M}^T \M$ by $U_R$ and
$\M {\M}^T$ by $U_L$, the rotation angles and signs can be computed:
\begin{eqnarray}
\label{eq:theta_LR}
\cos\theta_{L,R} = \sqrt{\frac{\sigma_+ \sigma_- + b^2-a^2
                   \pm \tilde{c}^2\mp c^2}{2\, \sigma_+ \sigma_-}}\,,&\qquad&
\sin\theta_{L,R} = \sqrt{\frac{\sigma_+ \sigma_- - b^2+a^2
                   \mp \tilde{c}^2\pm c^2}{2\, \sigma_+ \sigma_-}}\,,\\[1mm]
\epsilon_{_L} = {\rm sign}(a \tilde{c} + b c)\,,&\qquad&
\epsilon_{_R} = {\rm sign}(a c + b \tilde{c})\,. \label{eq:phi_LR}
\end{eqnarray}
In the final step of the computation the phase parameters $\alpha$ and
$\beta$ can be determined by inserting Eqs.\,(\ref{eq:theta_LR}) and
(\ref{eq:phi_LR}) into Eq.\,(\ref{app:svd}):
\begin{eqnarray}
\alpha  &=&  \sqrt{{\rm sign} \left[\, a(\sigma_+\sigma_- + b^2-a^2)
          -a(c^2+\tilde{c}^2)-2b c\tilde{c}\,\right]\, }\,,\\[1mm]
\beta   &=&  \sqrt{{\rm sign} \left[\, b(\sigma_+\sigma_- + b^2-a^2)
          +b(c^2+\tilde{c}^2)+2a c\tilde{c}\,\right]\, }\,,
\end{eqnarray}
up to an arbitrary overall sign of both parameters. If the smaller singular value
$m_1$ vanishes for ${\rm det}({\cal M}) =0$, the parameter $\alpha$ is undefined while
all the other angles are uniquely determined.

In the case $a=b=0$ and $\tilde{c}=c$, two singular values are degenerate with
$m_{1,2}=|c|$, and the unitary matrices $U_L$ and $U_R$ reduce to a single
unitary matrix $U$:
\begin{eqnarray}
  U \,=\, \left(\begin{array}{cc}
           1/\sqrt{2}  & -1/\sqrt{2} \\[2mm]
           1/\sqrt{2}  & \phantom{+}1/\sqrt{2}
           \end{array}\right) \cdot
     \left(\begin{array}{cc}
             1  & 0 \\[2mm]
             0  & i
           \end{array}\right)
  \,=\, \left(\begin{array}{cc}
           1/\sqrt{2}  & -i/\sqrt{2} \\[2mm]
           1/\sqrt{2}  & \phantom{+}i/\sqrt{2}
           \end{array}\right)\,,
\end{eqnarray}
corresponding to a $\pi/4$ rotation matrix and a phase matrix, which turns
the second eigenvalue positive.

\setcounter{equation}{0}
\setcounter{section}{2}
\def\thesection{\Alph{section}}
\renewcommand{\theequation}{{\rm \thesection.\arabic{equation}}}

\section*{Appendix B: The small-mixing approximation in the singular value
         decomposition}
\label{sec:small_mixing_svd}

In this second appendix, we provide details of the singular value decomposition
of the mass matrix{\footnote{The formalism applies also to general complex
matrices \cite{sychoi_haber}.}} in the small mixing approximation, in which the
coupling by two off-diagonal matrix blocks is weak and can be treated
perturbatively. For mathematical clarity, we present the solution for a general
$(N+M)\times (N+M)$ matrix in which the $N\times N$ and $M\times M$
submatrices are coupled weakly so that their mixing is small:
\begin{eqnarray} \label{matrixnm}
{\mathcal M}_{N+M} =
  \left(\begin{array}{cc}
  {\cal M}_N   &   X_{NM} \\[2mm]
   \widetilde{X}^T_{NM}    & {\cal M}_M
        \end{array}\right)\,.
\end{eqnarray}
To obtain the corresponding physical masses, we must
perform a singular value decomposition of
${\cal M}_{N+M}$:\footnote{In Eq.~(\ref{takaginm}), we use
primed subscripts to indicate that the corresponding  states
are continuously connected to the states of the unperturbed
block matrix, ${\rm diag}(\overline{\mathcal{M}}^D_N\,,\,
\overline{\mathcal{M}}^D_M)$, where the diagonal matrices
$\overline{\mathcal{M}}^D_N$ and $\overline{\mathcal{M}}^D_M$ are defined
in Eqs.~(\ref{dm1}) and (\ref{dm2}). \label{fn3}}
\begin{eqnarray} \label{takaginm}
L_{N+M}^T\, {\cal M}_{N+M}\, R_{N+M}  = {\rm
  diag}(m_{1'}\,,\,m_{2'}\,,\,\ldots\,,\, m_{N'+M'})\,,\qquad m_{k'}\geq 0\,,
\end{eqnarray}
where $L_{N+M}$ and $R_{N+M}$ are unitary.\footnote{When $N$ and $M$ are used in
subscripts of matrices, they refer to the dimension of the
corresponding square matrices.  For rectangular matrices, two
subscripts will be used.}
The non-negative diagonal elements $m_{k'}$ are called the singular values
of ${\M}_{N+M}$, which are defined as the non-negative square roots
of the eigenvalues of ${\M}_{N+M}^\dagger {\M}_{N+M}$ or, equivalently,
${\M}_{N+M} {{\M}^\dagger}_{N+M}$.

In Eq.$\,$(\ref{matrixnm}), ${\cal M}_N$ and ${\cal M}_M$ are $N\times N$ and
$M\times M$ symmetric submatrices
with singular values assumed to be generally larger than the matrix elements
of the $N\times M$ rectangular matrices, $X_{NM}$ and $\widetilde{X}_{NM}$.
In this case, one can treat the off-diagonal parts $X_{NM}$ and
$\widetilde{X}_{NM}$ as a perturbation as long as there are no accidental
near-degeneracies between the singular values of ${\cal M}_N$ and ${\cal M}_M$,
respectively.

\noindent {\bf (1)} In the first step, we separately perform
a singular value decomposition of ${\cal M}_N$ and
${\cal M}_M$:
\begin{eqnarray}
\overline{\cal M}^{D}_N &=& L^T_N {\cal M}_N R_N
= {\rm diag} (\overline{m}_{1'}, \ldots\, \overline{m}_{N'})\,, \label{dm1}\\[1mm]
\overline{\cal M}^{D}_M &=& L^T_M {\cal M}_M R_M
= {\rm diag} (\overline{m}_{N'+1'}, \dots, \overline{m}_{N'+M'})\,,\label{dm2}
\end{eqnarray}
where the diagonal elements $\overline m_{k'}$ are real and non-negative.
The ordering of the diagonal elements may conveniently be chosen according
to footnote \ref{fn3}.

Step {\bf (1)} results in a partial singular value decomposition of
${\cal M}_{N+M}$:
\begin{eqnarray}
\overline{\cal M}_{N+M}
\!\equiv\!
\left(\begin{array}{cc}
     L^T_N & \mathds{O} \\
     \mathds{O}^T &  L^T_M
\end{array}\right)\!
\left(\begin{array}{cc}
      {\cal M}_N           \! &\! X_{NM}  \\
      \widetilde{X}_{NM}^T \! &\! {\cal M}_M
      \end{array}\right)\!
\left(\begin{array}{cc}
      R_N          & \mathds{O} \\
      \mathds{O}^T & R_M
      \end{array}\right)
\!=\!
\left(\begin{array}{cc}
      \overline{\cal M}^{D}_{N}\! &\! L^T_N X_{NM} R_M \\
L^T_M \widetilde{X}_{NM}^T R_N \! &\! \overline{\cal M}^{D}_M
     \end{array}\right)
 \!\equiv\!
\left(\begin{array}{cc}
       \overline{\cal M}^{D}_{N} & Y_{NM} \\
          \widetilde{Y}_{NM}^T   & \overline{\cal M}^{D}_M
       \end{array}\right)\,,
\label{decal}
\end{eqnarray}
where $\mathds{O}$ is an $N\times M$ matrix of zeros.  The upper left and
lower right blocks of $\overline{\cal M}_{N+M}$ are diagonal with real
non-negative entries, but the upper right and lower left off-diagonal
blocks are non-zero.

\noindent {\bf (2)}
The ensuing $(N+M)\times (N+M)$ matrix, $\overline{\cal M}_{N+M}$, can be
subsequently block-diagonalized by performing an $(N+M)\times (N+M)$
singular value decomposition of $\overline{\cal M}_{N+M}$ in an approximate
expansion. Since the elements of the off-diagonal blocks in
$\overline{\cal M}_{N+M}$ are small compared to the diagonal elements
$\overline m_{k'}$, we may treat $Y_{NM}$ and $\widetilde{Y}_{NM}$ as
a perturbation.  More precisely, $Y_{NM}$ and $\widetilde{Y}_{NM}$
can be treated as a perturbation if
\begin{eqnarray} \label{pertcond}
\left|\frac{(Y_{NM})_{i'j'}}
{\overline m_{i'}-\overline m_{j'}}\right|\ll 1\quad \mbox{and}\quad
\left|\frac{(\widetilde{Y}_{NM})_{i'j'}}
{\overline m_{i'}-\overline m_{j'}}\right|\ll 1\,,
\end{eqnarray}
for all choices of $i'=1',\ldots,N'$ and $j'=N'+1'\,\ldots, N'+M'$.
These conditions, as can generally be anticipated, will naturally emerge
from the formalism below.

The perturbative block-diagonalization is accomplished by introducing two
$(N+M)\times (N+M)$ unitary matrices:
\begin{eqnarray}
&& {\cal L}_{N+M} \simeq
\left(\begin{array}{cc}
    \mathds{1}_{N\times N}
    - \frac{1}{2} \Omega_L \Omega^\dagger_L
  &  \Omega_L \\[1mm]
    -\Omega^\dagger_L & \mathds{1}_{M\times M}
   -\frac{1}{2} \Omega^\dagger_L \Omega_L
      \end{array} \right)\,,\\
&& {\cal R}_{N+M} \simeq
\left(\begin{array}{cc}
    \mathds{1}_{N\times N}
    - \frac{1}{2} \Omega_R \Omega^\dagger_R
  &  \Omega_R \\[1mm]
    -\Omega^\dagger_R & \mathds{1}_{M\times M}
   -\frac{1}{2} \Omega^\dagger_R \Omega_R
      \end{array} \right)\,,
\end{eqnarray}
where $\Omega_L$ and $\Omega_R$ are $N\times M$ complex matrices that vanish
when $X_{NM}$ and $\widetilde{X}_{NM}$ vanish and hence, like $X_{NM}$ and
$\widetilde{X}_{NM}$, are perturbatively small. Straightforward matrix
multiplication then yields:
\begin{eqnarray} \label{nbmnb}
{\cal L}^T_{N+M}
  \left(\begin{array}{cc} \overline{\cal M}_N^D & Y_{NM} \\
          \widetilde{Y}^T_{NM} & \overline{\cal M}_M^D\end{array}\right)
  {\cal R}_{N+M}
\,\approx\,
 \left(\begin{array}{cc}{\cal M}_N^{\prime\,D}  &
 Y_{NM} - \Omega^*_L\overline{\cal M}_M^D +\overline{\cal M}_N^D\Omega_R\\
 \widetilde{Y}^T_{NM}-\overline{\cal M}_M^D\Omega^\dagger_R
+\Omega^T_L\,\overline{\cal M}_N^D  &
 {\cal M}_M^{\prime\,D}\end{array}\right)\,,
\end{eqnarray}
where
\begin{eqnarray} \label{calbdef}
{\cal M}_N^{\prime\,D} &\equiv &
  \overline{\cal M}_N^D
  + \Omega^*_L \overline{\cal M}^D_M \Omega^\dagger_R
  -\Omega^*_L\widetilde{Y}^T_{NM}-Y_{NM}\Omega^\dagger_R
  -\onehalf \Omega^*_L\Omega^T_L \overline{\cal M}_N^D
  -\onehalf \overline{\cal M}_N^D \Omega_R\Omega^\dagger_R\,, \label{mnd}\\
{\cal M}_M^{\prime\,D} &\equiv &
 \overline{\cal M}_M^D+ \Omega^T_L \overline{\cal M}^D_N \Omega_R
  +\Omega^T_L Y_{NM} + \widetilde{Y}^T_{NM}\Omega_R
 -\onehalf \Omega^T_L\Omega^*_L \overline{\cal M}_M^D
 -\onehalf \overline{\cal M}_M^D \Omega^\dagger_R\Omega_R\,,\label{mmd}
 \end{eqnarray}
The block-diagonalization is achieved by demanding that
\begin{eqnarray}
&& Y_{NM}=\Omega^*_L \overline{\cal M}_M^D-\overline{\cal M}_N^D\Omega_R\,,
\label{calbeq1}\,,\\
&& \widetilde{Y}_{NM}=\Omega^*_R \overline{\cal M}_M^D
     -\overline{\cal M}_N^D\Omega_L\,. \label{calbeq2}
\end{eqnarray}
Inserting these relations in Eqs.$\,$(\ref{mnd}) and (\ref{mmd}) and
eliminating $Y_{NM}$ and $\widetilde{Y}_{NM}$, we obtain:
\begin{eqnarray}
{\cal M}_N^{\prime\,D} &= &
  \overline{\cal M}_N^D
  +\onehalf \overline{\cal M}_N^D\Omega_R\Omega^\dagger_R
  +\onehalf \Omega^*_L\Omega^T_L \overline{\cal M}_N^D
  -\Omega^*_L\overline{\cal M}_M^D \Omega^\dagger_R\,, \label{mnd2}\\[1mm]
 {\cal M}_M^{\prime\,D} &= &
 \overline{\cal M}_M^D
  +\onehalf \overline{\cal M}_M^D\Omega^\dagger_R\Omega_R
  +\onehalf \Omega^T_L\Omega^*_L \overline{\cal M}_M^D
  -\Omega^T_L\overline{\cal M}_N^D \Omega_R\,.\label{mmd2}
 \end{eqnarray}

The results above simplify somewhat when we recall that $\overline{\cal M}_N^D$
and $\overline{\cal M}_M^D$ are diagonal matrices [see Eq.$\,$(\ref{dm1}) and
(\ref{dm2})].  Combining the matrix elements of Eqs.$\,$(\ref{calbeq1}) and
(\ref{calbeq2}) yields two equations for the elements of $\Omega_{L}$ and
$\Omega_{R}$:
\begin{eqnarray}
\label{omegab}
&& \Omega_{Li'j'}\equiv \frac{1}{\overline{m}^2_{j'}-\overline{m}^2_{i'}}
     \left[\overline{m}_{i'} \widetilde{Y}_{NMi'j'}
          +Y^*_{NMi'j'}\overline{m}_{j'}\right]\,,\\
&& \Omega_{Ri'j'}\equiv \frac{1}{\overline{m}^2_{j'}-\overline{m}^2_{i'}}
     \left[\overline{m}_{i'} Y_{NMi'j'}
          +\widetilde{Y}^*_{NMi'j'}\overline{m}_{j'}\right]\,,
\end{eqnarray}
with $i'=1',\ldots,N'$ and $j'=N'+1'\,\ldots,N'+M'$.
Since the elements $\Omega_{Li'j'}$ and $\Omega_{Ri'j'}$ are the small parameters
of the perturbation expansion, the
perturbativity conditions previously given in Eq.$\,$(\ref{pertcond})
arise naturally.

At this stage, the result of the perturbative block diagonalization is:
\begin{eqnarray} \label{stage}
{\cal L}^T_{N+M}
\left(\begin{array}{cc} \overline{\cal M}_N^D & Y_{NM} \\
      \widetilde{Y}^T_{NM} &  \overline{\cal M}_M^D\end{array}\right)
{\cal R}_{N+M}
=
\left(\begin{array}{cc}
{\cal M}_N^{\prime\,D} & \mathds{O} \\
\mathds{O} & {\cal M}_M^{\prime\,D}
      \end{array}\right)\,,
\end{eqnarray}
up to third order in $\Omega$ in the off-diagonal blocks.
The ${\cal O}(\Omega^3)$ terms can be neglected consistently.
Also the re-diagonalization of the two diagonal blocks can be
omitted.
Though the off-diagonal elements of ${\cal M}_N^{\prime\,D}$ and
${\cal M}_M^{\prime\,D}$ are of ${\cal O}(\Omega^2)$, they
only effect, in the singular value decomposition, the corresponding
diagonal elements at ${\cal O}(\Omega^4)$, which we neglect in this
analysis. However, the diagonal elements of ${\cal M}_N^{\prime\,D}$ and
${\cal M}_M^{\prime\,D}$ also contain terms of ${\cal O}(\Omega^2)$,
which generate second-order shifts of the diagonal elements relative
to the $\overline m_{k'}$ obtained at step {\bf (1)}.  These
corrections are easily
obtained from the diagonal matrix elements of Eqs.$\,$(\ref{mnd2}) and
(\ref{mmd2}) after making use of Eq.$\,$(\ref{omegab}):
\begin{eqnarray}
m_{i'} &\simeq& \overline m_{i'}+\frac{1}{2}\sum_{j'=N'+1'}^{N'+M'}
\left\{\frac{\overline{m}_{i'}(|Y_{NMi'j'}|^2+|\widetilde{Y}_{NMi'j'}|^2)}{
       \overline{m}^2_{i'}-\overline{m}^2_{j'}}
     +2\frac{\overline{m}_{j'} Y_{NMi'j'}\widetilde{Y}_{NMi'j'}}{
       \overline{m}^2_{i'}-\overline{m}^2_{j'}}\right\}\,,
       \label{complexmi}\\[8pt]
m_{j'} &\simeq& \overline{m}_{j'}-\frac{1}{2}\sum_{i'=1'}^{N'}
\left\{\frac{\overline{m}_{j'}(|Y_{NMi'j'}|^2+|\widetilde{Y}_{NMi'j'}|^2)}{
       \overline{m}^2_{i'}-\overline{m}^2_{j'}}
     +2\frac{\overline{m}_{i'} Y_{NMi'j'}\widetilde{Y}_{NMi'j'}}{
       \overline{m}^2_{i'}-\overline{m}^2_{j'}}\right\}\,,\label{complexmj}
\end{eqnarray}
with $i'=1',..,N'$ and $j'=N'+1',..,N'+M'$. The shifted mass parameters
correspond to the physical mass values if the original mass matrix is
real.

\noindent
{\bf (3)} However, for complex mass matrices the shifted mass parameters
would in general be complex. These phases can be removed by substituting
$\mathcal{L} \to \mathcal{L} \mathcal{P}$ and $\mathcal{R} \to \mathcal{R}
\mathcal{P}$ with properly chosen phases
\begin{eqnarray}
{\cal P}
={\rm diag}(e^{-i\alpha_{1'}}\,,\,\ldots\,,\,e^{-i\alpha_{N'+M'}})\,.
\end{eqnarray}
Starting from Eqs.$\,$(\ref{complexmi}) and (\ref{complexmj}), one can evaluate
${\cal P}$ to second order in the perturbation $\Omega_{L,R}$. In particular,
for $\epsilon_{1,2}\ll a$, we have $a+\epsilon_1+i\epsilon_2\simeq
(a+\epsilon_1)e^{i\epsilon_2/a}$. From this result, we easily derive the
second-order expressions for the physical masses by just substituting
\begin{equation}
Y\tilde{Y} \to \Re{\rm e}{(Y\tilde{Y})} \,,
\end{equation}
while the phases are given by the imaginary part of $Y\tilde{Y}$,
\begin{eqnarray}
&& \alpha_{i'} \simeq \, \frac{1}{2} \sum_{j'=N'+1'}^{N'+M'}\,
     \frac{\overline{m}_{j'}}{
      \overline{m}_{i'}(\overline{m}_{i'}^2 -\overline{m}_{j'}^2)}
      \Im {\rm m}\left(Y_{NMi'j'}\widetilde{Y}_{NMi'j'}\right)\,,\\
&& \alpha_{j'} \simeq \, -\frac{1}{2}\, \sum_{i'=1'}^{N'}\,\,
     \frac{\overline{m}_{i'}}{
      \overline{m}_{j'}(\overline{m}_{i'}^2 -\overline{m}_{j'}^2)}
      \,\,\Im {\rm m}\left(Y_{NMi'j'}\widetilde{Y}_{NMi'j'}\right)\,,
\end{eqnarray}
with $i'=1',..,N'$ and $j'=N'+1',..,N'+M'$.

This completes the perturbative singular value decomposition of the mass
matrix for $N$-dimensional and $M$-dimensional subsystems of fermions weakly
coupled by an off-diagonal perturbation. Thus the physical masses
and the elements of the mixing matrices can be derived from the parameters
of the $N \times N$ and $M \times M$ subsystems
and the weak couplings $X_{NM},{\tilde{X}}_{NM}$ of the subsystems
[rotated to $Y_{NM},{\tilde{Y}}_{NM}$ finally].

As noted in Eq.$\,$(\ref{pertcond}),
the perturbation theory breaks down if any mass $\overline m_{i'}$ from the
$N$-dimensional subsystem is nearly degenerate with a corresponding mass
$\overline m_{j'}$ from the $M$-dimensional subsystem (if the
corresponding residues do not vanish). In this case the formalism developed
in Ref.~\cite{Haber} can be adopted to calculate the physical masses also in
the cross-over zones analytically.

\setcounter{equation}{0}
\setcounter{section}{3}
\def\thesection{\Alph{section}}
\renewcommand{\theequation}{{\rm \thesection.\arabic{equation}}}
\section*{Appendix C: Chargino, neutralino and scalar masses and mixing
                      elements by block-diagonalization}
\label{sec:appendix_c}

When the weak couplings among the gaugino and higgsino sectors are
switched on the mass eigenvalues and mixing parameters are
calculated using the block-diagonalization method adopting the
formulae in the preceding appendices.

\subsubsection*{Charginos}
Using the short-hand notation $m_{\tilde\chi^\pm_i}=m^\pm_i$, the chargino
mass eigenvalues are given approximately by
\begin{eqnarray}
m^\pm_1 &=& \overline{m}^\pm_1 +\frac{v^2}{2((\overline{m}^\pm_1)^2-\mu_c^2)}
        \left\{ \overline{m}^\pm_1\, \left[
                (\lambda_I v_u c_+ + g v_d\epsilon_+ s_+/\sqrt{2})^2
               +(\lambda_I v_d c_- - g v_u\epsilon_- s_-/\sqrt{2})^2\right]
               \right.\nonumber\\
        && \left. \hskip 3.8cm -2 \mu_c \alpha^2_c
                 (\lambda_I v_u c_+ + g v_d\epsilon_+ s_+/\sqrt{2})
                 (\lambda_I v_d c_- - g v_u\epsilon_- s_-/\sqrt{2})\right\}\,,
\end{eqnarray}
\begin{eqnarray}
m^\pm_2 &=& \overline{m}^\pm_2 +\frac{v^2}{2((\overline{m}^\pm_2)^2-\mu_c^2)}
        \left\{ \overline{m}^\pm_2\, \left[
                (\lambda_I v_u \epsilon_+ s_+ - g v_d c_+/\sqrt{2})^2
               +(\lambda_I v_d \epsilon_- s_- + g v_u c_-/\sqrt{2})^2\right]
               \right.\nonumber\\
        && \left. \hskip 3.8cm -2 \mu_c \beta^2_c
                 (\lambda_I v_u \epsilon_+ s_+ - g v_d c_+/\sqrt{2})
                 (\lambda_I v_d \epsilon_- s_- + g v_u c_-/\sqrt{2})\right\}
                 \,,
\end{eqnarray}
\begin{eqnarray}
m^\pm_3 &=& \mu_c - \frac{v^2}{2((\overline{m}^\pm_1)^2-\mu_c^2)}
       \left\{ \mu_c \left[
                (\lambda_I v_u c_+ + g v_d\epsilon_+ s_+/\sqrt{2})^2
               +(\lambda_I v_d c_- - g v_u\epsilon_- s_-/\sqrt{2})^2\right]
               \right.\nonumber\\
        && \left. \hskip 3.2cm -2 \overline{m}^\pm_1\,\alpha^2_c
                 (\lambda_I v_u c_+ + g v_d\epsilon_+ s_+/\sqrt{2})
                 (\lambda_I v_d c_- - g v_u\epsilon_- s_-/\sqrt{2})\right\}
                 \nonumber\\
        && \hskip 0.5 cm -\frac{v^2}{2((\overline{m}^\pm_2)^2-\mu^2)}
        \left\{ \mu_c \left[
                (\lambda_I v_u \epsilon_+ s_+ - g v_d c_+/\sqrt{2})^2
               +(\lambda_I v_d \epsilon_- s_- + g v_u c_-/\sqrt{2})^2\right]
               \right.\nonumber\\
        && \left. \hskip 3.2cm -2 \overline{m}^\pm_2\,\beta^2_c
                 (\lambda_I v_u \epsilon_+ s_+ - g v_d c_+/\sqrt{2})
                 (\lambda_I v_d \epsilon_- s_- - g v_u c_-/\sqrt{2})\right\}
                 \,,
\end{eqnarray}
where the signs $\epsilon_\pm$ and the phases $\alpha_\pm$
and $\beta_\pm$ are defined by
\begin{eqnarray}
\epsilon_\pm  &=& {\rm sign}[ (M'_2+M_2) M^D_2 \pm  g (M'_2-M_2) v_I]\,,
\label{eq:epspm} \\
\alpha_c    &=& \sqrt{{\rm sign}[ M'_2 (\sigma_+\sigma_- - M'^2_2+M^2_2)
                 -2 (M'_2+M_2) (M^D_2)^2 - 2 g^2 (M'_2-M_2) v_I]}\,,
\label{eq:phase_alpha_c}\\
\beta_c    &=& \sqrt{{\rm sign}[ M_2 (\sigma_+\sigma_- - M'^2_2+M^2_2)
                 +2 (M'_2+M_2) (M^D_2)^2 - 2 g^2 (M'_2-M_2) v_I]}\,,
\label{eq:phase_beta_c}
\end{eqnarray}
and the abbreviations $c_\pm =\cos\theta_\pm$ and $s_\pm=\sin\theta_\pm$,
and $c_{2\beta}=\cos2\beta$, $s_{2\beta}=\sin2\beta$ have been adopted.

\begin{figure}[t]
\epsfig{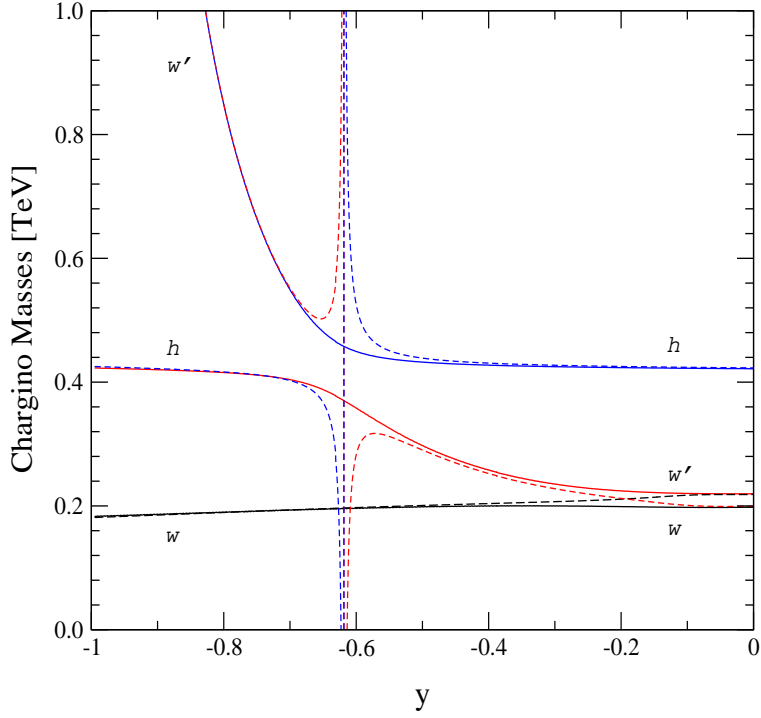}
\caption{\it Evolution of the approximate versus exact chargino masses as
             a function of the control parameter $y$ from the MSSM doublet
             ($y=-1$) to the Dirac ($y=0$) triplet along the path ${\cal P}_C$
             in Eq.$\,$(\ref{eq:chargino_path}) for $m=200$~GeV and
             $\tan\beta=5$. The singularity in the cross-over zone
             can be removed by using the specific formalism for degenerate
             states as developed in Ref.~\cite{Haber}.}
\label{fig:chaapp}
\end{figure}

The gaugino/higgsino mixing elements read in this approximation
\begin{eqnarray}
\begin{array}{lll}
U_{\pm 11} = \phantom{+}\alpha_c c_\pm \,,                & \quad
U_{\pm 12} = \phantom{+}\beta_c \epsilon_\pm s_\pm\,,     &\quad
U_{\pm 13} = \phantom{+}\alpha_c c_\pm \Omega_{\pm 13}
            +\beta_c \epsilon_\pm s_\pm \Omega_{\pm 23} \,,\\[1mm]
U_{\pm 21} =-\alpha_c \epsilon_\pm s_\pm \,,   &\quad
U_{\pm 22} = \phantom{+}\beta_c c_\pm\,,                  &\quad
U_{\pm 23} =-\alpha_c \epsilon_\pm s_\pm \Omega_{\pm 13}
            + \beta_c c_\pm \Omega_{\pm 23}\,,\\[1mm]
U_{\pm 31} =-\Omega^*_{\pm 13}\,,              &\quad
U_{\pm 32} =-\Omega^*_{\pm 23}\,,              &\quad
U_{\pm 33} = \phantom{+}1\,,
\end{array}
\end{eqnarray}
The matrix elements of the rectangular $\Omega_\pm$
matrices, which block-diagonalize the mass matrix are as follows
\begin{eqnarray}
\Omega_{+13}
 &=& \,\,\,\left[\overline{m}^\pm_1 \alpha_c
     (\lambda_I v_d c_- - g v_u \epsilon_- s_-/\sqrt{2})
    -\mu_c \alpha^*_c
     (\lambda_I v_u c_+ + g v_d \epsilon_+ s_+/\sqrt{2})\right]/
         (\mu_c^2-(\overline{m}^\pm_1)^2)\,,
     \nonumber\\[1mm]
\Omega_{+23}
&=& \,\,\,\left[\overline{m}^\pm_2 \beta_c
     (\lambda_I v_d \epsilon_- s_- + g v_u c_-/\sqrt{2})
    -\mu_c \beta^*_c
     (\lambda_I v_u \epsilon_+ s_+ - g v_d c_+/\sqrt{2})\right]/
         (\mu_c^2-(\overline{m}^\pm_2)^2)\,,
\end{eqnarray}
and
\begin{eqnarray}
\Omega_{-13}
 &=& \, -\left[\overline{m}^\pm_1 \alpha_c
     (\lambda_I v_u c_+ + g v_d \epsilon_+ s_+/\sqrt{2})
    -\mu_c \alpha^*_c
     (\lambda_I v_d c_- - g v_u \epsilon_- s_-/\sqrt{2})\right]/
         (\mu_c^2-(\overline{m}^\pm_1)^2)\,,
     \nonumber\\[1mm]
\Omega_{-23}
&=& \, -\left[\overline{m}^\pm_2 \beta_c
     (\lambda_I v_u \epsilon_+ s_+ - g v_d c_+/\sqrt{2})
    -\mu_c \beta^*_c
     (\lambda_I v_d \epsilon_- s_- + g v_u c_-/\sqrt{2})\right]/
         (\mu_c^2-(\overline{m}^\pm_2)^2)\,,
\end{eqnarray}
where the signs $\epsilon_{\pm}$ and the phase factors $\alpha_c$
and $\beta_c$ are defined in Eqs.$\,$(\ref{eq:epspm}) to (\ref{eq:phase_beta_c}).

As a numerical check for the analytic expansion we show
in Fig.$\,$\ref{fig:chaapp} the evolution of the approximate versus
exact chargino masses as a function of the control parameter $y$ from
the MSSM doublet ($y=-1$) to the Dirac ($y=0$) triplet along the path
${\cal P}_C$ in Eq.$\,$(\ref{eq:chargino_path}) for the same
parameter set used in Fig.$\,$\ref{fig:chargino}. The descending order of
the physical masses in the figure reflects, in obvious notation, the pattern
$w' \gg h > w$ in the MSSM limit. When the states $w'$ and $h$ become
degenerate near $y=-0.6$, the standard analytical expansion cannot be
applied any more. In this situation the mass spectrum must either be
obtained numerically or analytical expansions tailored specifically
for cross-over phenomena, see Ref.$\,$\cite{Haber}. On the other hand,
the ordering $h> w'> w$ is kept until the Dirac limit is reached. The
level-crossing phenomenon near $y=-0.2$ is due to the mixing between the
gaugino and higgsino sectors but not due to the $w'$-$w$ cross-over, as
$\overline{m}^\pm_2 > \overline{m}^\pm_1$ along the entire path.

\subsubsection*{Neutralinos}

With $m_{\tilde{\chi}^0_i}=m^0_i$ the neutralino mass eigenvalues
are given by
\begin{align}
&\hspace{-6em} \text{\uline{bino sector} :}    \nonumber \\[1mm]
m^0_1 = \overline{m}^0_1
      &-\frac{1}{4(\overline{m}^0_1+\mu)}
        \left[g' v_- s_1/\sqrt{2} - \lambda_Y v_+ c_1\right]^2
       -\frac{1}{4(\overline{m}^0_1-\mu)}
        \left[g' v_+ s_1/\sqrt{2} + \lambda_Y v_- c_1\right]^2 \,,
        \\[1mm]
m^0_2 = \overline{m}^0_2
      &-\frac{1}{4(\overline{m}^0_2-\mu)}
        \left[g' v_- c_1/\sqrt{2} + \lambda_Y v_+ s_1\right]^2
       -\frac{1}{4(\overline{m}^0_2+\mu)}
        \left[g' v_+ c_1/\sqrt{2} - \lambda_Y v_- s_1\right]^2 \,,
\end{align}
\begin{align}
&\hspace{-6em} \text{\uline{wino sector} :}     \nonumber \\[1mm]
m^0_3 = \overline{m}^0_3
      &-\frac{1}{4(\overline{m}^0_3+\mu)}
        \left[g\, v_- s_2/\sqrt{2} + \lambda_I\, v_+ c_2\right]^2
       -\frac{1}{4(\overline{m}^0_3-\mu)}
        \left[g\, v_+ s_2/\sqrt{2} - \lambda_I\, v_- c_2\right]^2 \,,
        \\
m^0_4 = \overline{m}^0_4
      &-\frac{1}{4(\overline{m}^0_4-\mu)}
        \left[g\, v_+ c_2/\sqrt{2} - \lambda_I\, v_+ s_2\right]^2
       -\frac{1}{4(\overline{m}^0_4+\mu)}
        \left[g\, v_+ c_2/\sqrt{2} + \lambda_I\, v_- s_2\right]^2 \,,
\end{align}
\begin{align}
&\hspace{-6em} \text{\uline{higgsino sector} :}   \nonumber \\
m^0_5 = \mu_n
      &+\frac{1}{4(\overline{m}^0_1+\mu)}
        \left[g' v_- s_1/\sqrt{2} - \lambda_Y v_+ c_1\right]^2
       -\frac{1}{4(\overline{m}^0_2-\mu)}
        \left[g' v_- c_1/\sqrt{2} + \lambda_Y v_+ s_1\right]^2
        \nonumber\\
      &+\frac{1}{4(\overline{m}_3+\mu)}
        \left[g v_+ s_2/\sqrt{2}\, +\, \lambda_I v_+ c_2\right]^2
       -\,\frac{1}{4(\overline{m}_4-\mu)}
        \left[g v_- c_2/\sqrt{2}\, -\, \lambda_I v_+ s_2\right]^2\,,
        \\
m^0_6 = \mu_n
      &-\frac{1}{4(\overline{m}^0_1-\mu)}
        \left[g' v_+ s_1/\sqrt{2} + \lambda_Y v_- c_1\right]^2
       +\frac{1}{4(\overline{m}^0_2+\mu)}
        \left[g' v_+ c_1/\sqrt{2} - \lambda_Y v_- s_1\right]^2
        \nonumber\\
      &-\frac{1}{4(\overline{m}^0_3-\mu)}
        \left[g v_{ud}^+ s_2/\sqrt{2} - \lambda_I v_{ud}^- c_2\right]^2
       +\frac{1}{4(\overline{m}^0_4+\mu)}
        \left[g v_+ c_2/\sqrt{2}\, +\, \lambda_I\, v_- s_2\right]^2\,,
\end{align}
where $v_\pm = v_u \pm v_d$.

The final $6\times 6$ diagonalization matrix reads approximately:
\begin{eqnarray}
U_N = \overline{U}_N\,
        \left(\begin{array}{cc}
            \mathds{1}_{4\times 4} &  \Omega_{4\times 2} \\
            -\Omega^\dagger_{4\times 2} & \mathds{1}_{2\times 2}
             \end{array}\right)\,,
\end{eqnarray}
where the elements of the rectangular matrix $\Omega_{4\times 2}$
are as follows:
\begin{eqnarray}
\text{\uline{bino sector} :} \;\; \Omega_{15}
 &=& \frac{-i}{2(\overline{m}^0_1+\mu)} \left[ g' v_- s_1/\sqrt{2}
            -\lambda_Y v_+ c_1\right]\,, \quad\,\,
\Omega_{16}
 = \frac{1}{2(\overline{m}^0_1-\mu)} \left[ g' v_+ s_1/\sqrt{2}
            +\lambda_Y v_- c_1\right]\,,
     \nonumber\\[1mm]
\Omega_{25}
  &=& \frac{1}{2(\overline{m}^0_2-\mu)} \left[ g' v_- c_1/\sqrt{2}
            -\lambda_Y v_- s_1\right]\,, \quad \,\,
\Omega_{26}
  = \frac{i}{2(\overline{m}^0_2+\mu)} \left[ g' v_+ c_1/\sqrt{2}
            -\lambda_Y v_- s_1\right]\,,
\quad
\end{eqnarray}
and
\begin{eqnarray}
\text{\uline{wino sector} :} \;\; \Omega_{35}
  &=& \frac{i}{2(\overline{m}^0_3+\mu)} \left[ g v_- s_2/\sqrt{2}
            +\lambda_I v_+ c_2\right]\,, \quad
\Omega_{36}
  = \frac{1}{2(\overline{m}^0_3-\mu)} \left[ g v_+ s_2/\sqrt{2}
            -\lambda_I v_- c_2\right]\,,
     \nonumber\\[1mm]
\Omega_{45}
  &=& \frac{-1}{2(\overline{m}^0_4-\mu)} \left[ g v_- c_2/\sqrt{2}
            -\lambda_I v_+ s_2\right]\,, \quad
\Omega_{46}
  = \frac{i}{2(\overline{m}^0_4+\mu)} \left[ g v_+ c_2/\sqrt{2}
            +\lambda_I v_- s_2\right]\,,
\end{eqnarray}
with abbreviations as before.

\subsubsection*{Scalar/Higgs Particles}

The block-diagonalization of the Higgs/scalar mass matrix when the
weak coupling between the Higgs and the sigma fields is included
gives the following results.

\noindent {\it (i) \uline{neutral pseudoscalars:}}

\noindent The physical pseudoscalar masses are given approximately
by
\begin{eqnarray}
M^2_{A_1} &=&  M^2_A
          -\frac{(M_Y-A_Y)^2 }{2(\tilde{m}'^2_Y-M^2_A)} \lambda^2_Y v^2
          -\frac{(M_I-A_I)^2 }{2(\tilde{m}'^2_I-M^2_A)} \lambda^2_I v^2\,,
          \\
M^2_{A_2} &=&  \tilde{m}^{\prime 2}_Y
          +\frac{(M_Y-A_Y)^2 }{2(\tilde{m}'^2_Y-M^2_A)} \lambda^2_Y v^2\,,
          \\
M^2_{A_3} &=&  \tilde{m}^{\prime 2}_I
          +\frac{(M_I-A_I)^2 }{2(\tilde{m}'^2_I-M^2_A)} \lambda^2_I v^2\,,
\end{eqnarray}
up to the order of $v^2/m^2_{I,Y}$,
while the physical pseudoscalar states are mixed according to the
relation $O^T_P\, {\cal M}^2_P\, O_P ={\rm diag} (M^2_{A_1},
M^2_{A_2}, M^2_{A_3})$ with the mixing elements given approximately
by
\begin{eqnarray}
&& O_{P11}=O_{P22}=O_{P33}= 1\,,\qquad
   O_{P12}=-O_{P21}
          =-\frac{(M_Y-A_Y)\lambda_Y v}{\sqrt{2}(\tilde{m}'^2_Y-M^2_A)}\,,
 \nonumber\\
&& O_{P13}=-O_{P31}=-\frac{(M_I-A_I)\lambda_I
v}{\sqrt{2}(\tilde{m}'^2_I-M^2_A)}\,,
    \quad
   O_{P23}=-O_{P32}
          =\frac{\lambda_Y \lambda_I v^2}{2(\tilde{m}'^2_I-\tilde{m}'^2_Y)}
    \,,
\end{eqnarray}
up to the order of $v/m_{I,Y}$.

\noindent {\it (ii) \uline{neutral scalars:}}

\noindent The block diagonalization is described by the $2\times 2$
matrix $\Omega_S$ with its elements
\begin{eqnarray}
\Omega_{S13} &=& -\frac{(2\tilde{m}^2_Y-\lambda^2_Y v^2) v_Y}{
                        (\tilde{m}^2_Y - m^2_Z) v} \,,\ \
\Omega_{S23} = \frac{\Delta_Y}{\tilde{m}^2_Y - M^2_A}\,,\nonumber\\
\Omega_{S14} &=& -\frac{(2\tilde{m}^2_I -\lambda^2_I\, v^2) v_I}{
                        (\tilde{m}^2_I - m^2_Z) v} \,,\ \
\Omega_{S24} = \frac{\Delta_I}{\tilde{m}^2_I - M^2_A}\,,
\end{eqnarray}
up to the order of $v/M_A, v/m_{Y,I}$.
Performing these subsequent transformations gives rise to the four
physical masses
\begin{eqnarray}
M^2_{S_1} &=& m^2_Z + \delta_H s_{2\beta} +\epsilon_H
                    -\frac{(\delta_H c_{2\beta} + \epsilon_H/t_\beta)^2}{
                                M^2_A- m^2_Z}
                    -\frac{(2\tilde{m}^2_Y-\lambda^2_Y v^2)^2 v^2_Y}{
                           (\tilde{m}^2_Y-m^2_Z) v^2}
                    -\frac{(2\tilde{m}^2_I-\lambda^2_I v^2)^2 v^2_I}{
                           (\tilde{m}^2_I-m^2_Z) v^2}\,, \\
M^2_{S_2} &=& M^2_A - \delta_H s_{2\beta} +\epsilon_H/t^2_\beta
                    +\frac{(\delta_H c_{2\beta} + \epsilon_H/t_\beta)^2}{
                                M^2_A- m^2_Z}
                    -\frac{\Delta^2_Y}{\tilde{m}^2_Y-M^2_A}
                    -\frac{\Delta^2_I}{\tilde{m}^2_I-M^2_A}\,,  \\
M^2_{S_3} &=& \tilde{m}^2_Y
    +\frac{(2 \tilde{m}^2_Y-\lambda^2_Y v^2)^2v_Y^2}{
           (\tilde{m}^2_Y-m^2_Z) v^2}
    + \frac{\Delta^2_Y}{\tilde{m}^2_Y-M^2_A}\,, \\
M^2_{S_4} &=& \tilde{m}^2_I
    +\frac{(2 \tilde{m}^2_I-\lambda^2_I v^2)^2v_I^2}{
           (\tilde{m}^2_I-m^2_Z) v^2}
    + \frac{\Delta^2_I}{\tilde{m}^2_I-M^2_A}\,,
\end{eqnarray}
up to the order of $v^2/M^2_A, v^2/m^2_{Y,I}$,
and the $4\times 4$ mixing matrix ${\cal O}_S$, connecting current
with mass eigenstates as ${\cal O}^T_S {\cal M}^2_S {\cal O}_S ={\rm
diag}(M^2_{S_1},\cdots,M^2_{S_4})$, with its elements:
\begin{eqnarray}
&& {\cal O}_{S11} = {\cal O}_{S22}={\cal O}_{S33}
                  = {\cal O}_{S44} =1\,,
    \qquad \ \
\, {\cal O}_{S12} = - {\cal O}_{S21} = s_h\,,\quad
   {\cal O}_{S34} = - {\cal O}_{S43} = 0\,,\nonumber\\[2mm]
&& {\cal O}_{S13} = -{\cal O}_{S31} =
-\frac{(2\tilde{m}^2_Y-\lambda^2_Y v^2) v_Y}{
                        (\tilde{m}^2_Y - m^2_Z) v}\,,\quad
\, {\cal O}_{S23} = -{\cal O}_{32} = \frac{\Delta_Y}{\tilde{m}^2_Y -
M^2_A}\,,
                \nonumber\\
&& {\cal O}_{S14} = -{\cal O}_{S41}
                  = -\frac{(2\tilde{m}^2_I\,-\,\lambda^2_I v^2)\, v_I}{
                        (\tilde{m}^2_I\, - m^2_Z) v}\,,\quad\ \
   {\cal O}_{S24} = -{\cal O}_{42}
                  = \frac{\Delta_I}{\tilde{m}^2_I - M^2_A}\,,
\end{eqnarray}
up to the order of $v/M_A, v/m_{Y,I}$
with the abbreviation $s_h = \sin\theta_h$.

\noindent {\it (iii) \uline{charged scalars:}}

\noindent
In the weak coupling limit the charged $H^\pm$ and
$s^\pm_{1,2}$ states are mixed by the $3\times 3$ matrix ${\cal
O}^\pm_S$ with components
\begin{eqnarray}
&& {\cal O}^\pm_{S11} = {\cal O}^\pm_{S22} = {\cal O}^\pm_{S33} = 1\,,
   \qquad
   {\cal O}^\pm_{S12} = -{\cal O}^\pm_{S21}
                      = \Delta_{1\pm}/(\tilde{m}'^2_I- M^2_A)\,,
    \nonumber\\[1mm]
&&  {\cal O}^\pm_{S13} = -{\cal O}^\pm_{S31}
                      = \Delta_{2\pm}/(\tilde{m}^2_I-M^2_A)\,,
   \qquad
   {\cal O}^\pm_{S23} = -{\cal O}^\pm_{S32} = 0\,,
\end{eqnarray}
up to the order of $v/M_A, v/m_{Y,I}$ to generate the physical charged
scalar masses
\begin{eqnarray}
M^2_{S^\pm_1} &=& \tilde{M}^2_{H^\pm}
                 -\Delta^2_{1\pm}/(\tilde{m}'^2_I-M^2_A)
                 -\Delta^2_{2\pm}/(\tilde{m}^2_I-M^2_A)\,,\\
M^2_{S^\pm_2} &=& \tilde{m}'^2_I + g^2 v^2_I
                   c^2_{2\beta}/4(\tilde{m}^2_I-\tilde{m}'^2_I)
                 + \Delta^2_{1\pm}/(\tilde{m}'^2_I-M^2_A)\,,\\
M^2_{S^\pm_3} &=& \rho \tilde{m}^2_I
                   c^2_{2\beta}/4(\tilde{m}^2_I-\tilde{m}'^2_I)
                 + \Delta^2_{2\pm}/(\tilde{m}^2_I-M^2_A)\,,
\end{eqnarray}
up to the order of $v^2/M^2_A, v^2/m^2_{Y,I}$.


\newpage
%

\end{document}